\documentclass[10pt]{article}

\usepackage[toc,page]{appendix}

\usepackage[utf8]{inputenc}

\usepackage[T1]{fontenc}

\input{glyphtounicode}
\usepackage[hyphens]{url}

\usepackage[top=1in, bottom=1.2in, left=1in, right=1in]{geometry}

\usepackage[tracking=true,kerning=true,protrusion=true,expansion=true,final]{microtype}
\AtBeginDocument{\frenchspacing}

\usepackage[dvipsnames]{xcolor}

\usepackage{enumitem}

\setlist[itemize,1]{leftmargin=\parindent,label={\upshape\textbullet}} 
\setlist[itemize,2]{leftmargin=2\parindent,label={\upshape\textendash}} 

\setlist[enumerate,1]{leftmargin=\parindent,label={\upshape\alph*.}} 
\setlist[enumerate,2]{leftmargin=2\parindent,label=\color{gray}{\upshape\roman*.}} 

\AtBeginDocument{\allowdisplaybreaks[1]}

\usepackage{graphicx}

\graphicspath{{./img/},{.}}

\usepackage{siunitx}
\usepackage{mathtools}
\usepackage{amsthm}
\usepackage{thmtools}
\usepackage{amssymb}
\usepackage[font=small,labelfont=bf,labelsep=space]{caption}
\usepackage[font=scriptsize]{subcaption}

\usepackage{enumerate}

\usepackage{newtxtext,newtxmath}

\usepackage[unicode=true,pdfusetitle, bookmarks=true,bookmarksnumbered=false,bookmarksopen=false, breaklinks=false,hidelinks,colorlinks=true,hypertexnames=false]{hyperref}
\hypersetup{colorlinks=true, linkcolor=MidnightBlue,citecolor=MidnightBlue, urlcolor=MidnightBlue}

\usepackage{booktabs}
\usepackage{tabularx}

\usepackage[capitalise]{cleveref}
\crefname{section}{Sec.}{Secs.}
\crefname{subsection}{Sec.}{Secs.}
\crefname{subsubsection}{Sec.}{Secs.}
\crefname{equation}{Eq.}{Eqs.}
\crefname{figure}{Fig.}{Figs.}
\crefname{table}{Table}{Tables}
\crefname{appendix}{App.}{Apps.}

\makeatletter
\AddToHook{cmd/appendix/before}{\def\cref@section@alias{appendix}\def\cref@subsection@alias{appendix}}
\makeatother

\usepackage{etoolbox}

\usepackage[small]{titlesec}

\titleformat*{\section}{\centering\bfseries}

\titleformat*{\subsection}{\bfseries}

\titleformat*{\subsubsection}{\itshape}

\titleformat{\paragraph}[runin]{\itshape}{}{}{}[.]
\titlespacing*{\paragraph}{0pt}{3.25ex plus 1ex minus .2ex}{0.25em}

\titlelabel{\thetitle\quad}

\DeclarePairedDelimiterX\abs[1]{\lvert}{\rvert}{
    \ifblank{#1}{\:\cdot\:}{#1}
}
\DeclarePairedDelimiterX\norm[1]{\lVert}{\rVert}{
    \ifblank{#1}{\:\cdot\:}{#1}
}
\DeclarePairedDelimiterX\anglebraket[1]{\langle}{\rangle}{
    \ifblank{#1}{\:\cdot\:}{#1}
}

\newtheorem{thm}{Theorem}[section]
\newtheorem{prop}[thm]{Proposition}
\newtheorem{cor}[thm]{Corollary}
\newtheorem{defn}[thm]{Definition}

\DeclareMathOperator*{\real}{Re}

\usepackage[doi=false,
eprint=false,
url=false,
isbn=false,
citestyle=numeric,
maxbibnames=20,
giveninits=true,
backend=biber]{biblatex}
\renewbibmacro{in:}{}

\usepackage{algorithm}
\usepackage{algpseudocode}

\usepackage{authblk}

\begin{document}

\title{\bfseries\Large Resonance-Driven Intermittency and Extreme Events in Turbulent Scalar Transport with a Mean Gradient}

\author[1]{Mustafa A. Mohamad\,\thanks{Corresponding author. Email: \texttt{mustafa.mohamad@ucalgary.ca}}}
\author[2]{Di Qi\,\thanks{Email: \texttt{qidi@purdue.edu}}}
\affil[1]{Department of Mechanical and Manufacturing Engineering, University of Calgary,  Calgary, AB, Canada}
\affil[2]{Department of Mathematics, Purdue University,  West Lafayette, IN, USA}
\date{\normalsize\today}

\maketitle


\begin{abstract}
    \noindent
    We study the statistical properties of passive tracer transport in turbulent flows with a mean gradient, emphasizing tracer intermittency and extreme events. An analytically tractable model is developed, coupling zonal and shear velocity components with both linear and nonlinear stochastic dynamics. Formulating the model in Fourier space, a simple explicit solution for the tracer invariant statistics is derived. Through this model we identify the resonance condition responsible for non-Gaussian behavior and bursts in the tracer. Resonant conditions,  that lead to a peak in the tracer variance, occur when the zonal flow and the shear flow phase speeds are equivalent. Numerical experiments across a range of regimes, including different energy spectra and zonal flow models, are performed to validate these findings and demonstrate how the velocity field and stochasticity determines tracer extremes. These results provide additional insight into the mechanisms underlying turbulent tracer transport, with implications for uncertainty quantification and data assimilation in geophysical and environmental applications.
\end{abstract}

\section{Introduction}
Turbulent transport of passive scalars represents a fundamental phenomenon in fluid dynamics. The physical law that describes the transport of a passive scalar $T_t(\boldsymbol x)$ (subscript denotes time dependence) is given by the advection-diffusion equation:
\begin{equation}\label{eq:turbfudd}
\frac{\partial T_t}{\partial t} + \boldsymbol v_t \cdot \nabla T_t =   \kappa \Delta T_t + S_t(\boldsymbol x), \qquad T_{t=0}(\boldsymbol x)= T_0(\boldsymbol x)
\end{equation}
where $\kappa>0$ is the molecular diffusivity constant, $\boldsymbol v_t$ is an incompressible velocity field satisfying $\nabla\cdot \boldsymbol v_t =0$, and $S_t(\boldsymbol  x)$ is a source term. The model is a linear equation, but is statistically nonlinear due to the advection flow. 

Passive tracers include physical tracers such as temperature  and chemical tracers  including solute concentration. They serve as essential diagnostic tools in environmental and geophysical sciences, and aid in understanding the mixing properties in engineering applications,  such  as  non-premixed turbulent combustion. While the advection-diffusion equation and turbulent mixing of passive scalars has been extensively studied since the works of \textcite{taylor1922}, \textcite{richardson1926}, and Kolmogorov~\cite{kolmogorov1941} among many others, understanding the statistical properties of tracer fields, particularly their intermittent behavior, remains an active area of interest~\cite{sreenivasan2019,warhaft2000}.

In this article, we focus on the statistical aspects of the tracer field, with particular emphasis on tracer intermittency and extreme events. These phenomena have significant consequences in practical applications including the spread of pollutants and hazardous chemicals in the air and atmosphere, the dispersion of anthropogenic contaminants in water bodies, and the behavior of Lagrangian tracers like measurement floats in the ocean that collect data~\cite{neelin2010}. Through analytical models and  simulation, we study the effects of intermittency for different velocity models and provide intuition on the physical features of the corresponding tracer fields.
The approach developed here extends upon an existing line of literature in turbulent diffusion, whereby simplified representations of the underlying velocity field are used to construct elementary models for turbulent diffusion. Early contributions extend back to deterministic models of time-dependent fields~\cite{marcoavellaneda1994} and periodic shear flows~\cite{bourlioux2002} to   recent works where more realistic stochastic representations have been assumed~\cite{majda2015}. Intermediate efforts have explored a range of in-between idealizations and flow regimes, including  uncorrelated velocity fields and white-noise limits of the shear flow, to eddy-diffusivity approximations of these models~\cite{majda2013,smith2005}. A comprehensive review  of passive scalar transport and related approximations is provided in~\textcite{majda1999}.

Building on this background  and results in~\cite{majda2015}, we study an analytically tractable model that captures key aspects of tracer intermittency with coupled zonal and shear velocity 
components. 
The shear flow satisfies a stochastic partial differential equation with a Gaussian structure and turbulent energy spectra, while the  zonal flow is modeled as a stochastic differential equation with nonlinear dynamics and non-Gaussian statistics.  
Through this formulation, we demonstrate a range of intermittency behavior that serve as valuable testbed regimes for uncertainty quantification (UQ) of non-Gaussian system~\cite{mohamad2015}  and data assimilation (DA) applications. These analytically tractable models with explicit tracer statistics are particularly relevant for developing filtering strategies in complex environmental systems with incomplete observations, such as in tracking chemical plumes, contaminants in oceans, and sea-ice modeling, as shown in recent works   that adopt a similar conceptual framework  in different contexts~\cite{lee2017,chen2023,mohamad2020,chen2021}. The challenges in filtering and predicting turbulent spatially extended signals primarily stem  from partial observations and model errors due to incomplete physics and resolution limits; as such,  an analytically tractable model   that captures essential features of realistic tracer transport provides a useful test model.
 
\subsection{Organization}

The paper is organized as follows: After introducing our key contributions in \cref{sec:contributions}, in \cref{sec:formulation} we present a detailed formulation of turbulent diffusion models with a mean gradient. \Cref{sec:general_properties} examines the general properties of these models, followed by \cref{sec:model_stat_soln} which examines  their statistical solutions. \Cref{sec:resonance} provides the resonance conditions and discusses their physical interpretation. In \cref{sec:regimes_single,sec:regimes_finite}, we present numerical results demonstrating various intermittency regimes, and we conclude with a discussion of implications and future directions in \cref{sec:conclusions}. Proofs of the major results in \cref{sec:model_stat_soln,sec:general_properties} are provided in~\cref{sec:appendixproofs}.

\subsection{Contributions}\label{sec:contributions}
Contributions in this paper include an analytically tractable model to study tracer intermittency, explicit tracer statistical solutions showing extreme events, and extensive numerical simulations displaying intermittency in different model regimes. We show a range of tracer intermittency scenarios that can be used for various studies in UQ and DA applications.

\section{Formulation of  turbulent diffusion models with a mean gradient}\label{sec:formulation}

In general, the transport of a passive tracer $T_t(\boldsymbol x)$ advected by an incompressible velocity field $\boldsymbol v_t(\boldsymbol x)$ is given  by
\begin{equation}\label{eq:trubdef}
\frac{\partial T_t}{\partial t} + \boldsymbol v_t \cdot \nabla T_t =  \kappa \Delta T_t + S_t(\boldsymbol x),\qquad \nabla \cdot \boldsymbol v_t = 0,
\end{equation}
where $\kappa$ is molecular diffusivity and  $S_t(\boldsymbol x)$ a tracer external source term. We study two-dimensional turbulent diffusion models where the passive tracer field has a known  background mean gradient $\boldsymbol \alpha = (\alpha_x,\alpha_y)$, so that the tracer field can be written as
\begin{equation}
T_t(\boldsymbol x) =  \boldsymbol \alpha \cdot \boldsymbol x + T_t'(\boldsymbol x) ,
\end{equation}
where the prime notation denotes fluctuations of the tracer field around the mean gradient term.

In the model we consider, the stochastic velocity field $\boldsymbol v_t$ is periodic in space with the form
\begin{equation}
\boldsymbol v_t(x) = (u_t, v_t(x)),
\end{equation}
which automatically satisfies the incompressibility condition. The spatially uniform horizontal velocity $u_t$  represents zonal cross sweeps, such as east-west zonal jets, and $v_t(x)$ is a shear flow along the $y$-axis, representing transverse waves, such as north-south Rossby waves.
  The equation for the tracer fluctuation term $T'$ using~\cref{eq:trubdef} is then given by
\begin{equation}\label{eq:fullmodel}
\frac{\partial T'_t}{\partial t} + u_t\frac{\partial T'_t}{\partial x} +  v_t(x)\frac{\partial T'_t}{\partial y}  =  \kappa \Delta T'_t - \alpha_x u_t - \alpha_y v_t(x) + S_t(\boldsymbol x) .
\end{equation}

For the simplified test model, we consider the  existence of a background mean gradient in the vertical direction, thus $\alpha_x \equiv 0$. Further, motivated from physical considerations,  we consider fluctuations that only depend on the $x$ variable alone so that
\begin{equation}
T_t(x,y)=  T'_t(x) + \alpha y,
\end{equation}
where we redefine $\alpha \equiv \alpha_y$. The fluctuations then satisfy the simplified model
\begin{equation}\label{eq:simplified_model}
\frac{\partial T_t}{\partial t} + u_t\frac{\partial T_t}{\partial x}    = \kappa \frac{\partial^2 T_t}{\partial x^2}  -d_T T_t  - \alpha  v_t(x) ,
\end{equation}
where we drop the prime notation from $T'$ and source terms $S_t(\boldsymbol x)\equiv  0$. The term with $d_T > 0$  is an explicit uniform damping term added to damp the zero mode that  arises from partial Fourier transform in the $y$ variable at non-zero modes in the general model in~\cref{eq:fullmodel}. This explicit damping term compensates for the lack of natural damping in the zero mode due to the absence of spatial $y$ derivatives in the simplified model~\cite{majda1999}.

We see that in~\cref{eq:simplified_model} the random velocity $v_t(x)$ drives fluctuations in the tracer field through the mean gradient $\alpha$.  These judicious  simplifications preserve key features of various inertial range statistics in turbulent diffusion, including intermittency, while yielding analytically tractable tracer solutions that facilitate rigorous mathematical analysis~\cite{majda2013}.

\subsection{Velocity field and  passive tracer model in Fourier space}

Next we formulate the velocity field for the passively advected tracer. We choose a  general stochastic representation in order to capture the range of patterns that appear in general turbulent signals. There are two components to the velocity field $\boldsymbol{v}_t = (u_t, v_t(x))$, a zonal component $u_t$ and a spatially dependent shear term $v_t(x)$.

The spatially uniform zonal  flow, i.e. the cross sweep, satisfies the nonlinear  stochastic diffusion equation:
\begin{equation}\label{eq:crosszonal}
du_t = f(u_t) \, dt + \sigma(u_t)\,dW_t,
\end{equation}
where $W_t$ is a real Wiener process. The velocity  $u_t$ can be decomposed into
\begin{equation}
u_t = \overline{u} + u_t',
\end{equation}
consisting of an ensemble mean $\overline{u}$ and a fluctuating component $u_t'$.

The shear velocity $v_t(x)$ satisfies a stochastic partial differential  equation of the form
\begin{equation}
    \frac{\partial v_t}{\partial t} + P\biggl(\frac{\partial}{\partial x}, u_t \biggr)v_t =  \dot W_v(x,t),
\end{equation}
where $P$ is a linear operator that combines both dispersive and dissipative effects acting on $v_t$, coupled with the zonal flow   $u_t$.  The spatially dependent shear flow $v_t(x)$ is modeled by the following  stochastically forced dissipative advection equation,  where the cross sweep dependence $u_t$ enters linearly,
\begin{equation}\label{eq:shearflowmodel}
\frac{\partial v_t}{\partial t} =   u_t R_1\biggl(\frac{\partial}{\partial x}\biggr)v_t +  R_2\biggl(\frac{\partial}{\partial x}\biggr)v_t - \gamma_v  \biggl(\frac{\partial}{\partial x}\biggr)v_t +\dot W_v(x,t).
\end{equation}
Here the linear operators $R_1,R_2,\gamma_v$ are defined   through their image on   Fourier modes:
\begin{equation}
 R_1 \biggl(\frac{\partial}{\partial x}\biggr) =  ia_k e^{ikx} ,\qquad R_2\biggl(\frac{\partial}{\partial x}\biggr)= ib_ke^{ikx}, \qquad \gamma_v\biggl(\frac{\partial}{\partial x}\biggr)  = \gamma_{v,k}  e^{ikx},
\end{equation}
such that $\gamma_v$ is a positive definite linear operator  $\gamma_{v,k}>0$ representing dissipation, and $R_1,R_2$ are linear operators that represent both internal effects of $u_t$ on $v_t$ and wavelike effects, respectively, so that the real-valued dispersion relation is given by:
\begin{equation}
\omega_{v,k} =  a_k u_t + b_k.
\end{equation}

With the above description a summary of the  simplified turbulent diffusion model  in physical space is given by
\begin{align}
du_t &= f(u_t) \, dt + \sigma(u_t)\,dW_t,\\
\frac{\partial v_t}{\partial t} &= u_t R_1\biggl(\frac{\partial}{\partial x}\biggr)v_t +  R_2\biggl(\frac{\partial}{\partial x}\biggr)v_t - \gamma_v  \biggl(\frac{\partial}{\partial x}\biggr)v_t +\dot W_v(x,t),\\
\frac{\partial T_t}{\partial t} &= - u_t \frac{\partial T_t}{\partial x} -  d_T T_t + \kappa \frac{\partial^2 T_t}{\partial x^2}  - \alpha v_t.
\end{align}
Note, since the equations for $v_t$ and $T_t$ are linear,   we employ the following Fourier expansion (the conjugating  Fourier modes  ensure $T_t(x)\in \mathbb R$ and $v_t(x)\in \mathbb R$)
\begin{equation}
T_t(x) = \sum_{k} \widehat{T}_{k,t} e^{ikx}, \quad \widehat T_{-k,t} = \widehat T_{k,t}^*, \quad\text{ and }\quad v_t(x) = \sum_{k} \hat{v}_{k,t} e^{ikx}, \quad \hat v_{-k,t} = \hat v_{k,t}^*,
\end{equation}
to write the explicit equation for each wavenumber to write the model in Fourier space. 
\begin{defn}
The turbulent shear model in Fourier space can be formulated as
    \begin{align}\label{eq:Fourierspacemodel}
        du_t &= f(u_t) \, dt + \sigma(u_t)\,dW_t,\\
        d\hat{v}_{k,t} &= (-\gamma_{v,k} + i \omega_{v,k}) \hat v_{k,t} \, dt + \sigma_{v,k} \, dB_{k,t},  \label{eq:flucfourer}\\
        d\widehat{T}_{k,t} &=(-\gamma_{T,k} + i \omega_{T,k}) \widehat T_{k,t}\,dt -\alpha \hat v_{k,t}\, dt,
        \end{align}
        where
        \begin{equation}
        \gamma_{T,k} = d_T +\kappa k^2, \qquad  \omega_{v,k}(t) = a_k u_t + b_k, \qquad \omega_{T,k}(t)  = -u_t k.
        \end{equation}
\end{defn}
The noise in~\cref{eq:flucfourer} is a complex Wiener process, $B_{k,t} = (B_{k,t}^1 + i B_{k,t}^2)/\sqrt{2}$, with $B^i_{k,t}$ being independent, real Wiener processes, such that   $W_v(x,t) = \sum_k  B_{k,t} e^{ikx}$. Also, in order for $v_t$ to be real-valued, we require $\hat v_{-k,t} = \hat v_{k,t}^*$, which is enforced through the constraints on:
\begin{equation}
\gamma_{v,k} =\gamma_{v,-k}, \quad a_k = -   a_{-k} , \quad b_k = -  b_{-k}, \quad B_{k,t} =   B_{-k,t}^*  ,
\end{equation}
and the real-valued constraint for $T_t$ is   automatically satisfied.

\subsection{Shear flow velocity field models}
The stochastic  zonal cross sweep dynamics in~\cref{eq:crosszonal} and the shear flow in~\cref{eq:shearflowmodel} can model a wide range of interesting turbulent flows. For the shear flow, several relevant models include random flows, non-dispersive waves, and quasi-geostrophic (QG) baroclinic 1.5 layer flows:
\begin{itemize}
    \item {Random flows:} 
    \begin{equation}\label{eq:randomvelocity}
    \gamma_{v,k} = d_v + \nu k^2,\quad a_k= b_k = 0 ,
    \end{equation}
    where $\nu$ is the flow viscosity.
    \item {Non-dispersive waves:}  
    \begin{equation}
    \gamma_{v,k} = d_v + \nu k^2,\quad a_k= 0 ,\quad b_k = -ck,
    \end{equation}
    with wave speed $c$. In this model, zonal flow is uncoupled from the shear flow since $a_k= 0$. This model is commonly encountered in the engineering community.
    \item {Quasi-geostrophic (QG)  $\beta$-plane  flows:} This correlated Rossby model~\cite{majda2013a,vallis2006} has parameters  
    \begin{equation}\label{eq:qgmodel}
        \gamma_{v,k} = d_v + \nu k^2, \quad a_k = k\Bigl(\frac{F}{k^2}-1\Bigr), \quad b_k = \frac{\beta k}{  k^2 + F},
    \end{equation}
    where $F=L_R^{-2}$ and $L_R$ is the deformation radius of Rossby waves, $\beta$ represents rotation due to Coriolis forcing. This dispersive model captures essential features of baroclinic Rossby waves and plays a central role in atmosphere-ocean dynamics.
\end{itemize}

The prescribed energy spectrum $E_{v,k}$ for the shear flow sets the strength of the white noise forcing $\sigma_{v,k}$ for each wavenumber for $v_k$. In~\cref{sec:general_properties}, we show that the statistics of the shear flow is  Gaussian, with energy spectra given by
\begin{equation}
E_{v,k} = \frac{\sigma^2_{v,k}}{2\gamma_{k,v}},
\end{equation}
so that the noise for the $k$th mode is set by $\sigma_k = \sqrt{2\gamma_{v,k} E_{v,k}}$. Example variance spectra for the shear flow  include equipartition (white noise),    Kolmogorov spectrum, and a combined spectrum with equipartition for the large scale modes and a Kolmogorov spectrum for the small scales:
\begin{itemize}
    \item {Equipartition spectrum (white noise)}
    \begin{equation}\label{eq:quipartition}
E_{v,k} = E_0, \quad \text{for all $k$}.
\end{equation}
    \item {Kolmogorov spectrum}
    \begin{equation}\label{eq:Kolmoogrov}
    E_{v,k} = E_0 \abs{k}^{-5/3}.
    \end{equation}
    \item {Combined spectrum:}
    \begin{equation}\label{eq:combineddd}
E_{v,k} = \begin{cases}
E_0,  &\abs{k} \le k_0,\\
E_0 \bigr|\tfrac{k}{k_0}\bigl|^{-5/3} ,    &\abs{k} > k_0,\end{cases}
    \end{equation}
    which mimics realistic energy spectra  for large scale waves.
\end{itemize}
To  investigate tracer intermittency in representative models, we analyze various shear flow configurations and their corresponding energy spectra.

\subsection{Zonal flow velocity models}

The zonal cross sweep velocity is decomposed into a constant  mean $\overline{u}$ and a stochastic fluctuating term $u_t'$ around the mean, $u_t = \overline{u} + u_t'$. Here we discuss various types of models for the zonal flow and  their  statistical properties. In~\cref{appendixa:zonal} further details are provided.

\subsubsection{Linear  zonal model}\label{subsec:linearou}
The simplest stochastic zonal flow model is a forced Ornstein–Uhlenbeck (OU) type process given by
\begin{equation}\label{eq:ou}
du_t = (-\gamma_u u + f) \, dt + \sigma_u \, dW_t,
\end{equation}
with constant forcing $f$.  The steady-state mean, variance and the invariant probability density function for such a linear model are easily obtained from the associated Fokker-Planck equation and are given by, respectively,
\begin{equation}\label{eq:outstats}
    \overline{u}=  \mathbb E(\abs{u_\infty}) = \frac{f}{\gamma_u} , \quad E_u = \mathbb E(\abs{u_\infty}^2) =  \frac{\sigma_u^2}{2\gamma_u} , \quad p_u = \mathcal N(\overline{u}, E_u),
\end{equation}
where $\mathcal N(\mu,\Gamma)$ denotes a real-valued Gaussian with mean $\mu$ and variance $\Gamma$. It is possible to consider time dependent forcing leading to non-constant mean flows, however we refrain from this generalization. With constant forcing, the zonal flow fluctuations are simply offset by $\overline{u}$.

\subsubsection{Non-linear zonal model}\label{ssub:non_linear_models}

To capture the inherent non-Gaussianity and multiscale dynamics of geophysical flows, we extend our analysis to a more general class of stochastic models for zonal jet dynamics, characterized by cubic nonlinearity and correlated additive-multiplicative (CAM) noise structure:
\begin{equation}\label{eq:cubmodel}
d u_t  =  ( a  u_t + b u_t^2 - c u_t^3 +  f)\,dt  + (A- B u_t) \,d W_2   + \sigma_u \,d W_1,
\end{equation}
This system represents the simplest example of dynamics derived from low-frequency reductions of large-scale climate dynamics  and is the  normal form for scalar stochastic climate models obtained via the stochastic mode reduction strategy~\cite{majda2009}. We require  $c>0$ to ensure mean stability (cubic damping term), and   $W_1, W_2$ are independent Wiener processes, where the term  $(A- B u_t) d W_2$ is referred to as correlated additive and multiplicative (CAM) noise. 

For the special case with zero CAM noise, i.e. $A=B=0$, \cref{eq:cubmodel} is a standard gradient stochastic differential equation
\begin{equation}
d x_t = -\nabla V(x_t) \,dt + \sigma \, dW_t,
\end{equation}
with potential $V(x_t)$ and the stationary distribution $p(x) = N_0 e^{-2 V(x)/ \sigma^2},$ where $N_0$ is a normalization constant. The explicit form of the potential for~\cref{eq:cubmodel} is given by
\begin{equation}
V_u(x) = - fx - \frac{a}{2} x^2 - \frac{b}{3} x^3 + \frac{c}{4}x^4 .
\end{equation}

The stationary probability measure for the general form with CAM noise,  
can be shown to be given by
\begin{equation}
p_u(u) = \frac{N_0}{((B x - A)^2 + \sigma_u^2)^{a_1}} \exp\biggl( d \arctan\biggl(\frac{Bx - A}{\sigma_u} \biggr)\biggr) \exp\biggl( \frac{-c_1 x^2 + b_1 x}{B^4}\biggr),
\end{equation}
where $N_0$ is a normalization constant. The coefficients $a_1,b_1,c_1,d$ are provided in~\cref{appendixa:campdf}.

In remainder of the article we concentrate on models with $A=0$:
\begin{equation}\label{eq:cubmodelsimpler}
d u_t  =  ( a  u_t + b u_t^2 - c u_t^3 +  f)\,dt  +    B u_t  \,d W_2   + \sigma_u \,d W_1,
\end{equation}
as it retains the main features interesting features that occur from multiplicative noise. This model with $b=c=0$ and   $a = -\gamma_u$ the model reduces to the  OU process in~\cref{subsec:linearou} when $B=0$.

\paragraph{Numerical experiments}

We present several test cases to demonstrate the dynamics of nonlinear zonal flow across different parameter regimes. In  these experiments, we maintain the additive noise at a moderate level, $\sigma_u = 1$, and identify prototypical behavior by fixing $c=1$ and $b=0$.
Based on the stability analysis of the nonlinear cubic model in $(a,f)$ parameter space (see~\cref{appendixa:zonal}), we investigate two distinct scenarios.

\begin{figure}[htb!]
    \centering
    \smallskip\footnotesize{$f = 0,\quad B = 0$}\\ 
    \includegraphics[scale=0.395]{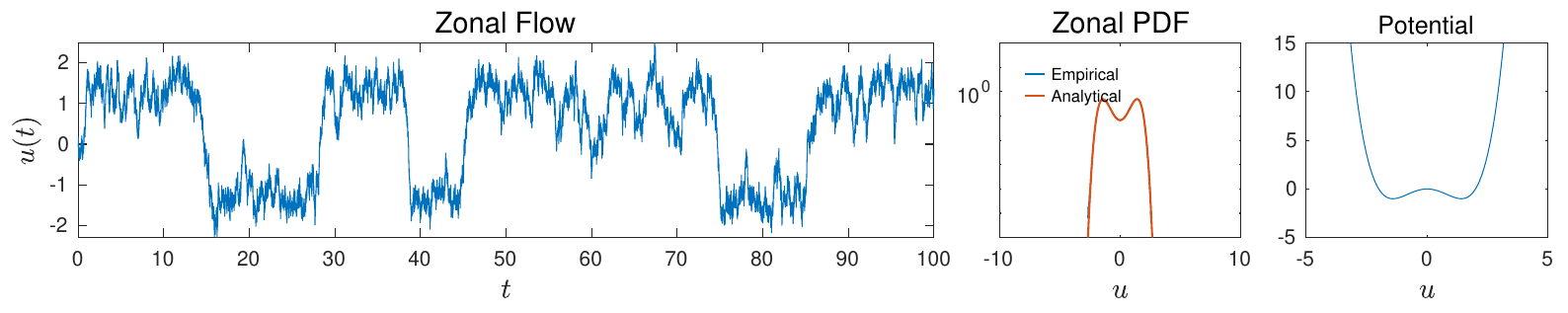}
    \\\smallskip\footnotesize{$f = 1.0,\quad B = 0$}\\ 
    \includegraphics[scale=0.395]{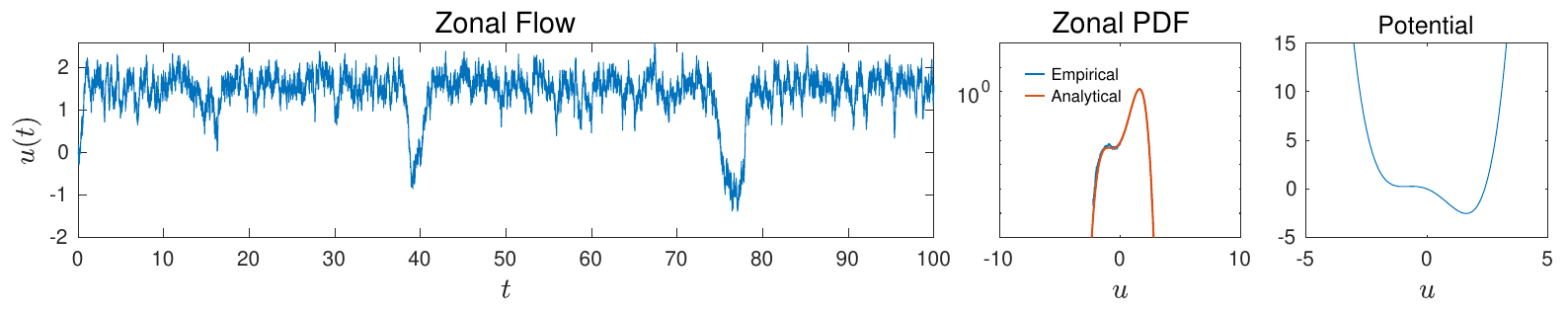}
    \caption{Nonlinear zonal flow dynamics with zero multiplicative  noise $ B=0$: realization (left), zonal PDF (center), and  zonal potential.}
    \label{fig:testregcases}
    \vspace{1em}
    \centering
    \footnotesize{$f = 0.0,\quad B =2.5$}\\ 
    \includegraphics[scale=0.395]{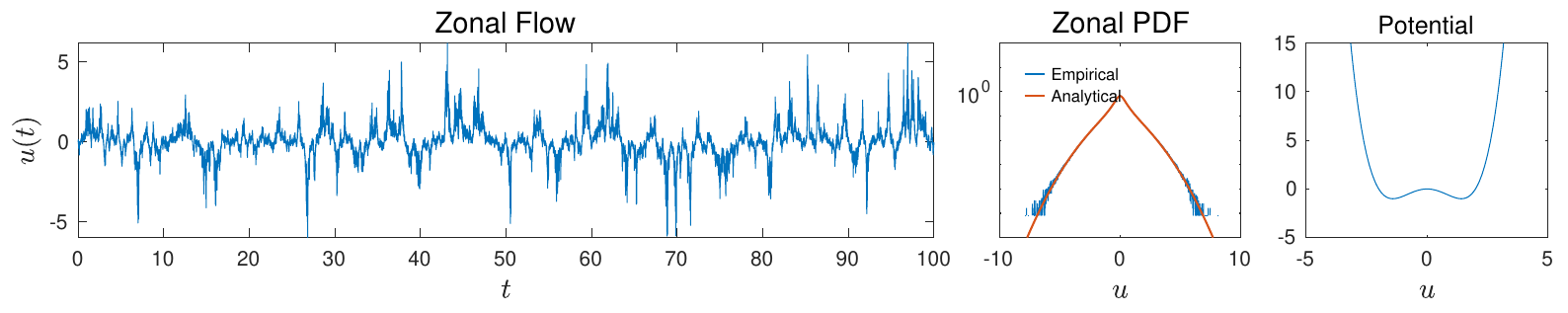}
    \\\smallskip\footnotesize{$f = 1.0,\quad B =2.5$}\\ 
    \includegraphics[scale=0.395]{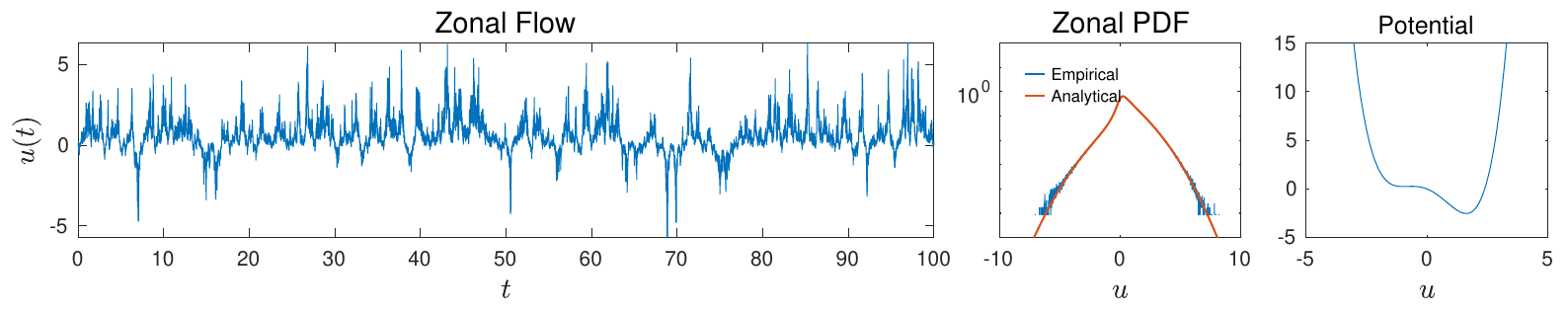}
    \caption{Nonlinear zonal flow dynamics with  multiplicative noise, $  B=2.5$: realization (left), zonal PDF (center), and  zonal potential. }
    \label{fig:testregcasesstrongmulp}
\end{figure}

In the first case, we set the multiplicative noise to zero ($B=0$), as shown in~\cref{fig:testregcases}. The regime with $a=2$ and $f=0$ exhibits two metastable fixed points with stochastic switching between them. The transition frequency between these states depends on the system parameters and can be  controlled. In the second test case, we set the additive forcing to $f=1.0$, which places the system outside the bistable regime. Here, the dynamics demonstrate non-Gaussian behavior with positive skewness, where the locally quadratic potential shape dominates the PDF, though it remains approximately Gaussian for moderate values of $\sigma_u$.

In~\cref{fig:testregcasesstrongmulp}, we examine the same test cases in $(a,f)$ parameter space, but set $B=2.5$ to demonstrate the distinctive effects of multiplicative noise. The numerical experiments reveal that strong multiplicative noise 
inhibits the switching behavior characteristic of the double-well potential observed in the absence of multiplicative noise. This  occurs because the multiplicative noise accelerates fluctuations beyond the stable equilibria, where the system subsequently experiences strong damping that drives it back toward the origin, which becomes the effective fixed point of the dynamics.
Consequently, the system predominantly resides near zero, where the effect of multiplicative noise is minimal. This behavior produces stationary probability distributions that are unimodal with pronounced non-Gaussianity and skewness when    $f\neq 0$.

\section{Tracer model general properties and trajectory solution}\label{sec:general_properties}
The tracer model in~\cref{eq:Fourierspacemodel} has two fundamental properties. First, correlation between different Fourier modes occurs exclusively through the zonal flow $u_t$. Second, the dynamics of $\hat v_{k,t}$ and $\widehat T_{k,t}$ are linear and \emph{conditionally Gaussian} given a fixed realization of $u_t$. This conditional Gaussianity can be exploited for efficient filtering and prediction (see~\cite{liptser2001,chen2016}),  and is used here to  analytically determine the solution of $\widehat T_{k,t}$, including its limiting stationary distribution.

A  noteworthy characteristic of this system is that \emph{it possesses no positive Lyapunov exponents, yet exhibits intermittent non-Gaussian solutions and extreme events}---a signature of systems containing intermittent instabilities. This property can be verified from~\cref{eq:Fourierspacemodel} by observing that the system is positively damped ($\lambda_{T,k}, \lambda_{v,k} > 0$), so checking Lyapunov stability is trivial. These  properties will be demonstrated through numerical experiments presented in subsequent sections.

For simplicity, we can assume $\hat v_{k,0}$ and $\widehat T_{k,0}$ are initialized from zero. By  integration, we have the shear flow trajectory solution
\begin{equation}\label{eq:shearfouriersol}
    \hat v_{k,s} = \int_{0}^s \exp(-\gamma_{v,k}(s-r) + i \omega_{v,k}[r,s])\sigma_{v,k} \, d B_{k,r}.
\end{equation}
The expression $X[r,s]$ is used to denote the integral $X[r,s]\coloneqq \int_r^s X_u du$, thus  $\omega_{v,k}[r,s]$  represents the accumulated phase. We see that $\hat v_{k,s}$ is a complex Gaussian with mean and variance, respectively,
\begin{equation}
\mathbb E ( \hat v_{k,s}) = 0,  \qquad  \mathbb E ( \abs{\hat v_{k,s}}^2) =E_{v_k}(1 - e^{-2 \gamma_{v,k} s}), \quad \text{where } E_{v_k} =  \frac{\sigma^2_{v,k}}{2\gamma_{k,v}}.
\end{equation}
In the long time limit $s\to \infty$,  the shear flow  converges to a Gaussian probability measure $\pi_{\hat v_k} = \mathcal{CN} (0, E_{v_k})$, where  $\mathcal{CN}(\mu,\Gamma)$ denotes a  complex Gaussian with mean $\mu$ and variance $\Gamma$.

Similarly, we can integrate   the equation for $\widehat T_{k,t}$ using   the result~\cref{eq:shearfouriersol}.
\begin{prop}\label{prop:trajsol}
    The exact trajectory solution of the tracer model is given by
    \begin{align}
        \widehat T_{k,t} &=  \int_0^t -\alpha \exp(-\gamma_{T,k}(t-s) + i \omega_{T,k}[s,t])\hat v_{k,s} \, ds \\
        &=  {\int_0^t   \int_r^t -\alpha \sigma_{v,k} \exp(-\gamma_{T,k}(t-s) -\gamma_{v,k}(s-r) +  + i \omega_{T,k}[s,t] +  i \omega_{v,k}[r,s])   \, ds \, d B_{k,r}.}
    \end{align}
\end{prop}

From the trajectory solution, conditioned on a zonal flow trajectory  $u_t$, we find that $\widehat T_{k,t}$  is a complex Gaussian random variable $\mathcal{CN} (0,\Sigma_{k,t\mid u})$, with zero mean and variance  given by the following result.
\begin{prop}\label{prop:condvar}
The conditional variance of a trajectory solution is given by
    \begin{align}
        \Sigma_{k,t\mid u} &=   \alpha^2 \sigma_{v,k}^2 \int_0^t \biggl\lvert \int_r^t  \exp(-\gamma_{T,k}(t-s) -\gamma_{v,k}(s-r) +  + i \omega_{T,k}[s,t] +  i \omega_{v,k}[r,s])   \, ds  \biggr\rvert^2  \, dr \\
          &{=\alpha^2 \sigma_{v,k}^2 \int_0^t  \exp(-2\gamma_{v,k} (t-r) )  \biggl\lvert   \int_r^t  \exp( -\gamma_{R,k} (t-s) + i \omega_{R,k}[s,t]  )  \, ds  \biggr\rvert^2  \, dr},\label{eq:altintegral}
        \end{align}
where $\gamma_{R,k} \coloneqq \gamma_{T,k} - \gamma_{v,k}$ and  
\[
    \omega_{R,k} \coloneqq \omega_{T,k} - \omega_{v,k} = -(a_k+k)u_t - b_k  .
\]
Alternatively, we can express the variance as:
\begin{equation}
{\Sigma_{k,t\mid u} = \alpha^2 \sigma_{v,k}^2 \int_0^t  \exp(-2\gamma_{v,k} (t-r) )  \biggl\lvert   \int_r^t  \exp( -\gamma_{R,k} (t-s) + i \omega_{R,k}[s,t]  )  \, ds  \biggr\rvert^2  \, dr.}
\end{equation}
\end{prop}
\begin{cor}\label{cor:varbound}
An upper bound on the conditional variance is given by
\begin{equation}
      \Sigma_{k,t|u} \leq \frac{\alpha^2 \sigma_{v,k}^2 }{\gamma_{R,k}^2}\biggl(\frac{1}{2 \gamma_{v,k}}  +  \frac{1}{2 \gamma_{T,k}} \biggr)
\end{equation}
\end{cor}

\section{Tracer model statistical solutions}\label{sec:model_stat_soln}

In this section, we analyze the statistical properties of the tracer model in~\cref{eq:Fourierspacemodel} and reveal the resonance mechanism driving tracer intermittency. We begin by introducing a multiscale formulation that captures the slow evolution of velocity fields relative to tracer dynamics (\cref{sec:multiscale}), then present numerical methods for integrating the resulting stiff system (\cref{sec:numerics}). Building on the trajectory solution from~\cref{prop:trajsol}, we derive the stationary distribution of the tracer field and show how it depends on a conditional variance that can amplify dramatically  in~\cref{sec:limit_dist}. Most importantly, we identify that this amplification occurs through a resonance mechanism when phase speeds align, providing a quantitative explanation for extreme events and non-Gaussian statistics in the tracer field (\cref{sec:resonance}).

\subsection{Multiscale formulation}\label{sec:multiscale}
We consider  the case where the velocity field $(u_t, v_t(x))$ evolves slowly compared to the advection and diffusion processes. This assumption is a natural condition for the dynamics of atmosphere-ocean systems, where large-scale flows typically vary on slower timescales compared to the small-scale turbulent motions they influence. To incorporate this separation of scales, we scale the governing equations in Fourier space for the zonal and shear flow dynamics by a small parameter $\epsilon$.

Under this formulation, the Fourier space model for the cross sweeps and shear flow are  scaled by $\epsilon$, so that~\cref{eq:Fourierspacemodel} takes the form
\begin{align}
du_t &=  \epsilon f(u_t) \, dt + \sqrt{\epsilon}  \sigma(u_t)\,dW_t,\\
d\hat{v}_{k,t} &= (-  \epsilon \gamma_{v,k} + i \omega_{v,k}) \hat v_{k,t} \, dt +  \sqrt{\epsilon} \sigma_{v,k} \, dB_{k,t}, \\
d\widehat{T}_{k,t} &=(-\gamma_{T,k} + i \omega_{T,k}) \widehat T_{k,t}\,dt -\alpha \hat v_{k,t}\, dt.
\end{align}
The frequency  $\omega_{v,k}$ is not scaled by $\epsilon$, since it  represents   internal wavelike effects of the cross sweeps on $v_t$ (which should be on the same scale), and the equation for $\widehat T_{k,t}$ is exactly as before,  but here the advection term due to the shear flow is slowly varying.

To study the dynamics on a long timescale, we consider the rescaled time  $t'= \epsilon t$. Substitution into the governing equations (and dropping   primes for clarity) gives  
\begin{defn}\label{cor:rescaled}
    On long timescales the turbulent shear model under slowly varying velocity fields is given by
\begin{align}\label{eq:multiscaleeq}
du_t &=    f(u_t) \, dt +   \sigma(u_t)\,dW_t,\\
d\hat{v}_{k,t} &= (-  \gamma_{v,k} + i \epsilon^{-1}  \omega_{v,k}) \hat v_{k,t} \, dt +   \sigma_{v,k} \, dB_{k,t},   \\
d\widehat{T}_{k,t} &= \epsilon^{-1} (-\gamma_{T,k} + i \omega_{T,k}) \widehat T_{k,t}\,dt -\epsilon^{-1} \alpha \hat v_{k,t}\, dt,
\end{align}
where the time dependent  frequencies are given by
\begin{equation}
    \omega_{v,k} = a_k u_t  + b_k, \quad \omega_{T,k}  = -u_t k .
\end{equation}
\end{defn}
This rescaled system reveals a  separation of timescales. As $\epsilon$ approaches zero, the velocity field $(u_t, v_t)$ evolves much more slowly than the tracer field. This separation allows us to treat the velocity field as approximately constant when analyzing the rapid fluctuations in the tracer dynamics, while  capturing the  long-term evolution of the flow structure.

\subsection{Numerical integration}\label{sec:numerics}
Integration of the multiscale tracer model in~\cref{eq:multiscaleeq} requires   special care due to stiffness. The zonal flow $u_t$ is integrated using an explicit Euler-Mayurama scheme, while $\hat v_{k,t}$ and $\widehat T_{k,t}$ are updated using an exact exponential-integrator scheme. The exponential integrator exactly handles the stiff linear terms containing $\epsilon^{-1}$,  avoiding numerical instability that would arise from explicit methods. The  updates for  step $\Delta$ are given as follows: 
\begin{align}
u_{t+\Delta} &= u_t + f(u_t)\Delta + \sigma (u_t)\sqrt{ \Delta } \, w_t  \\
\hat v_{k,t+\Delta} &= \exp\bigl((-\gamma_{v,k} + i \epsilon^{-1} \omega_{v,k}(t))\Delta\bigr) \hat v_{k,t}  + \sigma_{v,k}  \sqrt{ \Delta\exp\bigl((-\gamma_{v,k} + i \epsilon^{-1} \omega_{v,k}(t))\Delta\bigr)} \frac{b_t^1 + i b_t^2}{\sqrt{2}}\\
\widehat{T}_{k,t+\Delta} &= \exp\bigl(\epsilon^{-1} (-\gamma_{T,k} + i \omega_{T,k})\Delta \bigr)\widehat{T}_{k,t} - \epsilon^{-1} \alpha \Delta \exp\bigl(\epsilon^{-1} (-\gamma_{T,k} + i \omega_{T,k})\Delta \bigr) \hat v_{k,t}\\
\shortintertext{where}
\gamma_{T,k} &=  d_T +  \kappa k^2 , \quad \omega_{v,k} =  a_k u_t + b_k, \quad \omega_{T,k}   = -u_t k  ,
\end{align}
and $w_t$, $b_t^1$,$b_t^2$ are independent standard  normal random variables.

\subsection{Limiting distribution for tracer statistics}\label{sec:limit_dist}

An approximate analytical  result for the  stationary distribution for the tracer statistics can be derived by analyzing  the steady-state conditional variance $\Sigma_{k,t\mid u}$. Since the tracer trajectory is a conditional Gaussian integral, given $u_t$, its full distribution can be expressed using  the law of total probability. The stationary distribution of the real part of the tracer mode $\real(\widehat T_{k})$ is then
\begin{equation}
    p(x) = \int \frac{1}{\sqrt{\pi \widetilde \Sigma_k(u)}} \exp\biggl( -\frac{x^2}{\widetilde \Sigma_k(u)} \biggr) p_u(u) \, du.
\end{equation}
Where, $\widetilde \Sigma_{k}(u)$ is the stationary value of the  conditional variance   $\Sigma_{k,t\mid u}$.
\begin{prop}\label{prop:tracermodepdf}
Under slowly varying velocity fields, the conditional tracer variance converges to the stationary value
\begin{equation}
    \widetilde \Sigma_{k}(u) = \frac{\alpha^2 E_{v,k}}{\gamma_{T,k}^2 + \omega_{R,k}(u)^2} . 
    \end{equation}
    In the stationary limit,    $u$ is treated as a static parameter sampled from its  steady state distribution.
\end{prop}
The steady-state distribution of the passive scalar is obtained by the same approach,  additionally  summing over all wavenumbers.
\begin{thm}
    The stationary  distribution of the tracer field $T(x)$ for the model in~\cref{eq:multiscaleeq} is given by:
    \begin{equation}\label{eq:tracerpdf}
    p(\lambda) = \int \frac{1}{\sqrt{2 \pi \widetilde \Sigma(u)}} \exp\biggl( -\frac{\lambda^2}{2 \widetilde \Sigma(u)} \biggr) p_u(u) \, du,
    \end{equation}
    where
    \begin{equation}\label{eq:condvar_limit}
    \widetilde \Sigma(u) = \sum_{k \in \mathbb{N}} \frac{\alpha^2 E_{v,k}}{\gamma_{T,k}^2 + \omega_{R,k}(u)^2} 
    \end{equation}
\end{thm}

\subsection{Intermittency and extreme events through 
resonance}\label{sec:resonance}
Extreme events in the turbulent tracer field are linked to peaks in the conditional variance. Inspecting~\cref{eq:condvar_limit} we see that the conditional variance reaches its maximum when $\omega_{R,k} \coloneqq \omega_{T,k} - \omega_{v,k} = 0$, which corresponds to a resonant condition when the phase speeds of the zonal flow, shear flow, and tracer field align, i.e.  $\omega_{T,k} = \omega_{v,k}$
This resonance leads to bursts in the tracer field variance, occurring when $\omega_{R,k} = 0$ or
\begin{align}\label{eq:rescond}
      u_t' + \overline{u} = u_{\text{res},k} &\coloneqq -\frac{b_k}{a_k +k} \\
   \quad u_t' = u'_{\text{res},k} &\coloneqq -\frac{b_k}{a_k +k} - \overline{u}
   \end{align}
which define the resonant phase speeds. \emph{When the zonal flow fluctuations  $u_t'$  crosses the phase speed threshold $u'_{\text{res},k}$ the wavenumbers $\pm k$ are excited, producing an intermittent burst.} Unlike intermittency in unstable systems—where finite-time instabilities yield heavy-tailed statistics and bursts—this mechanism is resonance-driven: fluctuations in the zonal flow trigger resonance, amplifying the conditional variance and causing non-Gaussian tracer statistics. 

For deterministic periodic shears this 'resonance' driven intermittency was first noted in~\cite{bourlioux2002} and was linked to a physical interpretation of 'blocked' and 'un-blocked' streamlines. In this interpretation, when the zonal flow is $u\approx 0$ the shear flow is unblocked leading to strong convective transport of the tracer along the direction parallel to the mean scalar gradient and strong mixing by diffusion. Conversely, when $u\neq 0$, the transverse sweeps are blocked and transport along the gradient is minimal. 
The resonance condition~\cref{eq:rescond} can be interpreted as a generalization of this result to stochastic zonal and shear flows.

Understanding how the zonal and shear flows affect tracer statistics is crucial, particularly the role of nonlinearity in the zonal flow. While zonal fluctuations do not change the resonant phase speeds---these are set by the wave dynamics of the shear flow and zonal mean---they do influence how often the system crosses resonance, thus modulating tracer statistics (see~\cref{eq:tracerpdf}). This means that the   statistics of the nonlinear  zonal flow   can act to either  enhance turbulent tracer transport through  increased intermittency or reduce intermittency relative to a linear (Gaussian) flow model. This underscores the importance of the zonal flow's stochasticity in the tracer field intermittency, and has implications for linearization approaches.

Although the shear flow does not directly affect the frequency that resonance is reached---the zonal flow statistics determine this---wavelike effects in the shear modify the resonant phase speed values. This influences how often the zonal flow crosses these thresholds. In a purely random shear flow with no wavelike effects, where $a_k = b_k = 0$ (see~\cref{eq:randomvelocity}), the resonant speeds collapse to a single value:
\begin{equation}
    u'_{\text{res}} = -\overline{u};
    \end{equation}
    A similar synchronization appears in non-dispersive advection, where $a_k=0$ and $b_k=-ck$:
\begin{equation}
u'_{\text{res}} = c -\overline{u}.
\end{equation}
In both cases, crossing the resonance threshold excites all modes simultaneously, leading to stronger intermittency, as every excited mode contributes to the tracer field. This also produces finer-scale structures during extreme events due to the excitation of higher-wavenumber modes.

In contrast, dispersive shear flows yield multiple resonant phase speed thresholds (one for each wavenumber k), whereas purely random shear flows and non-dispersive advection synchronize these thresholds, exciting all scales at once. This distinction strongly influences the nature of intermittency and the structure of extreme tracer events.

\section{Numerical experiments and regimes for single-mode systems}\label{sec:regimes_single}

We now perform numerical experiments to examine the effect of zonal and shear flows on tracer intermittency and extreme events across various regimes.
We examine a single Fourier mode, i.e. $k=1$, and assume the shear flow is described   by the   $\beta$-plane QG   model in~\cref{eq:qgmodel}.   Unless stated otherwise,  the following system  parameters are fixed
\begin{equation}
 \epsilon = 0.010,\quad   d_T = 0.1,\quad  \kappa = 0.001,\quad  d_v = 0.6,\quad  \nu = 0.1,\quad \alpha = 1,\quad  \beta = 8.91,\quad  F = 2.5.
 \end{equation}

\subsection{Stochastic zonal mean flow with linear dynamics}\label{ssub:stochastic_zonal_mean_flow_with_linear_dynamics}

Consider a  zonal flow described by the linear stochastic model in~\cref{eq:ou} with statistics in~\cref{eq:outstats}. We consider a case where the eastward zonal jet has the following parameters:  $E_u = 0.5$ (with $\gamma_u = 1$ and $\sigma_u = 1$). The forcing is set to $f=0.4431$, such that $u'_{\text{res}} = -1$, and so fluctuations crossing this threshold occur with probability $p(u < u'_{\text{res}}) = 0.0228$. At resonance (when $u = u'_{\text{res}}$), the conditional Gaussian variance increases dramatically, with $\widetilde \Sigma(u'_{\text{res}}) > 87\widetilde \Sigma(\overline{u})$, indicating an 87-fold amplification of the tracer variance at resonance compared to   mean zonal flow conditions.

\begin{figure}[htbp]   
    \centering
    \footnotesize ${\epsilon = 0.10}$\\\smallskip 
    \includegraphics[scale=0.31]{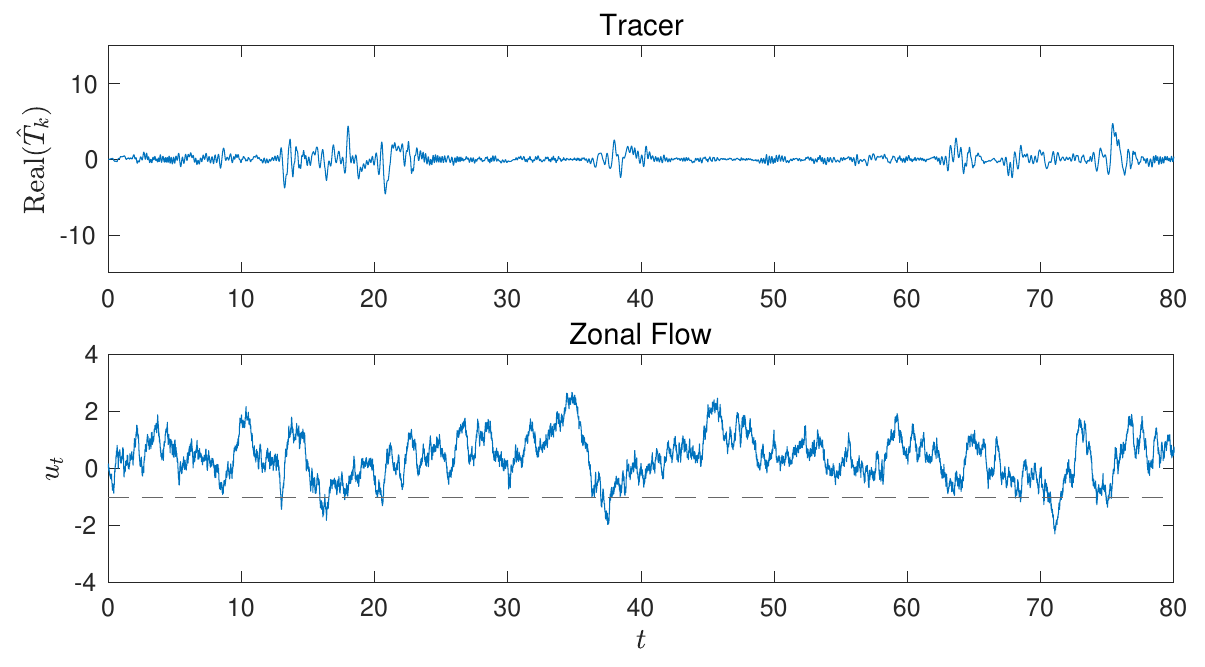}
    \includegraphics[scale=0.31]{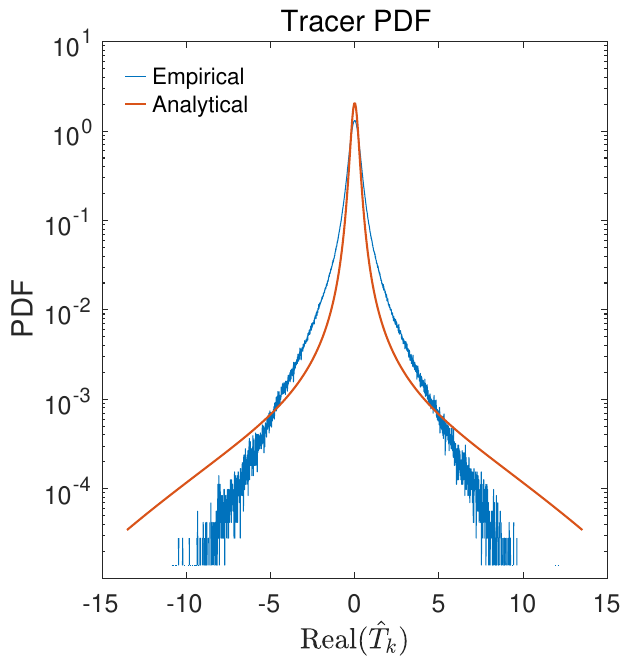}\\
    \smallskip 
    \footnotesize${\epsilon = 0.010}$\\\smallskip
    \includegraphics[scale=0.31]{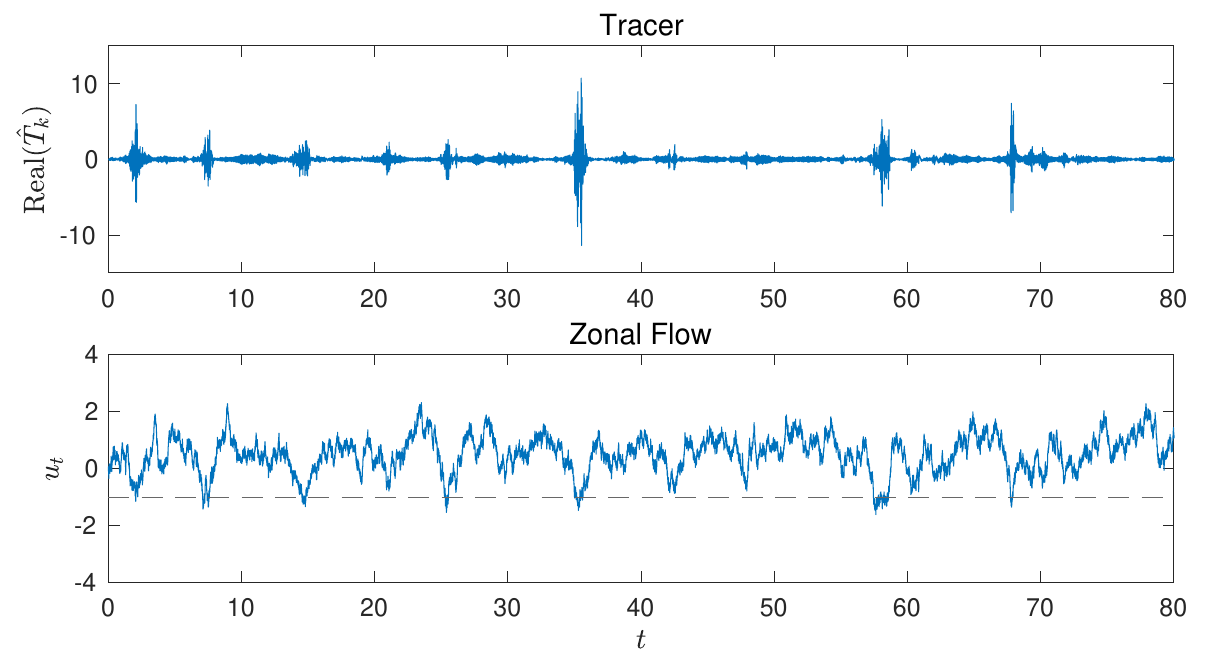}
    \includegraphics[scale=0.31]{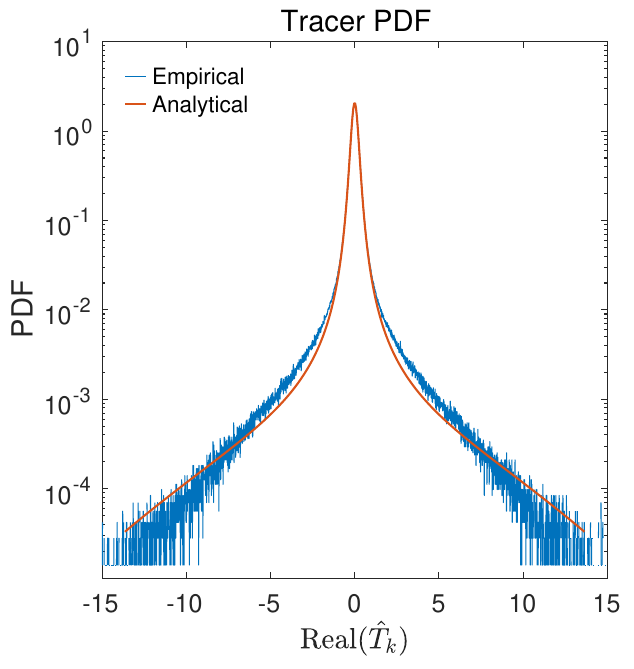}\\
    \smallskip 
    \footnotesize${\epsilon = 0.001}$\\\smallskip
    \includegraphics[scale=0.31]{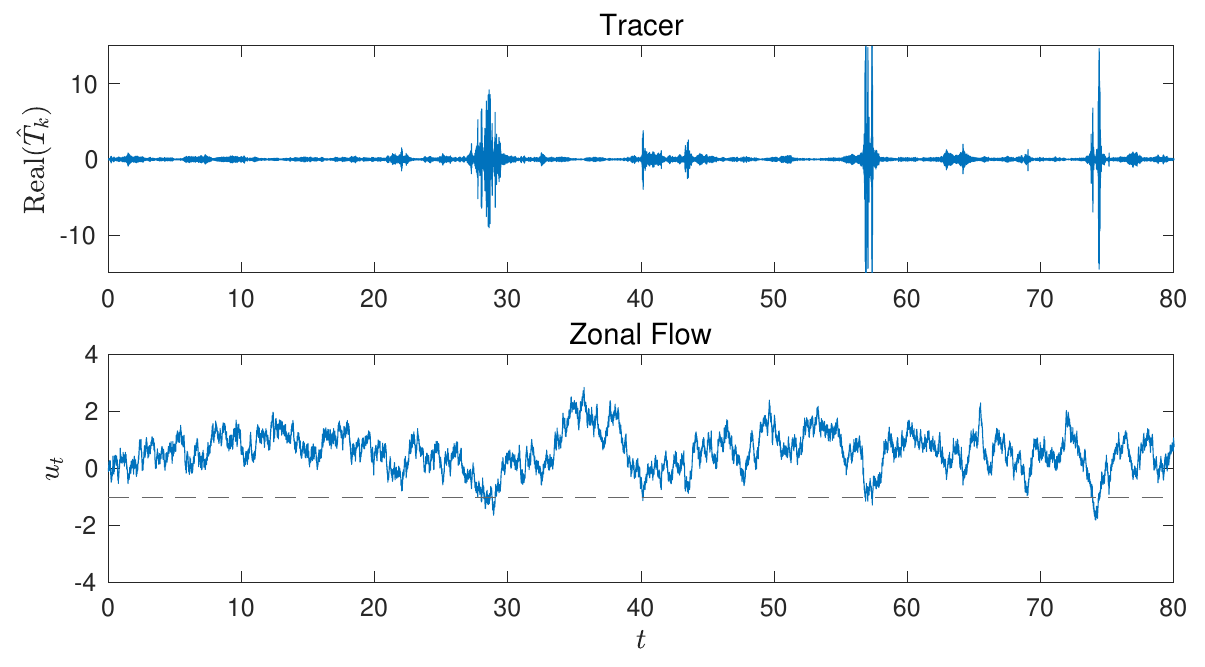}
    \includegraphics[scale=0.31]{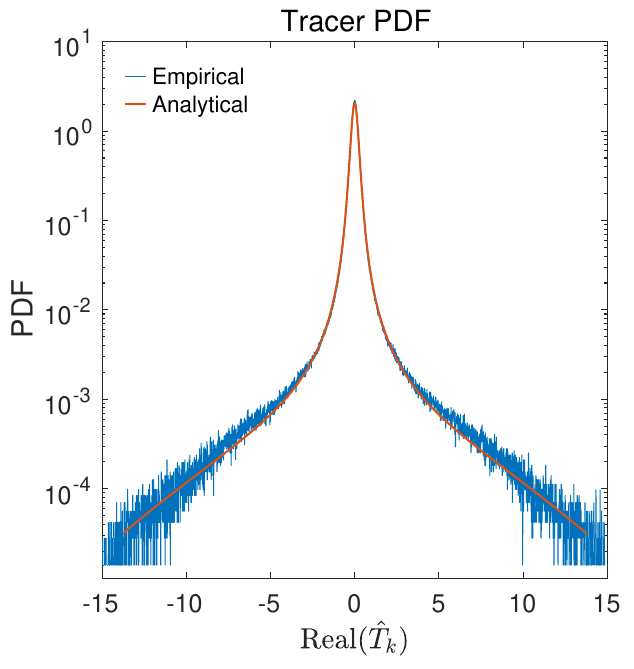}\\
    \caption{Sample realizations and corresponding equilibrium PDFs and analytical result. Model: single mode,   $\beta$-plane QG flow. Linear zonal flow with  $E_u = 0.5$ ($\gamma_u = 1,\sigma_u = 1$). Dashed line in $u_t$ plot is the resonance threshold $u_{\text{res},k} = -1.0$.}
    \label{fig:sample_realizations_and_the_corresponding_equilibrium_pdfs_and_their_analytical_limit_model_single_mode_k_1_beta_plane_q_g_flow_a_k_k_3_k_2_f_b_k_beta_k_k_2_f_parameters_overline_u}
\end{figure}

In~\cref{fig:sample_realizations_and_the_corresponding_equilibrium_pdfs_and_their_analytical_limit_model_single_mode_k_1_beta_plane_q_g_flow_a_k_k_3_k_2_f_b_k_beta_k_k_2_f_parameters_overline_u} we    plot the limiting equilibrium PDF along with the   histogram of the time series and the corresponding  realizations of the tracer mode for various $\epsilon$.
At any fixed time, the tracer distribution is Gaussian, however the variance is time-dependent and shoots at zero crossings of the frequency $\omega_R$ or equivalently when the zonal flow fluctuations $u_t'$ crosses $u'_{\text{res}}$.  Furthermore, as $\epsilon$ tends to zero intermittency is enhanced since the slowly varying zonal flow  $u_t$ spends a longer period of  time in the resonance regime leading to larger extreme events. We note the close agreement between the analytical result and the histogram of the time series as $\epsilon$ tends to zero.

\subsection{Stochastic zonal  flow with nonlinear dynamics}\label{ssub:stochastic_zonal_mean_flow_with_nonlinear_dynamics}

\begin{figure}[htbp]
    \centering
    \footnotesize {Nonlinear Zonal Flow\\\smallskip $f = 1.0,\quad B =0$}\\\smallskip 
    \includegraphics[scale=0.31]{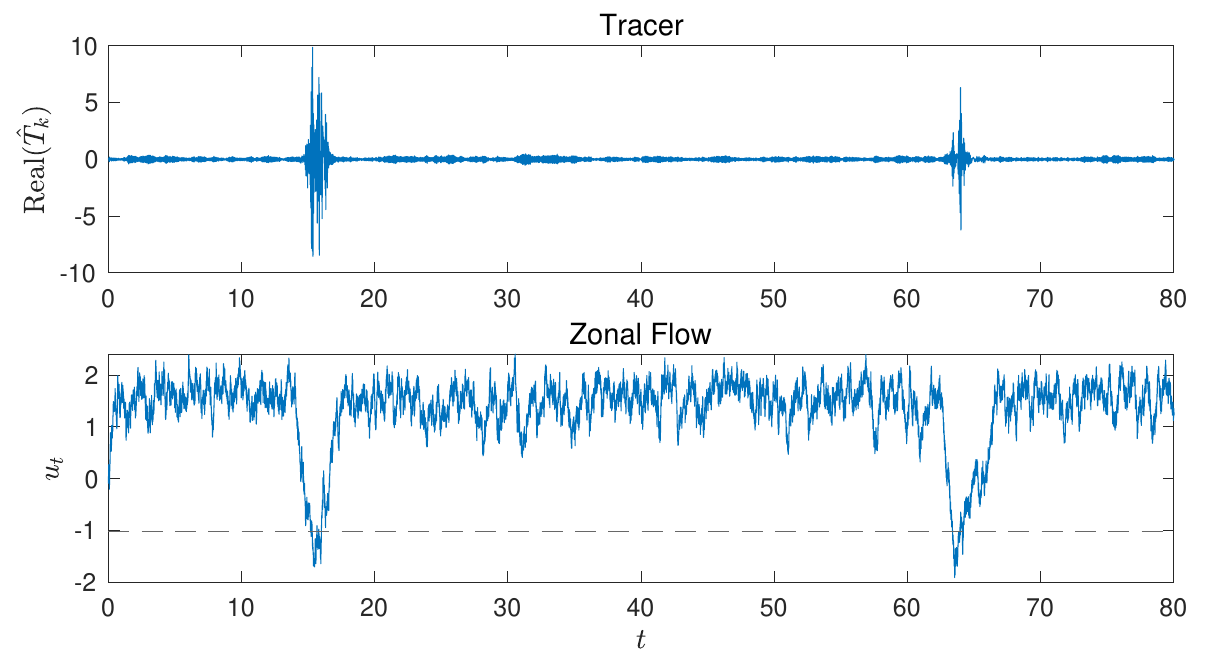}
    \includegraphics[scale=0.31]{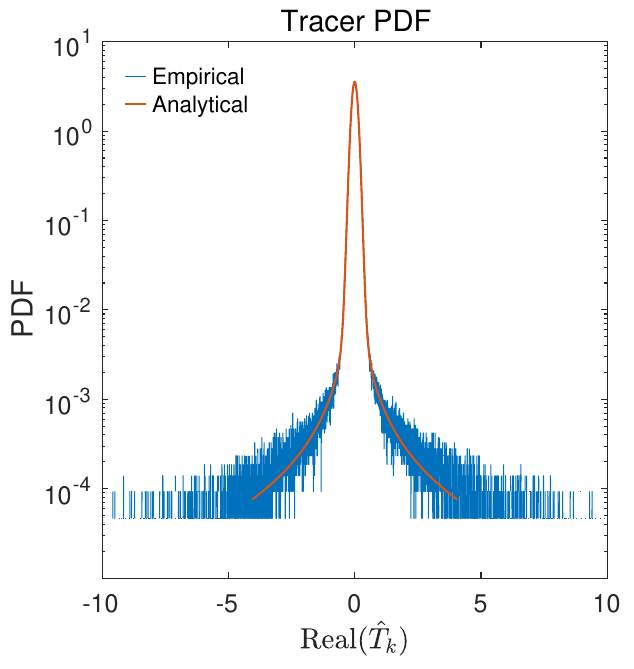}
    \includegraphics[scale=0.31]{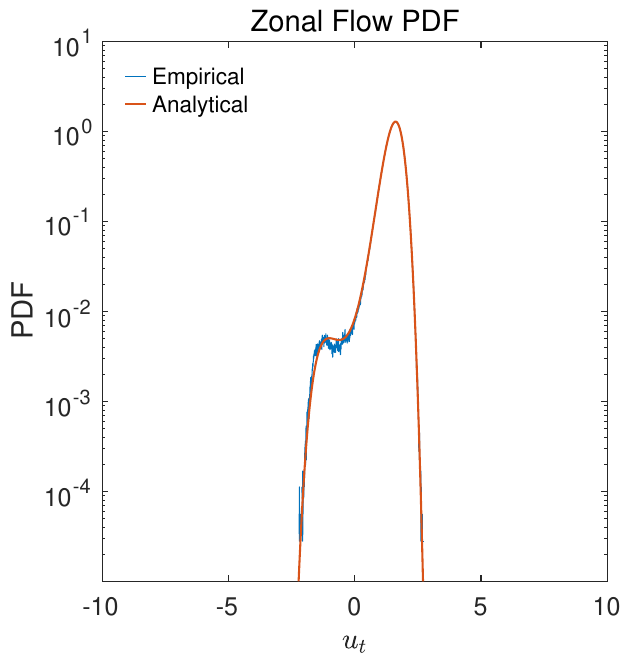}\\
    \smallskip 
    \footnotesize {$f = 0,\quad B =0$}\\\smallskip   
    \includegraphics[scale=0.31]{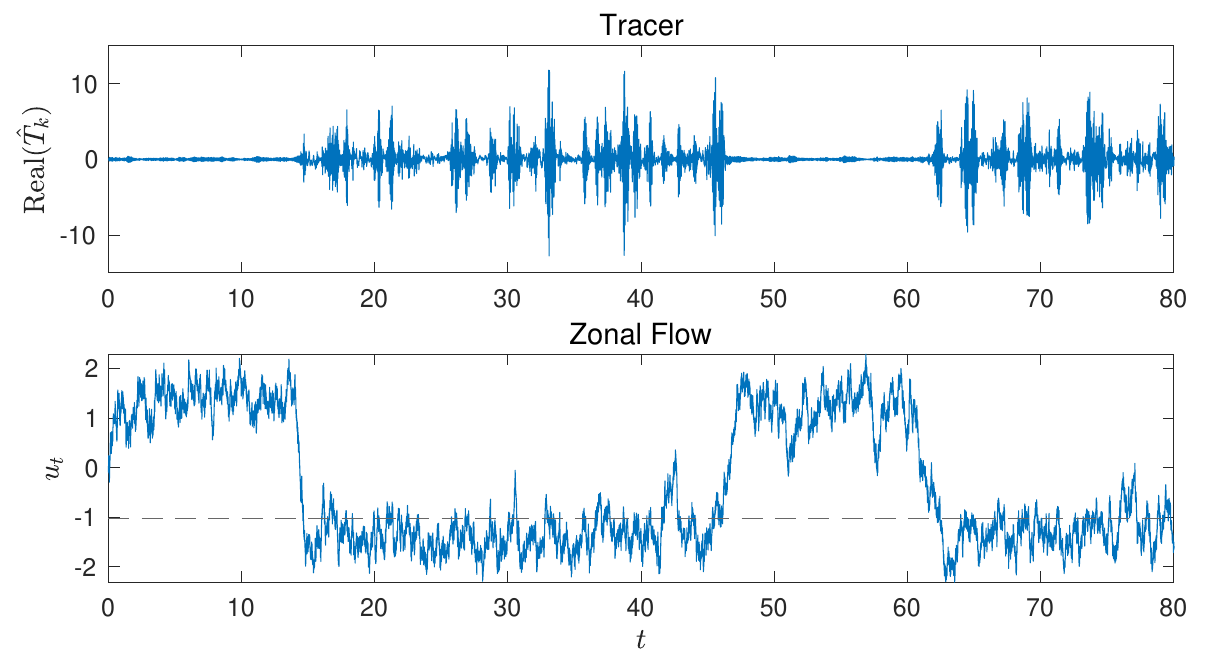}
    \includegraphics[scale=0.31]{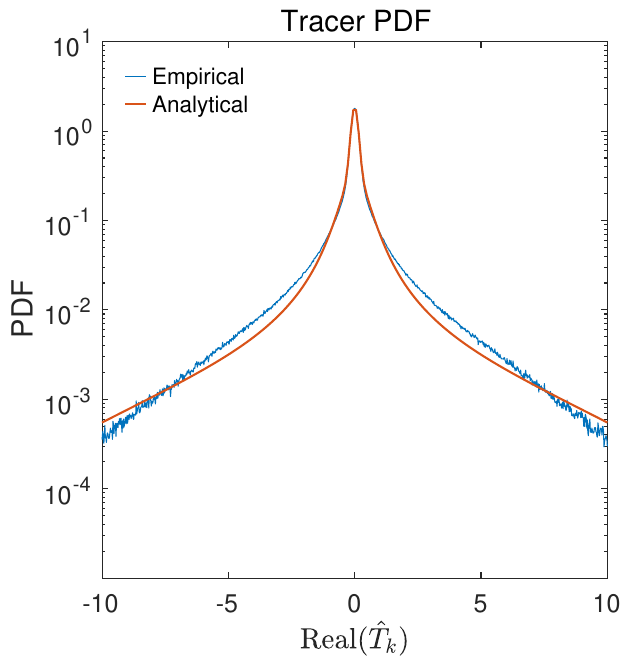}
    \includegraphics[scale=0.31]{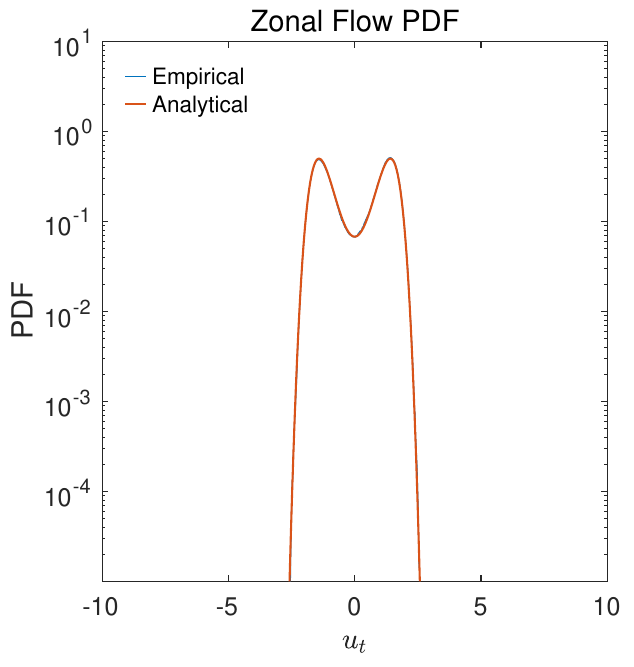}\\
    \caption{Sample realizations,  corresponding equilibrium PDFs, and analytical results. Model: single mode,   $\beta$-plane QG flow. Nonlinear zonal flow with different parameters. Dashed line in $u_t$ plot is the resonance threshold $u_{\text{res},k}$.}
    \label{fig:sample_realizations_and_the_corresponding_equilibrium_pdfs_and_their_analytical_limit_cubic_nocam}
\end{figure}

\begin{figure}[htbp]
    \centering
    \footnotesize {Nonlinear Zonal Flow\\\smallskip $f = 1.0,\quad B =2.5$}\\\smallskip 
    \includegraphics[scale=0.31]{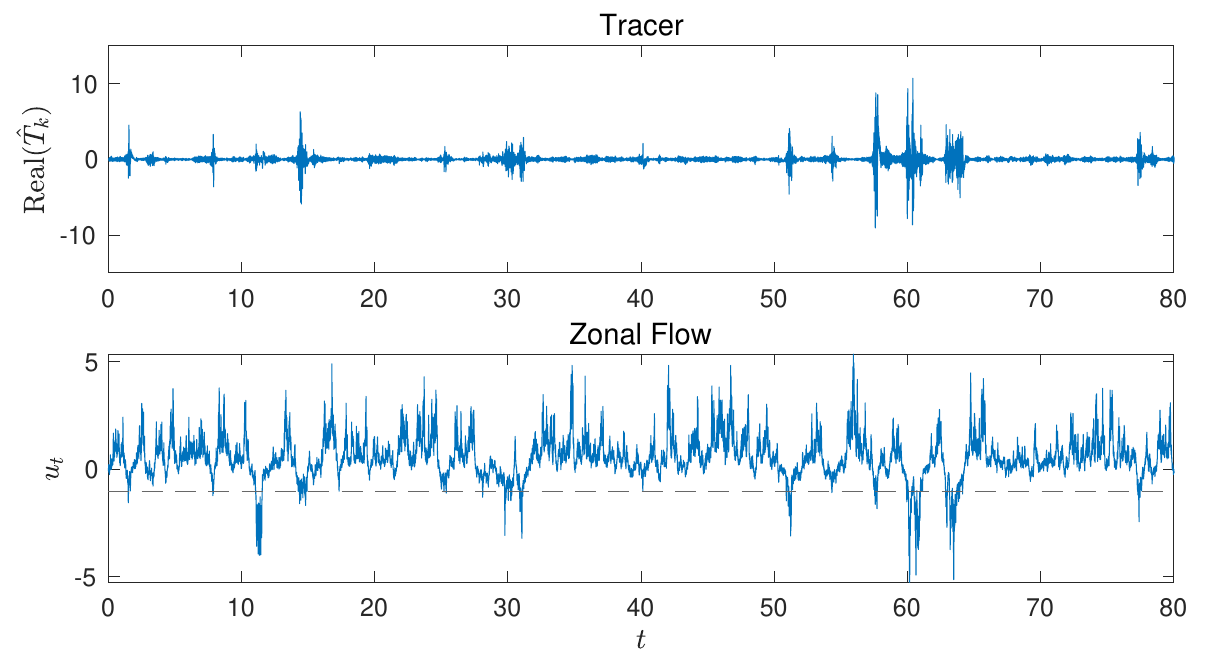}
    \includegraphics[scale=0.31]{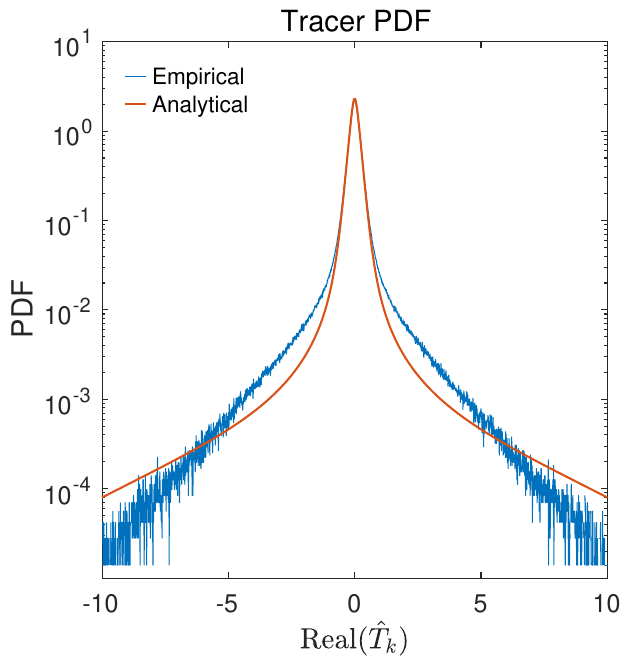}
    \includegraphics[scale=0.31]{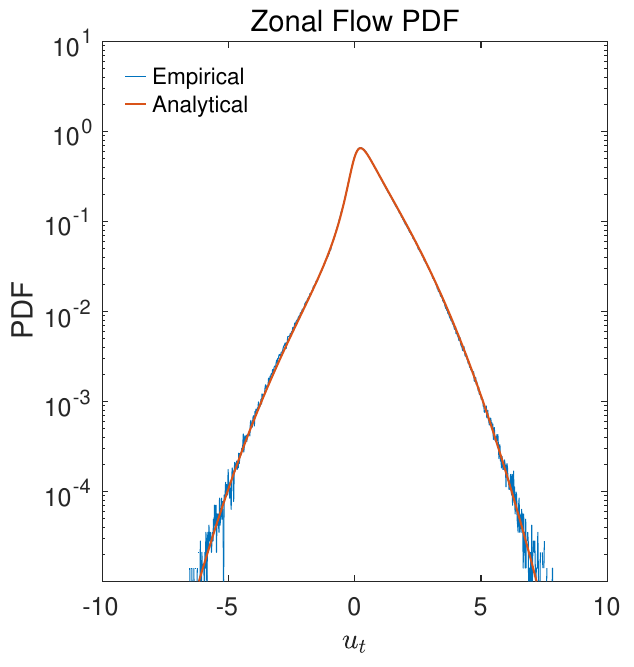}\\
    \smallskip 
    \footnotesize {$f = 0,\quad B = 2.5$}\\\smallskip 
    \includegraphics[scale=0.31]{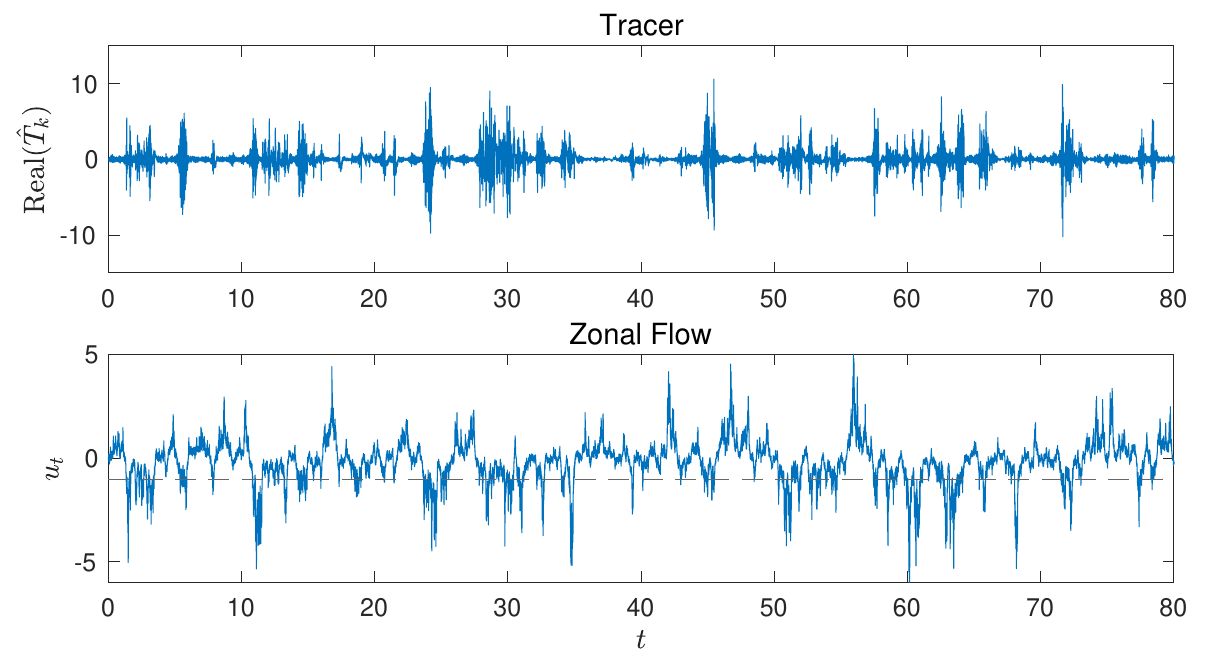}
    \includegraphics[scale=0.31]{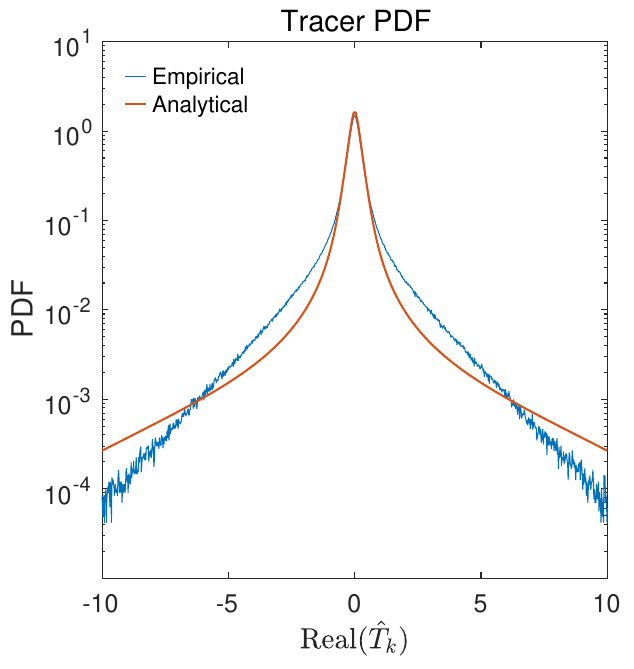}
    \includegraphics[scale=0.31]{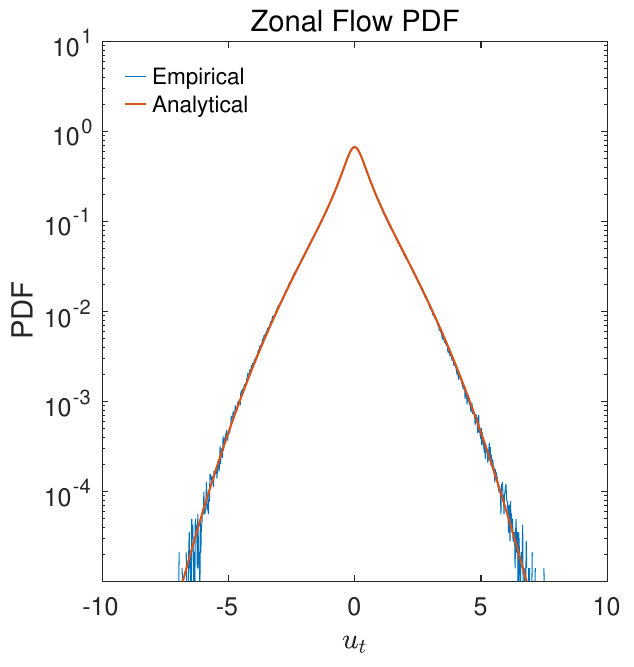}\\
    \caption{Sample realizations,  corresponding equilibrium PDFs, and analytical results. Model: single mode,  $\beta$-plane QG flow.  
    Nonlinear zonal flow with different parameters and multiplicative noise. Dashed line in $u_t$ plot is the resonance threshold~$u_{\text{res},k}$.}
    \label{fig:sample_realizations_and_the_corresponding_equilibrium_pdfs_and_their_analytical_limit_cubic_cam}
\end{figure}

We now consider the nonlinear model in~\cref{eq:cubmodel}.
Motivated by the discussion and the regimes presented  in~\cref{ssub:non_linear_models} we consider several representative cases with  interesting statistics  for the zonal flow, including cases with   zero multiplicative noise $B=0$ in~\cref{fig:sample_realizations_and_the_corresponding_equilibrium_pdfs_and_their_analytical_limit_cubic_nocam} and strong multiplicative noise $B = 2.5$ in~\cref{fig:sample_realizations_and_the_corresponding_equilibrium_pdfs_and_their_analytical_limit_cubic_cam}, with zonal flows that correspond to those in~\cref{fig:testregcases,fig:testregcasesstrongmulp}. As in~\cref{ssub:non_linear_models}, we set $\sigma_u = 1$ and  set  $c = 1$ and $b=0$ throughout the analysis.

Some important points that these cases demonstrate in the single Fourier mode case is that \emph{ strong nonlinearity and non-Gaussian statistics in the zonal flow, such as bimodal distributions or  skewed heavy-tailed statistics, do not necessarily lead to enhanced tracer intermittency.} 
In fact, it is possible to observe a zonal flow with strongly non-Gaussian features compared to a linear case with Gaussian statistics, yet the tracer field PDF is nearly identical. 
This can be understood by examining the threshold crossing frequency at the resonance value $u_{\text{res},k}$ (shown as dashed lines in the $u_t$ plots). This crossing frequency is the critical factor determining tracer statistics and,  notably, it is not uniquely determined by the zonal flow's statistical distribution or dynamics. Consequently, two zonal flows with dramatically different PDFs  can produce nearly identical tracer statistics if they happen to cross the resonance threshold with similar frequency; this is a distinctive feature of this single mode case.

 Other interesting observations include the `on-off' type intermittency regime in the double well zonal flow test case.
 As an aside, note that the analytical limiting tracer formulas for the experiments with strong multiplicative noise do not agree as well to the cases with $B==0$
 This is expected since large multiplicative noise leads to a diffusion process that has a shorter timescale and thus the timescale separation between the zonal flow and tracer modes is decreased.


\section{Numerical experiments and regimes for multi-mode systems}\label{sec:regimes_finite}

We  now consider   finitely many Fourier modes and their effect on the  distribution of the tracer field. Recall  $T_t(x) = \sum_{k} \widehat{T}_{k,t} e^{ikx},$ with $\widehat T_{k,t} = \widehat T_{-k,t}^*$, which  means that a  finite number of modes have a combined effect on the tracer field statistics and intermittency.  This is more clearly understood  by looking at the formula for the conditional variance of the tracer field, which is simply the sum of the conditional variance for each mode~\labelcref{eq:condvar_limit}:
\begin{equation}
\widetilde\Sigma(u) = \sum_{k\in N}  \widetilde\Sigma_{k}(u) =  \sum_{k\in N}    \frac{\alpha^2 E_{v,k}}{\gamma_{T,k}^2 + \omega_{R,k}(u)^2}.
\end{equation}
In the finitely many Fourier mode scenario  we have   more interesting dynamics compared to the single Fourier mode case, since the total variance is a sum of the conditional variance for each mode, which can have their variance  peak at different  zonal phase speeds.  The fact that different modes may have different resonance values  has an impact on the overall nature  of  extreme events in the tracer field. These points will be demonstrated in numerical experiments.

The energy spectrum of the shear flow $v_t(x)$ is set to either equipartition~\labelcref{eq:quipartition} or a Kolmogorov spectrum~\labelcref{eq:Kolmoogrov}:
\begin{equation}
E_{k,v} = E_0  \quad \text{(equipartition)} \qquad  E_{k,v} = E_0 \abs{k}^{-5/3} \quad \text{(Kolmogorov)} .
\end{equation}
We normalize the total energy to be equivalent between both spectra.
We consider these two cases to demonstrate the effects of the energy level of the shear flow on extreme events. Under equipartition, each mode $\hat{v}_{k}$ has equal energy. When $u = u_{\text{res},k}$ (resonance crossings occur), all excited modes contribute with similar intensity. This induces bursts of comparable magnitude across all excited modes $\hat{T}_{k}$, though smaller-scale modes contribute less to extreme events due to selective damping. In contrast, under a Kolmogorov spectrum, smaller scales in the shear flow possess progressively less energy, resulting in diminished contributions to the tracer field statistics when excited. Consequently, the statistics of extreme events in the tracer field are dominated by the largest, most energetic scales.

\subsection{Numerical experiments}
As in the single Fourier case, we    consider the    $\beta$-plane QG flow model in~\cref{eq:qgmodel} as a representative wavelike, dispersive shear flow, with the  same    parameters
\begin{equation}
 \epsilon = 0.010,\quad   d_T = 0.1,\quad  \kappa = 0.001,\quad  d_v = 0.6,\quad  \nu = 0.1,\quad \alpha = 1,\quad  \beta = 8.91,\quad  F = 2.5,
\end{equation}
for the set of wavenumbers $\abs{k}  \leq 5$.

\subsubsection{Stochastic zonal   flow with linear dynamics}
\begin{figure}[htbp]
    \centering
    \begin{minipage}{.49\textwidth}
        \centering
    \subcaption{Equipartition spectrum}\vspace{0.5em}
    \includegraphics[scale=0.395]{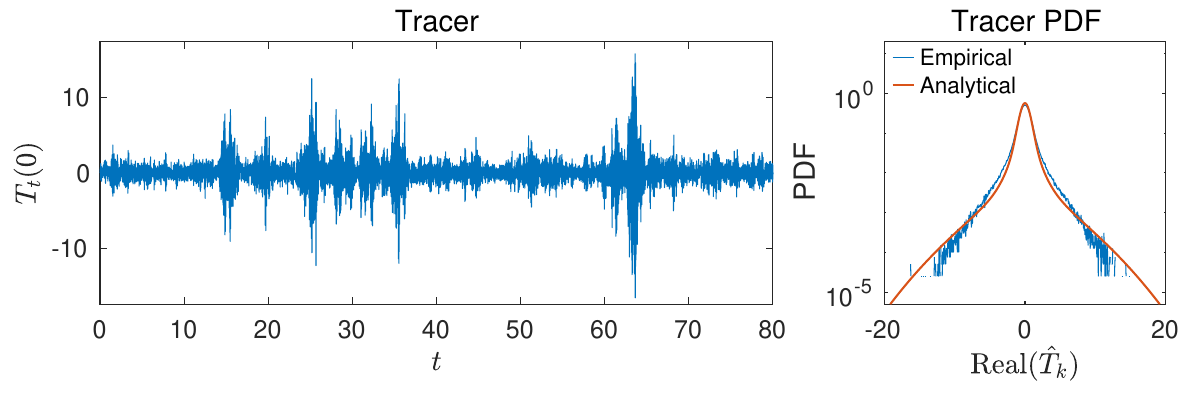}\\
    \includegraphics[scale=0.395]{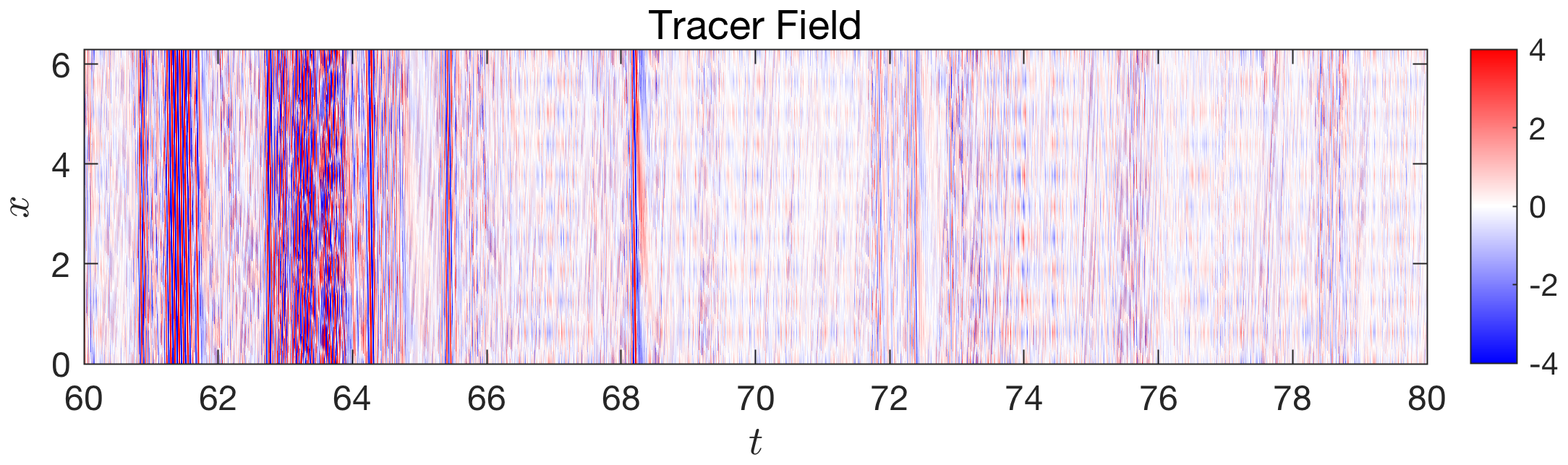}\\
    \includegraphics[scale=0.395]{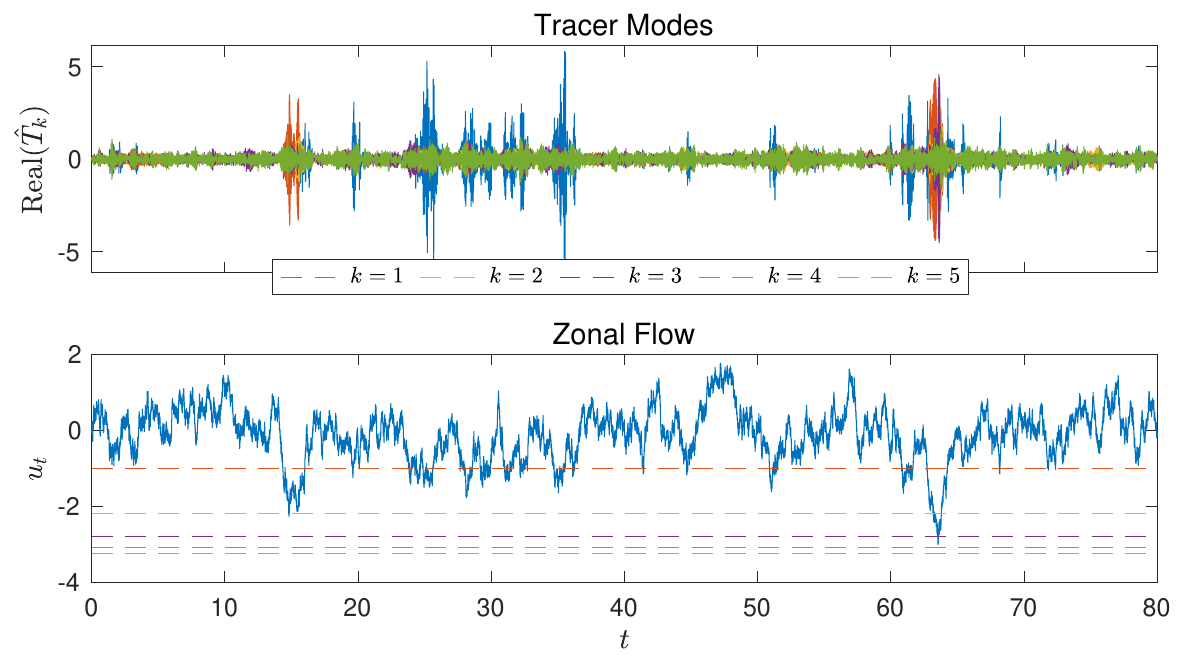}\\
    \includegraphics[scale=0.395]{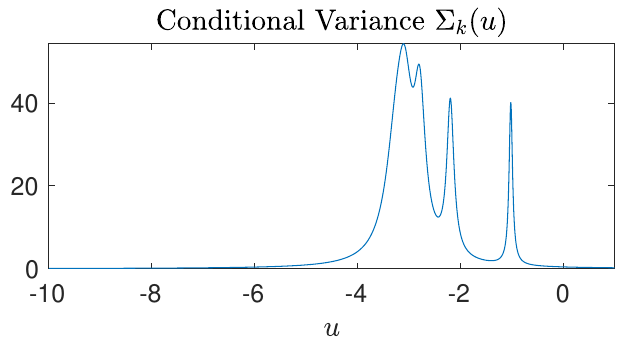}\\
    \end{minipage}
    \begin{minipage}{.49\textwidth}
        \centering
    \subcaption{Kolmogorov spectrum}\vspace{0.5em}
    \includegraphics[scale=0.395]{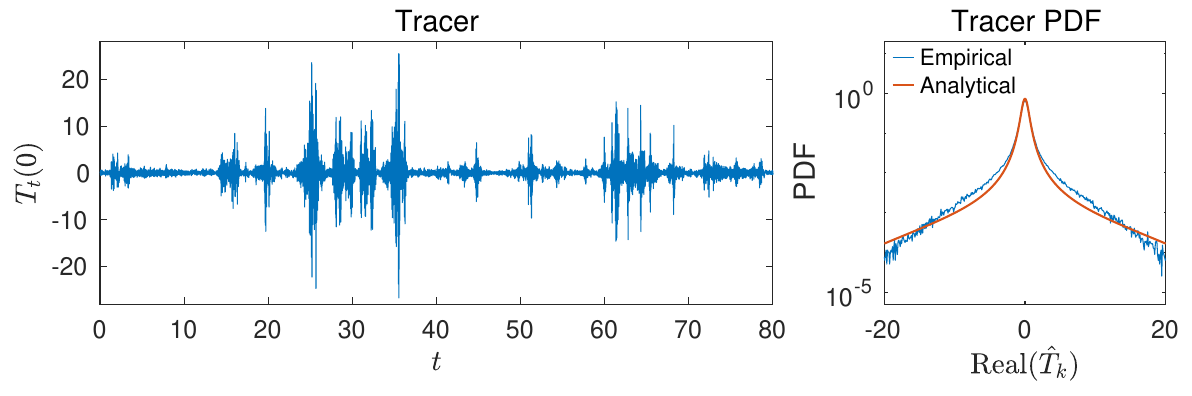}\\
    \includegraphics[scale=0.395]{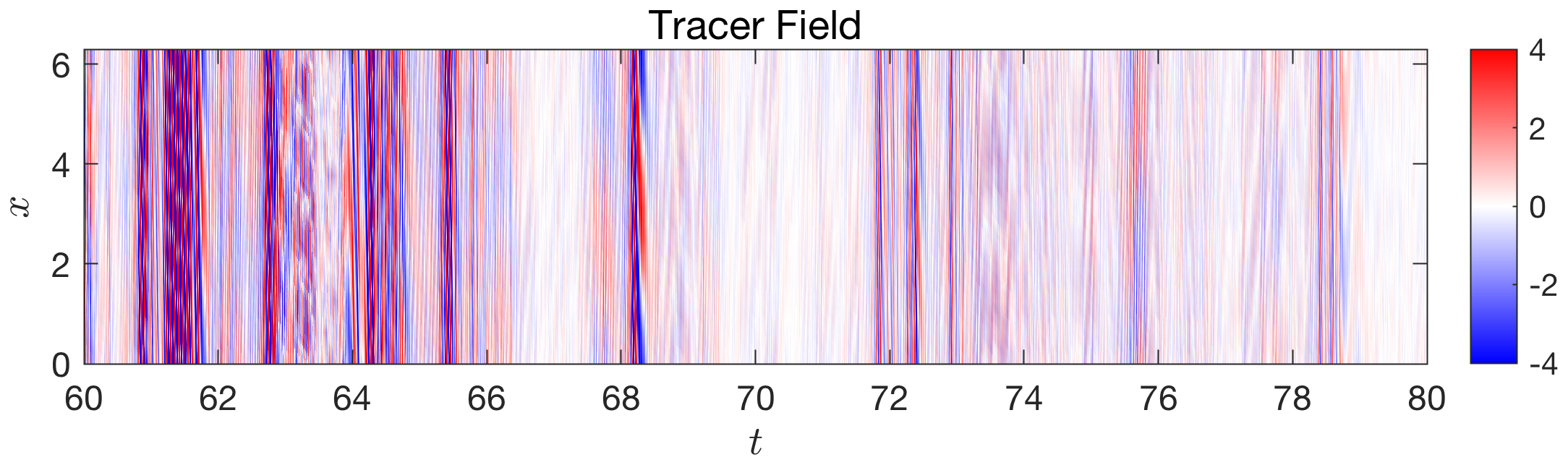}\\
    \includegraphics[scale=0.395]{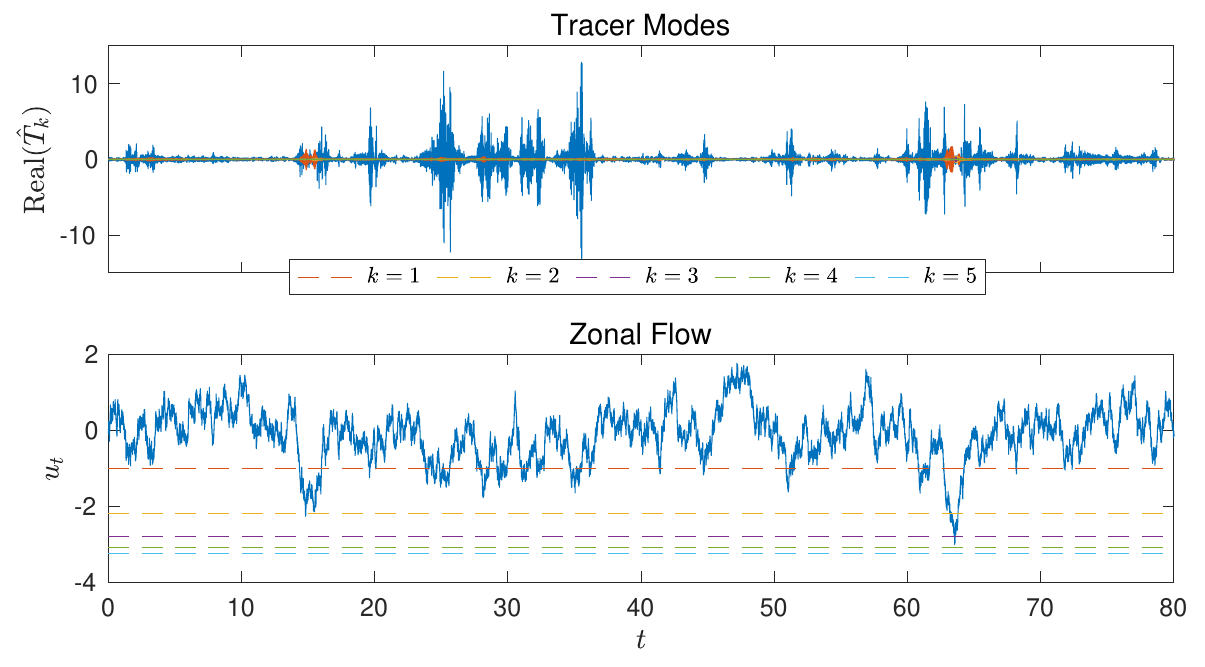}\\
    \includegraphics[scale=0.395]{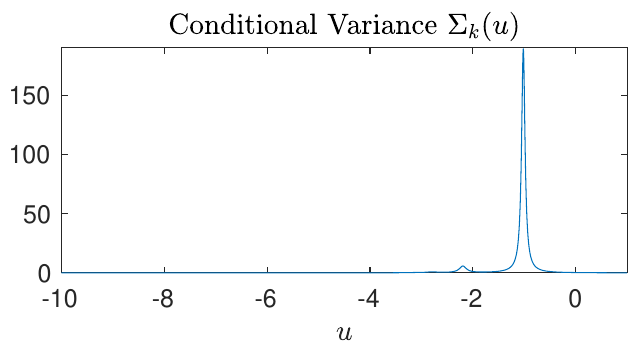}
    \end{minipage}
    \caption{Multiple mode, $\beta$-plane QG flow. Linear zonal flow with  $E_u = 0.5$ ($\gamma_u = 1,\sigma_u = 1$).}
    \label{fig:model_multiple_mode_beta_plane_q_g_flow_linear_zonal_fluctuations_equip}
\end{figure}

In~\cref{fig:model_multiple_mode_beta_plane_q_g_flow_linear_zonal_fluctuations_equip} we compare the results under linear zonal fluctuations with an equipartition spectrum and Kolmogorov spectrum. Observe the multiple resonance thresholds $u_{\text{res},k}$, which are plotted as dashed lines in the figure showing the zonal flow trajectory $u_t$. All bursts in the tracer time series are triggered by resonance crossings of $u_t$. The high frequency modes have thresholds that are further from the mean of $u_t$ and are increasingly rare to cross; consequently, their contribution to the overall tracer statistics is lower.

\emph{Under equipartition, tracer dynamics involve more active high-frequency modes and the tracer field exhibits finer-scale spatial features compared to the Kolmogorov case.} This occurs because all modes receive equal energy
 and contribute comparably whenever their thresholds are crossed. This behavior is clearly reflected in the conditional variance $\widetilde{\Sigma}$(u), which exhibits multiple peaks corresponding to each mode's resonance threshold, including prominent peaks for high-frequency modes. In contrast, under a Kolmogorov spectrum, high-frequency modes carry less energy, thus contributing less to the tracer dynamics. This leads us to the following important point: \emph{the tracer field statistics are primarily controlled by the most energetic modes that cross the resonance thresholds most frequently.}

\subsubsection{Stochastic zonal   flow with nonlinear dynamics}

\begin{figure}[htbp!]
    \centering
    \begin{minipage}{.49\textwidth}\centering
    \subcaption{Equipartition: $f = 1.0$}\vspace{0.5em}
    \includegraphics[scale=0.395]{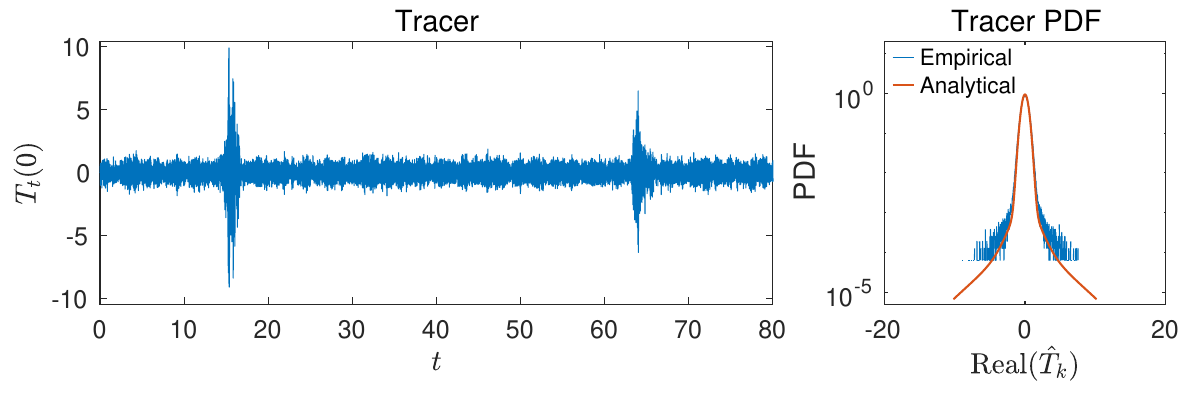}\\
    \includegraphics[scale=0.395]{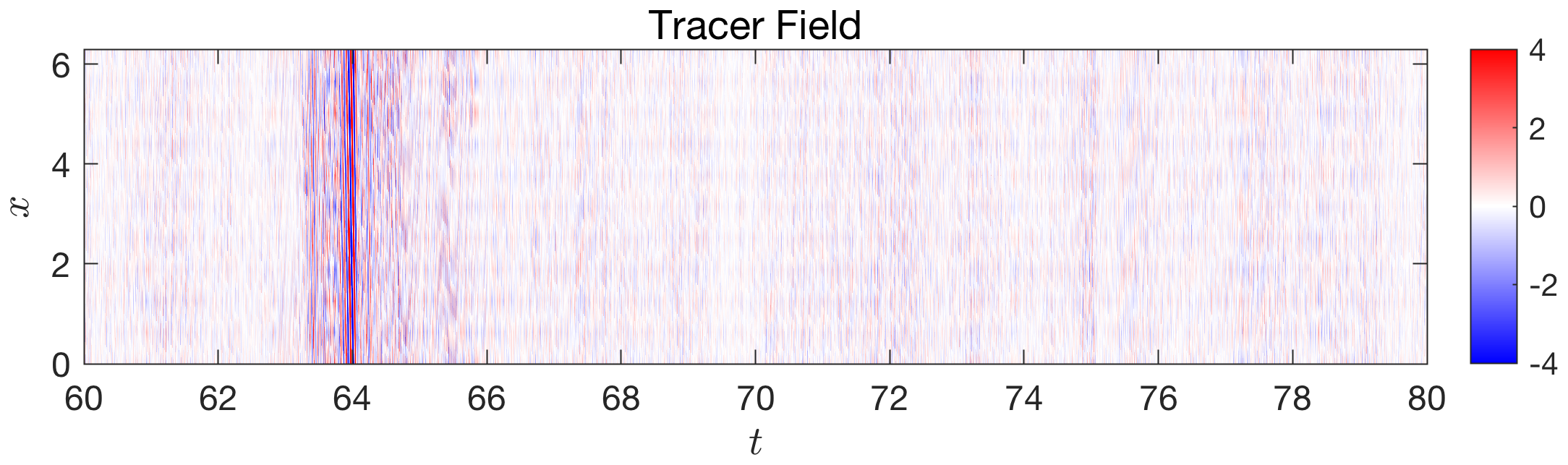}\\
    \includegraphics[scale=0.395]{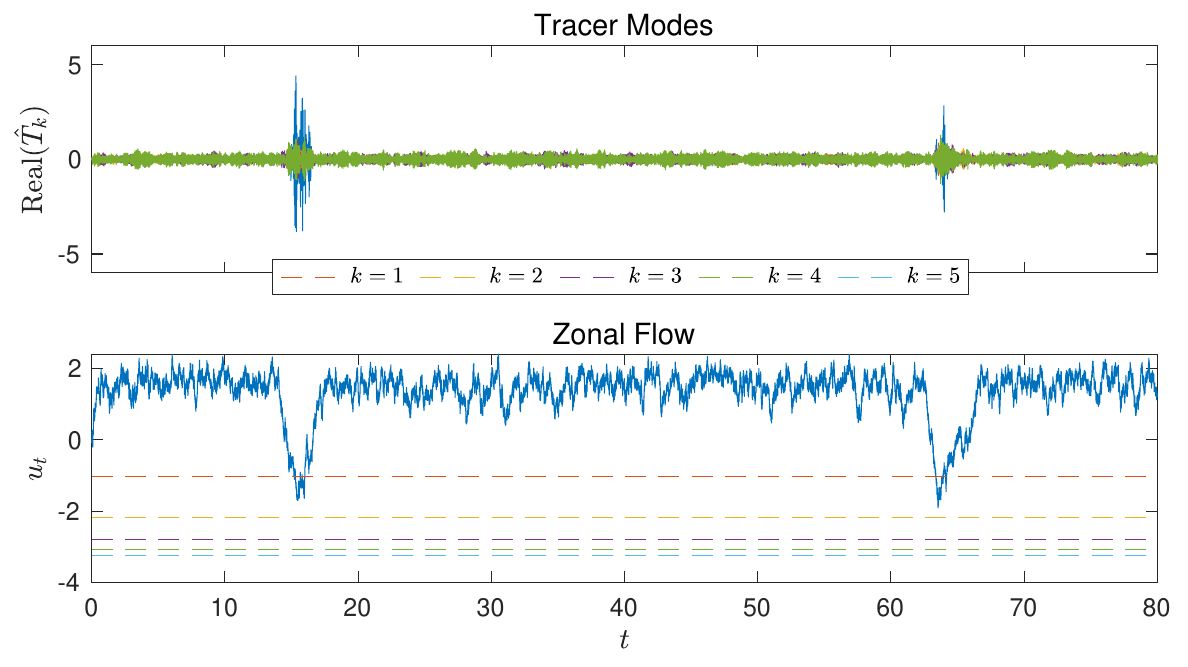}
    \end{minipage}
    \begin{minipage}{.49\textwidth}\centering
    \subcaption{Equipartition: $f = 0.0$}\vspace{0.5em}
    \includegraphics[scale=0.395]{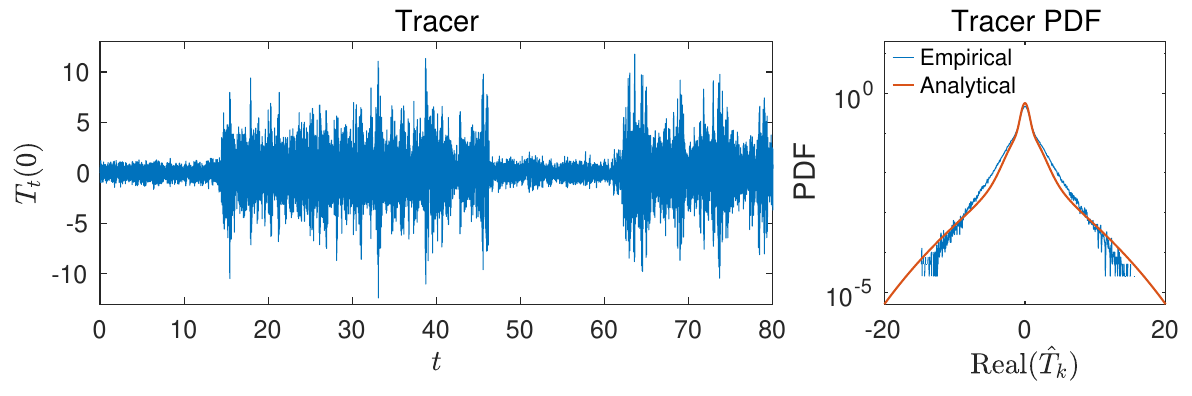}\\
    \includegraphics[scale=0.395]{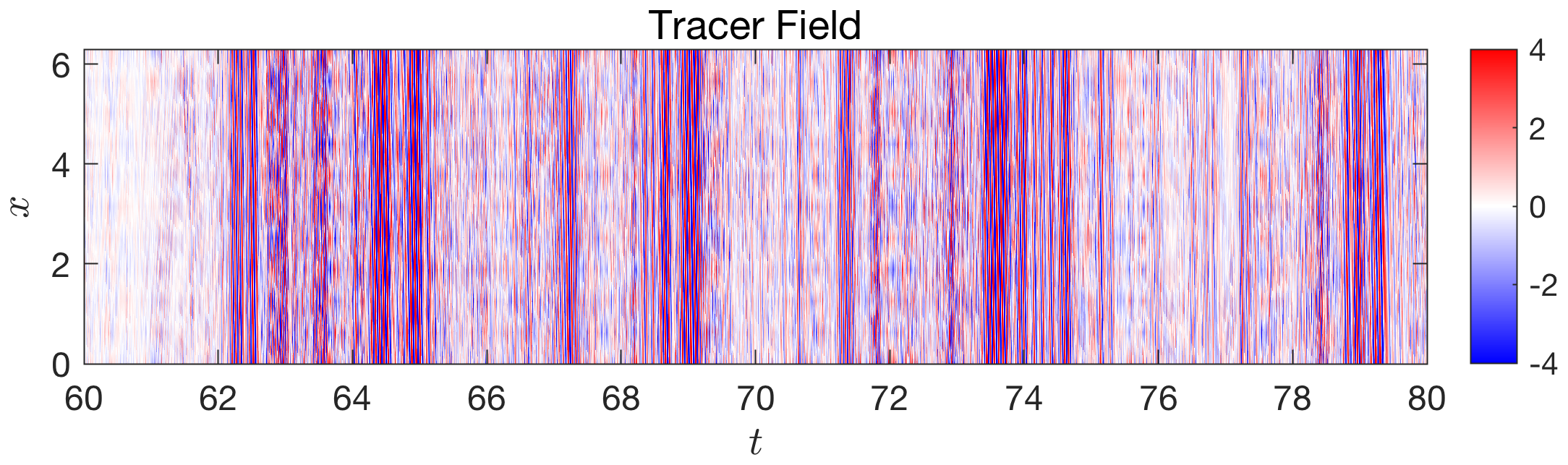}\\
    \includegraphics[scale=0.395]{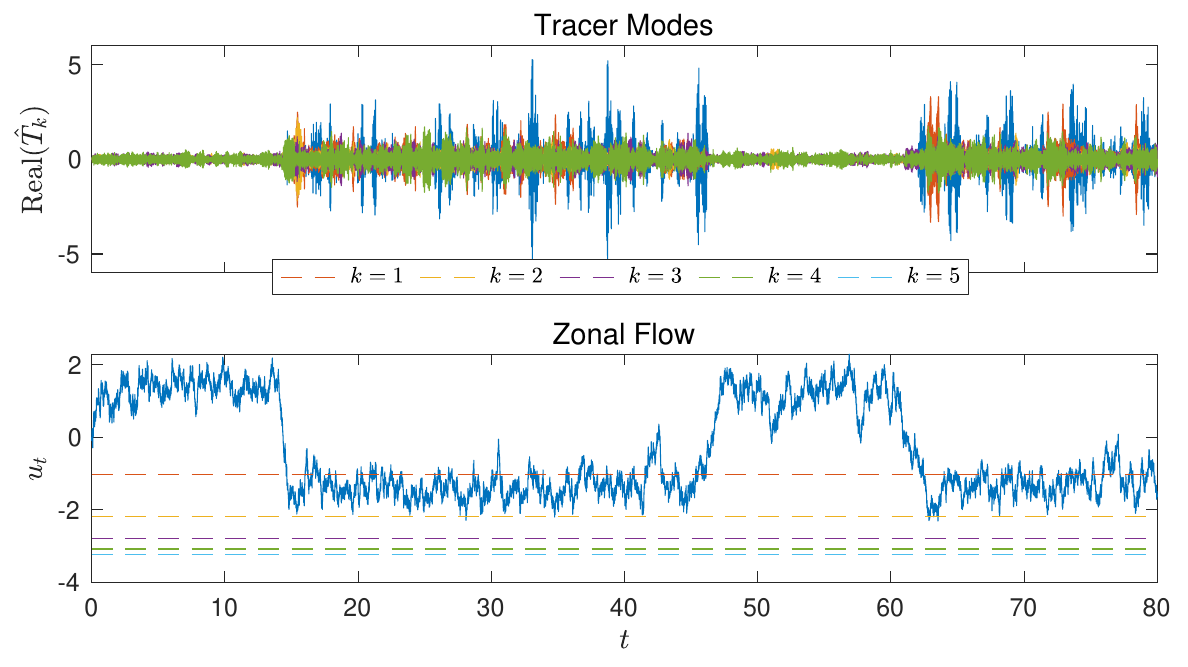}
    \end{minipage}
    \caption{Multiple mode, $\beta$-plane QG flow. Nonlinear zonal flow under zero multiplicative noise. }
    \label{fig:multiple_mode_beta_plane_qg_flow_nonlinear_zonal_fluctuations_with_zero_multiplicative_noise_and_equipartition_shear_spectrum_nk5_cases}
\end{figure}

\begin{figure}[htbp!]
    \centering
    \begin{minipage}{.49\textwidth}\centering
    \subcaption{Equipartition: $f = 1.0$}\vspace{0.5em}
    \includegraphics[scale=0.395]{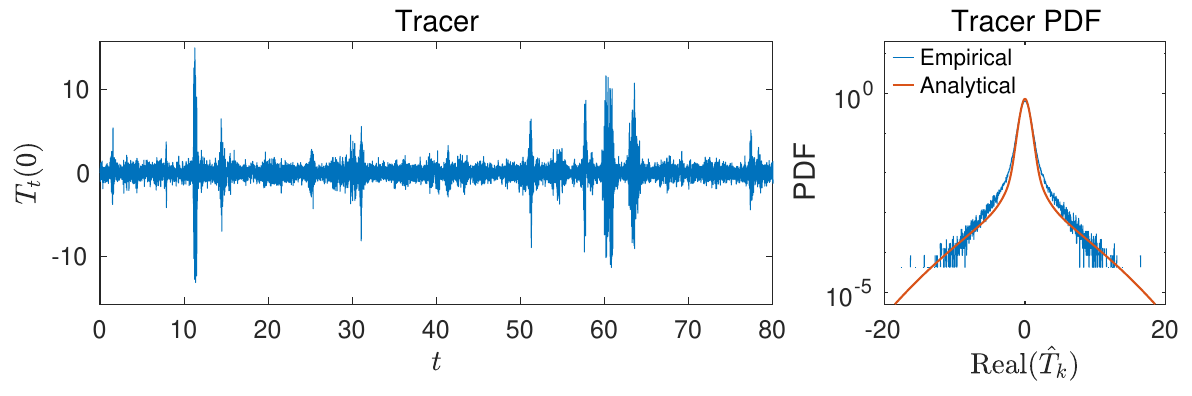}\\
    \includegraphics[scale=0.395]{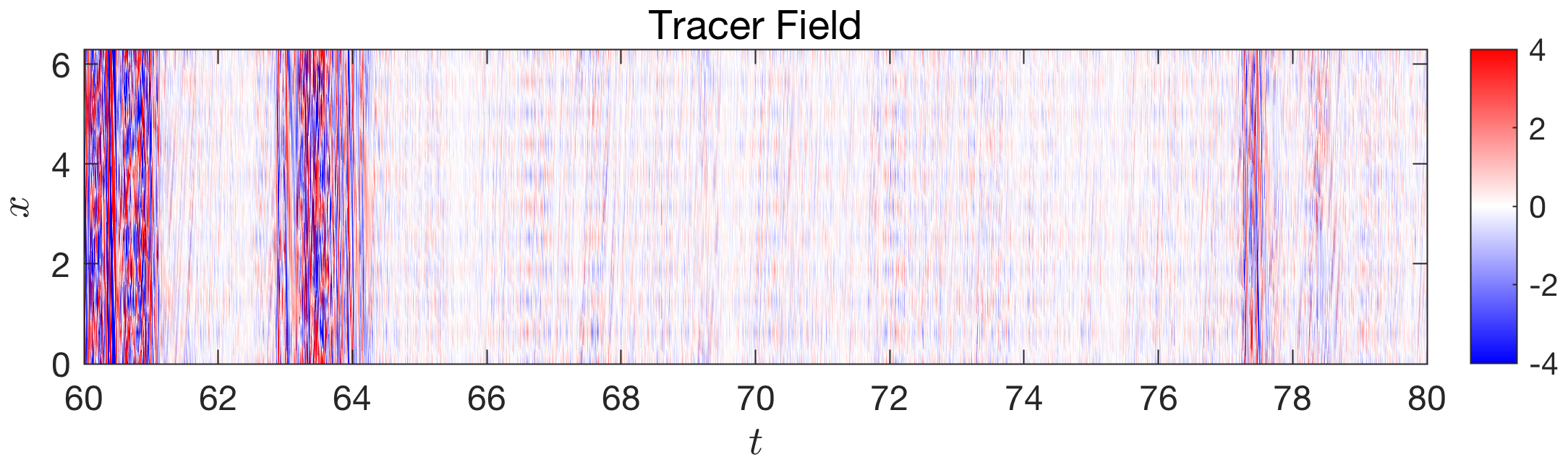}\\
    \includegraphics[scale=0.395]{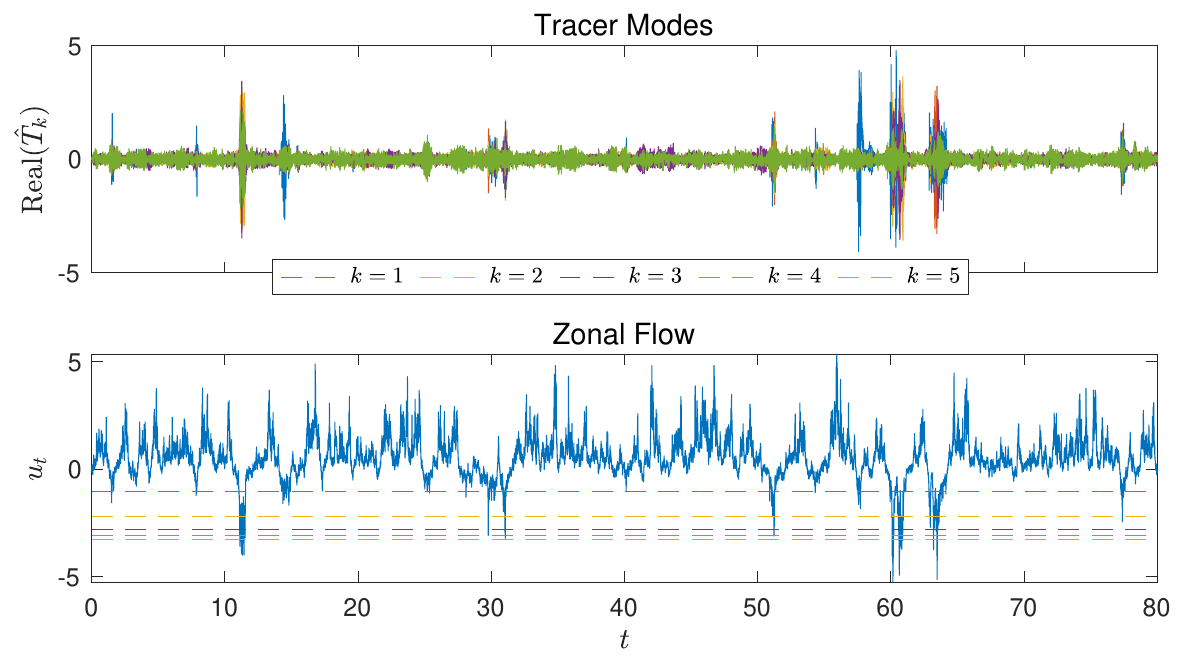}
    \end{minipage}
    \begin{minipage}{.49\textwidth}\centering
    \subcaption{Equipartition: $f = 0.0$}\vspace{0.5em}
    \includegraphics[scale=0.395]{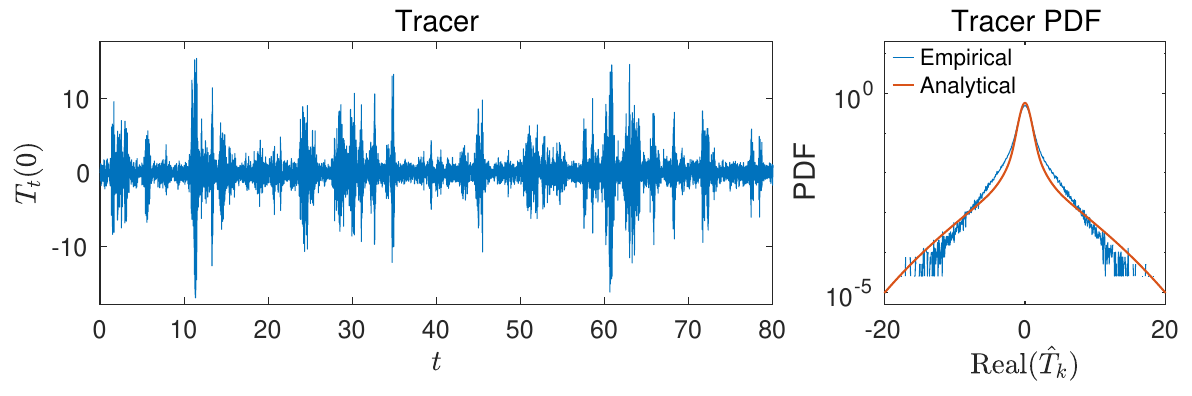}\\
    \includegraphics[scale=0.395]{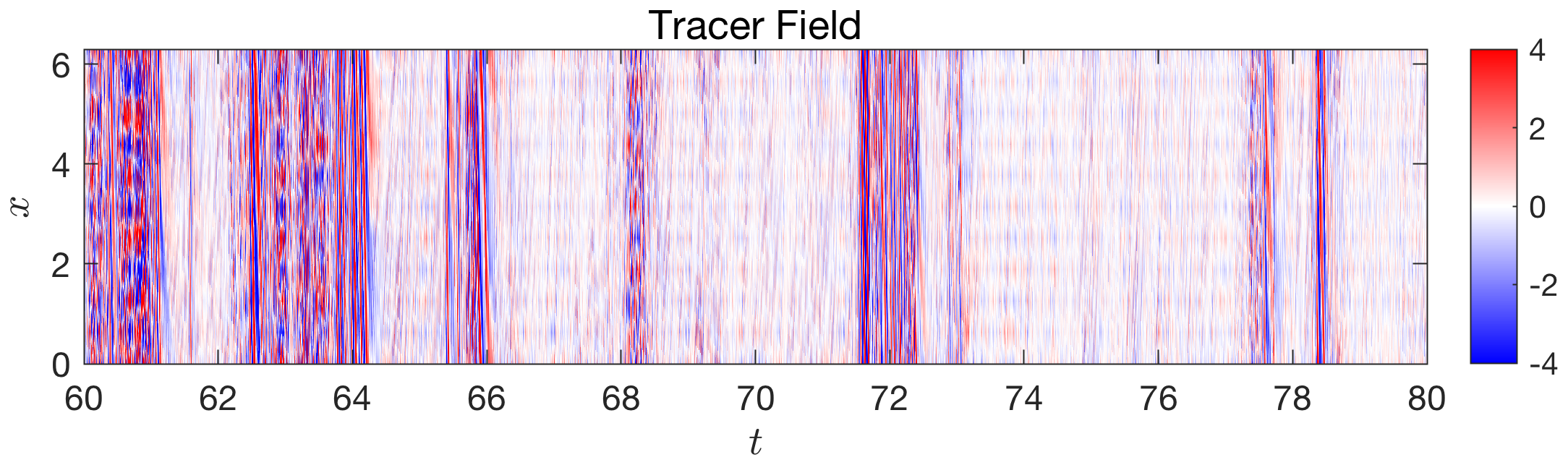}\\
    \includegraphics[scale=0.395]{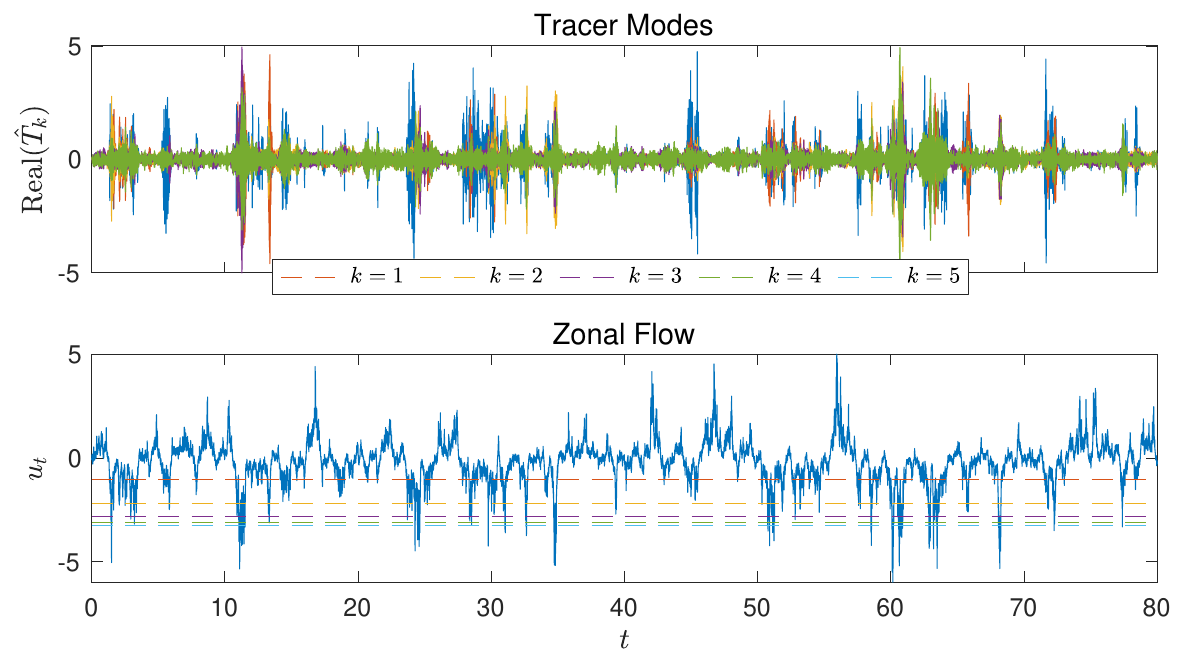}
    \end{minipage}
    \caption{Multiple mode, $\beta$-plane QG flow. Nonlinear zonal flow under multiplicative noise.}
    \label{fig:multiple_mode_beta_plane_qg_flow_nonlinear_zonal_fluctuations_with_multiplicative_noise_and_equipartition_shear_spectrum_nk5_cases}
\end{figure}

We now investigate cases involving nonlinear zonal flows, as described in~\cref{ssub:non_linear_models}. Matching the single-mode analysis, we explore both \emph{additive-only} noise  (with zero multiplicative component,  $B = 0$, see~\cref{fig:multiple_mode_beta_plane_qg_flow_nonlinear_zonal_fluctuations_with_zero_multiplicative_noise_and_equipartition_shear_spectrum_nk5_cases}) and \emph{strongly nonlinear regimes} (with multiplicative noise $ B = 2.5$, see~\cref{fig:multiple_mode_beta_plane_qg_flow_nonlinear_zonal_fluctuations_with_multiplicative_noise_and_equipartition_shear_spectrum_nk5_cases}). As before we fix $\sigma_u = 1$ and  consider  cubic nonlinearities    with $c = 1$ and $b=0$.

As observed in the single-mode case~\cref{ssub:stochastic_zonal_mean_flow_with_nonlinear_dynamics}, non-Gaussian zonal flow statistics and nonlinear dynamics can significantly influence tracer intermittency and extreme event generation. This remains true in the multi-mode setting, though now the response is further modulated by the \emph{non-synchronized resonance thresholds} across wavenumbers, typical of dispersive systems.

Notably, we observe that nonlinear zonal dynamics can either \emph{enhance or suppress tracer intermittency}, depending on the skewness and kurtosis of the zonal flow distribution. For instance, as seen in~\cref{fig:multiple_mode_beta_plane_qg_flow_nonlinear_zonal_fluctuations_with_multiplicative_noise_and_equipartition_shear_spectrum_nk5_cases}, skewness toward resonance thresholds, induced by asymmetric forcing, leads to \emph{enhanced intermittency}, while skewness away from the resonant range suppresses it. Similarly, large kurtosis increases the probability of reaching multiple resonance thresholds (especially for higher wavenumbers), increasing tracer burst magnitude. Conversely, distributions with low kurtosis and skewness directed away from resonant values (a `non-resonant' forcing scenario) yield \emph{suppressed tracer variability}. These effects are most pronounced under an equipartition spectrum, where all modes are energetically active.

In~\cref{fig:multiple_mode_beta_plane_qg_flow_nonlinear_zonal_fluctuations_with_zero_multiplicative_noise_and_equipartition_shear_spectrum_nk5_cases}, we show cases with purely additive noise. Similar to the single-mode results, we find that in certain parameter regimes, \emph{higher wavenumber modes can be excited `from below'}. That is, tracer activity emerges first in the smallest scales before propagating to larger scales. This phenomenon, particularly evident in the case with the double-well zonal flow, contrasts the typical scenario where large-scale modes dominate first. Additionally, these regimes exhibit a form of \emph{on-off intermittency}, driven by the zonal flow transitioning between the two metastable states.

While in the single-mode setting we noted that \emph{statistical similarity} could occur between linear and nonlinear cases, provided the resonance crossing frequency matched, this equivalence generally breaks down in multi-mode, dispersive flows. The reason is that resonance thresholds are no longer in general synchronized, and nonlinearities affect each mode differently. Therefore, the cumulative tracer response differes from the linear case, especially under equipartition. In contrast, in non-dispersive or random flows, where the phase speeds are synchronized, matching resonance frequencies may still result in comparable statistics.

\subsubsection{Influence of shear flow structure on tracer intermittency}

To understand how wave dynamics affect tracer intermittency, we compare three classes of shear flows under equipartition energy spectra, each with distinct resonance characteristics:
\begin{itemize}
\item {Random shear flows:}
\begin{equation}
 a_k= b_k = 0  \quad\implies\quad u_{\text{res}} = 0
\end{equation}

\item {Advection:}
\begin{equation}
  a_k= 0 ,\quad b_k=-ck \quad\implies\quad u_{\text{res}} = c
\end{equation}

\item {Quasi-geostrophic $\beta$-plane flows:}
\begin{equation}
    a_k = k\Bigl(\frac{F}{k^2}-1\Bigr), \quad b_k = \frac{\beta k}{  k^2 + F} \quad\implies\quad u_{\text{res}} =-\frac{\beta k^2}{F (k^2 + F)}
\end{equation}
\end{itemize}

For advective flows, we set $c = 1.0183$ to match the $k=1$ resonance threshold with QG parameters $\beta = 8.91$, $F = 2.5$, for fair comparison between advective and dispersive flows. 
For linear zonal dynamics, we match the resonance crossing rate by adjusting the forcing parameter $f$. However, for nonlinear zonal flows  the dynamics are inherently coupled to the equilibrium structure, making it impossible to shift resonance crossing rates without potentially altering the dynamical regime.  
Rather than artificially modifying the nonlinear parameters to match crossing rates,   we examine how these different resonance   patterns interact under identical zonal flow realizations. This approach demonstrates how  the interplay between zonal flow statistics and resonance threshold locations determines tracer intermittency.
The comparative analysis reveals several key insights (see~\cref{sec:supplementary_numerical_simulations} for details):
\begin{itemize}
\item \emph{Nonlinear zonal flow effects:} Nonlinear zonal flows in `on-off' regimes generate intermittent bursts with enhanced persistence compared to linear zonal flows. This behavior results from the   interplay between metastable zonal dynamics and mode-dependent resonance thresholds, with the effects being most pronounced when zonal flow transitions occur near resonance values.

\item \emph{Dispersive vs. non-dispersive shear flows:} Dispersive flows exhibit fewer small-scale spatial features compared to their non-dispersive counterparts. This occurs because higher wavenumber modes in dispersive systems have increasingly rare resonance threshold crossings, resulting in lower probabilities for extreme events in the tracer PDF.

\item \emph{Advective vs. random shear flows:} Advective flows produce more coherent spatio-temporal extreme events with prominent oscillatory behavior due to their non-zero wave speeds. While the limiting analytical tracer PDF predictions, as $\epsilon \to 0$, remain identical, finite-$\epsilon$ effects in advective flows show enhanced probability mass for intermediate magnitude fluctuations and exponential-like tail behavior in highly intermittent regimes.
\end{itemize}

In summary, nonlinear zonal dynamics in multi-mode systems reveal \emph{qualitatively distinct regimes} compared to linear stochastic forcing. The tracer field statistics become   sensitive to the \emph{structure of the zonal flow distribution}, and new phenomena such as \emph{non-resonant suppression, multi-scale on-off intermittency}, and \emph{asymmetric excitation} arise, especially under dispersive and equipartition conditions.


\section{Conclusion}\label{sec:conclusions}
This study reveals the critical mechanisms through which stochastic zonal and shear flows produce tracer intermittency in turbulent diffusion with a mean gradient. By making assumptions that  preserve key physical mechanisms of tracer transport, following previous literature, we derived analytically tractable pathwise solutions and explicit expressions for the tracer PDF. Following this, a simplified analytical approximation was derived for the conditional variance of the tracer field, under a slowly varying velocity field, which provides a closed form equation for the tracer PDF that was validated through numerical experiments. From these analytical results and numerical experiments, we demonstrated several key velocity field features that determine how non-Gaussianity and extreme events  arise in tracer fields.

The primary result reveals that resonance, through phase speed alignment between the zonal and shear flow, rather than transient instabilities, are responsible for the observed tracer intermittency. When the phase speeds of zonal flow fluctuations cross specific   thresholds, determined by the underlying wave dynamics of the velocity field, dramatic amplification of tracer variance occurs. This resonance-driven mechanism represents a   pathway to turbulent intermittency that differs from finite-time Lyapunov instabilities. 
The analytical framework demonstrates that the conditional variance of the tracer field peaks when $\omega_{R,k} = 0$, corresponding to resonance conditions between the zonal flow, shear flow, and tracer field. This provides a quantitative explanation  of extreme events and allows for prediction of intermittency if the flow characteristics are known.
 
 Importantly, we identified significant differences in tracer behavior across flow regimes. Dispersive flows with wavelike features exhibit separated resonance thresholds across wavenumbers, leading to sequential excitation of modes and smoother extreme events. In contrast, random shear flows and non-dispersive waves synchronize these thresholds, exciting all scales simultaneously, producing sharper, more intermittent, structures with enhanced small-scale features.  Our comparison of equipartition and Kolmogorov energy spectra show that the spatial structure of extreme events is strongly influenced by the distribution of energy in the shear flow. Under equipartition, multiple peaks in the conditional variance lead to stronger intermittency with pronounced small-scale features, whereas the Kolmogorov spectrum produces more large-scale dominated extreme events.

 The nonlinear dynamics of zonal flow is crucial in modulating intermittency. Despite being statistically non-Gaussian, nonlinear zonal flows do not necessarily enhance tracer intermittency; rather, their effect depends on how frequently they cross resonance thresholds. This challenges linearization approaches and highlights the importance of accurately capturing zonal flow statistics in turbulent transport models.

 The results in this paper have  implications for modeling and prediction of tracer transport in geophysical and environmental applications. The identified resonance mechanism provides a simple   basis for understanding tracer bursts in systems ranging from atmospheric pollutant transport to oceanic mixing. Furthermore, the demonstrated sensitivity to flow characteristics shows that accurate representation of zonal and shear flow statistics is essential for reliable prediction of extreme tracer burst events.
 Future work could extend this approach to three-dimensional flows involving vertical shear   and   use the model for various data assimilation (DA) and uncertainty quantification (UQ) applications. The analytical tractability of our approach makes it particularly valuable for developing and testing various DA and UQ schemes  that can capture non-Gaussian statistics of tracer intermittency while remaining computationally efficient.

\subsection*{Acknowledgments}
M.A.M is partially supported by an NSERC Discovery Grant and Alberta Innovates. D.Q. is partially supported by ONR Grant N00014-24-1-2192, and NSF Grant DMS-2407361. The authors would also like to dedicate this paper to the memory of Andrew J. Majda whose contributions  made this work possible.

\printbibliography

@article{bourlioux2002,
  title = {Elementary Models with Probability Distribution Function Intermittency for Passive Scalars with a Mean Gradient},
  author = {Bourlioux, A. and Majda, A. J.},
  date = {2002-02-01},
  journaltitle = {Physics of Fluids},
  shortjournal = {Physics of Fluids},
  volume = {14},
  number = {2},
  pages = {881--897},
  issn = {1070-6631},
  doi = {10.1063/1.1430736}
}

@article{chen2016,
  title = {Filtering Nonlinear Turbulent Dynamical Systems through Conditional {{Gaussian}} Statistics},
  author = {Chen, Nan and Majda, Andrew J.},
  date = {2016-12-01},
  journaltitle = {Monthly Weather Review},
  volume = {144},
  number = {12},
  pages = {4885--4917},
  issn = {0027-0644, 1520-0493},
  doi = {10.1175/MWR-D-15-0437.1},
  langid = {english}
}

@article{chen2021,
  title = {Lagrangian {{Data Assimilation}} and {{Parameter Estimation}} of an {{Idealized Sea Ice Discrete Element Model}}},
  author = {Chen, Nan and Fu, Shubin and Manucharyan, Georgy},
  date = {2021-10},
  journaltitle = {Journal of Advances in Modeling Earth Systems},
  shortjournal = {J Adv Model Earth Syst},
  volume = {13},
  number = {10},
  pages = {e2021MS002513},
  issn = {1942-2466, 1942-2466},
  doi = {10.1029/2021MS002513},
  langid = {english}
}

@article{chen2023,
  title = {Uncertainty Quantification of Nonlinear {{Lagrangian}} Data Assimilation Using Linear Stochastic Forecast Models},
  author = {Chen, Nan and Fu, Shubin},
  date = {2023-10},
  journaltitle = {Physica D: Nonlinear Phenomena},
  shortjournal = {Physica D: Nonlinear Phenomena},
  volume = {452},
  pages = {133784},
  issn = {01672789},
  doi = {10.1016/j.physd.2023.133784},
  langid = {english}
}

@article{janson2010,
  title = {Roots of Polynomials of Degrees 3 and 4},
  author = {Janson, Svante},
  date = {2010-09-13},
  eprint = {1009.2373},
  eprinttype = {arXiv},
  eprintclass = {math},
  doi = {10.48550/arXiv.1009.2373},
  journal = {arXiv:1009.2373}
}

@article{kolmogorov1941,
  title = {The Local Structure of Turbulence in Incompressible Viscous Fluid for Very Large {{Reynolds}}},
  author = {Kolmogorov, Andrey Nikolaevich},
  date = {1941},
  journaltitle = {Numbers. In Dokl. Akad. Nauk SSSR},
  volume = {30},
  pages = {301}
}

@article{lee2017,
  title = {Stochastic {{Superparameterization}} and {{Multiscale Filtering}} of {{Turbulent Tracers}}},
  author = {Lee, Yoonsang and Majda, Andrew J. and Qi, Di},
  date = {2017-01},
  journaltitle = {Multiscale Modeling \& Simulation},
  shortjournal = {Multiscale Model. Simul.},
  volume = {15},
  number = {1},
  pages = {215--234},
  issn = {1540-3459, 1540-3467},
  doi = {10.1137/16M1080239},
  langid = {english}
}

@book{liptser2001,
  title = {Statistics of Random Processes {{II}}. {{Applications}}},
  author = {Liptser, Robert S. and Shiryaev, Albert N.},
  date = {2001},
  series = {Stochastic Modelling and Applied Probability},
  edition = {2},
  number = {6},
  publisher = {Springer-Verlag Berlin Heidelberg},
  doi = {10.1007/978-3-662-10028-8},
  isbn = {978-3-540-63928-2}
}

@article{majda1999,
  title = {Simplified Models for Turbulent Diffusion: {{Theory}}, Numerical Modelling, and Physical Phenomena},
  shorttitle = {Simplified Models for Turbulent Diffusion},
  author = {Majda, Andrew J. and Kramer, Peter R.},
  date = {1999-06},
  journaltitle = {Physics Reports},
  shortjournal = {Physics Reports},
  volume = {314},
  number = {4--5},
  pages = {237--574},
  issn = {03701573},
  doi = {10.1016/S0370-1573(98)00083-0},
  langid = {english}
}

@article{majda2009,
  title = {Normal Forms for Reduced Stochastic Climate Models},
  author = {Majda, Andrew J. and Franzke, Christian and Crommelin, Daan},
  date = {2009-03-10},
  journaltitle = {Proceedings of the National Academy of Sciences},
  shortjournal = {Proc. Natl. Acad. Sci. U.S.A.},
  volume = {106},
  number = {10},
  pages = {3649--3653},
  issn = {0027-8424, 1091-6490},
  doi = {10.1073/pnas.0900173106},
  langid = {english}
}

@article{majda2013,
  title = {Elementary Models for Turbulent Diffusion with Complex Physical Features: Eddy Diffusivity, Spectrum and Intermittency},
  shorttitle = {Elementary Models for Turbulent Diffusion with Complex Physical Features},
  author = {Majda, Andrew J. and Gershgorin, Boris},
  date = {2013-01-13},
  journaltitle = {Philosophical Transactions of the Royal Society A: Mathematical, Physical and Engineering Sciences},
  shortjournal = {Phil. Trans. R. Soc. A.},
  volume = {371},
  number = {1982},
  pages = {20120184},
  issn = {1364-503X, 1471-2962},
  doi = {10.1098/rsta.2012.0184},
  langid = {english}
}

@article{majda2013a,
  title = {Elementary Models for Turbulent Diffusion with Complex Physical Features: Eddy Diffusivity, Spectrum and Intermittency},
  shorttitle = {Elementary Models for Turbulent Diffusion with Complex Physical Features},
  author = {Majda, Andrew J. and Gershgorin, Boris},
  date = {2013-01-13},
  journaltitle = {Philosophical Transactions of the Royal Society A: Mathematical, Physical and Engineering Sciences},
  volume = {371},
  number = {1982},
  pages = {20120184},
  publisher = {Royal Society},
  doi = {10.1098/rsta.2012.0184}
}

@article{majda2015,
  title = {Intermittency in Turbulent Diffusion Models with a Mean Gradient},
  author = {Majda, Andrew J and Tong, Xin T},
  date = {2015-10-01},
  journaltitle = {Nonlinearity},
  shortjournal = {Nonlinearity},
  volume = {28},
  number = {11},
  pages = {4171--4208},
  issn = {0951-7715, 1361-6544},
  doi = {10.1088/0951-7715/28/11/4171}
}

@article{marcoavellaneda1994,
  title = {Simple Examples with Features of Renormalization for Turbulent Transport},
  author = {{Marco Avellaneda} and Majda, Andrew J.},
  date = {1994-02-15},
  journaltitle = {Philosophical Transactions of the Royal Society of London. Series A: Physical and Engineering Sciences},
  shortjournal = {Phil. Trans. R. Soc. Lond. A},
  volume = {346},
  number = {1679},
  pages = {205--233},
  issn = {0962-8428, 2054-0299},
  doi = {10.1098/rsta.1994.0019},
  langid = {english}
}

@article{mohamad2015,
  title = {Probabilistic {{Description}} of {{Extreme Events}} in {{Intermittently Unstable Dynamical Systems Excited}} by {{Correlated Stochastic Processes}}},
  author = {Mohamad, Mustafa A. and Sapsis, Themistoklis P.},
  date = {2015-01},
  journaltitle = {SIAM/ASA Journal on Uncertainty Quantification},
  shortjournal = {SIAM/ASA J. Uncertainty Quantification},
  volume = {3},
  number = {1},
  pages = {709--736},
  issn = {2166-2525},
  doi = {10.1137/140978235},
  langid = {english}
}

@article{mohamad2020,
  title = {Recovering the {{Eulerian}} Energy Spectrum from Noisy {{Lagrangian}} Tracers},
  author = {Mohamad, Mustafa A. and Majda, Andrew J.},
  date = {2020-02-01},
  journaltitle = {Physica D: Nonlinear Phenomena},
  shortjournal = {Physica D: Nonlinear Phenomena},
  volume = {403},
  pages = {132374},
  issn = {0167-2789},
  doi = {10.1016/j.physd.2020.132374}
}

@article{neelin2010,
  title = {Long Tails in Deep Columns of Natural and Anthropogenic Tropospheric Tracers},
  author = {Neelin, J. David and Lintner, Benjamin R. and Tian, Baijun and Li, Qinbin and Zhang, Li and Patra, Prabir K. and Chahine, Moustafa T. and Stechmann, Samuel N.},
  date = {2010-03-01},
  journaltitle = {Geophysical Research Letters},
  volume = {37},
  number = {5},
  publisher = {John Wiley \& Sons, Ltd},
  issn = {1944-8007},
  doi = {10.1029/2009GL041726},
  langid = {english}
}

@article{richardson1926,
  title = {Atmospheric Diffusion Shown on a Distance-Neighbour Graph},
  author = {Richardson, Lewis Fry},
  date = {1926-04},
  journaltitle = {Proceedings of the Royal Society of London. Series A, Containing Papers of a Mathematical and Physical Character},
  volume = {110},
  number = {756},
  pages = {709--737},
  doi = {10.1098/rspa.1926.0043}
}

@article{smith2005,
  title = {Tracer Transport along and across Coherent Jets in Two-Dimensional Turbulent Flow},
  author = {Smith, K. Shafer},
  date = {2005-12},
  journaltitle = {Journal of Fluid Mechanics},
  volume = {544},
  pages = {133--142},
  publisher = {Cambridge University Press},
  issn = {1469-7645, 0022-1120},
  doi = {10.1017/S0022112005006750},
  langid = {english}
}

@article{sreenivasan2019,
  title = {Turbulent Mixing: {{A}} Perspective},
  shorttitle = {Turbulent Mixing},
  author = {Sreenivasan, Katepalli R.},
  date = {2019-09-10},
  journaltitle = {Proceedings of the National Academy of Sciences},
  volume = {116},
  number = {37},
  pages = {18175--18183},
  publisher = {Proceedings of the National Academy of Sciences},
  doi = {10.1073/pnas.1800463115}
}

@article{taylor1922,
  title = {Diffusion by Continuous Movements},
  author = {Taylor, G. I.},
  date = {1922},
  journaltitle = {Proceedings of the London Mathematical Society},
  volume = {s2-20},
  number = {1},
  eprint = {https://londmathsoc.onlinelibrary.wiley.com/doi/pdf/10.1112/plms/s2-20.1.196},
  pages = {196--212},
  doi = {10.1112/plms/s2-20.1.196}
}

@book{vallis2006,
  title = {Atmospheric and Oceanic Fluid Dynamics: Fundamentals and Large-Scale Circulation},
  shorttitle = {Atmospheric and Oceanic Fluid Dynamics},
  author = {Vallis, Geoffrey K.},
  date = {2006},
  publisher = {Cambridge university press},
  location = {Cambridge},
  isbn = {978-0-521-84969-2},
  langid = {english}
}

@article{warhaft2000,
  title = {Passive {{Scalars}} in {{Turbulent Flows}}},
  author = {Warhaft, Z.},
  date = {2000-01-01},
  journaltitle = {Annual Review of Fluid Mechanics},
  volume = {32},
  pages = {203--240},
  publisher = {Annual Reviews},
  issn = {0066-4189, 1545-4479},
  doi = {10.1146/annurev.fluid.32.1.203},
  issue = {Volume 32, 2000},
  langid = {english}
}

@article{yuan2011,
  title = {Invariant Measures and Asymptotic {{Gaussian}} Bounds for Normal Forms of Stochastic Climate Model},
  author = {Yuan, Yuan and Majda, Andrew J.},
  date = {2011-05},
  journaltitle = {Chinese Annals of Mathematics, Series B},
  shortjournal = {Chin. Ann. Math. Ser. B},
  volume = {32},
  number = {3},
  pages = {343--368},
  issn = {0252-9599, 1860-6261},
  doi = {10.1007/s11401-011-0647-2},
  langid = {english}
}

\clearpage
\appendix

\renewcommand{\thefigure}{\thesection.\arabic{figure}}
\setcounter{figure}{0}

\renewcommand{\theequation}{\thesection.\arabic{equation}}
\setcounter{equation}{0}

\section{Zonal Flow Model Details}

\subsection{Dynamical regimes}\label{appendixa:zonal}
\begin{figure}[htbp!]
\centering
    \includegraphics[width=0.325\textwidth]{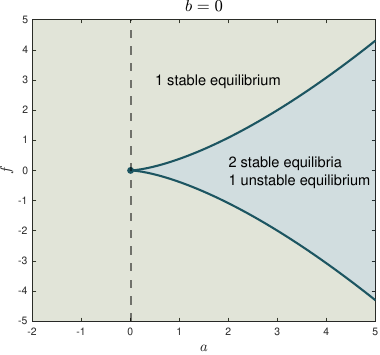}
    \hspace{0.05\textwidth}
    \includegraphics[width=0.325\textwidth]{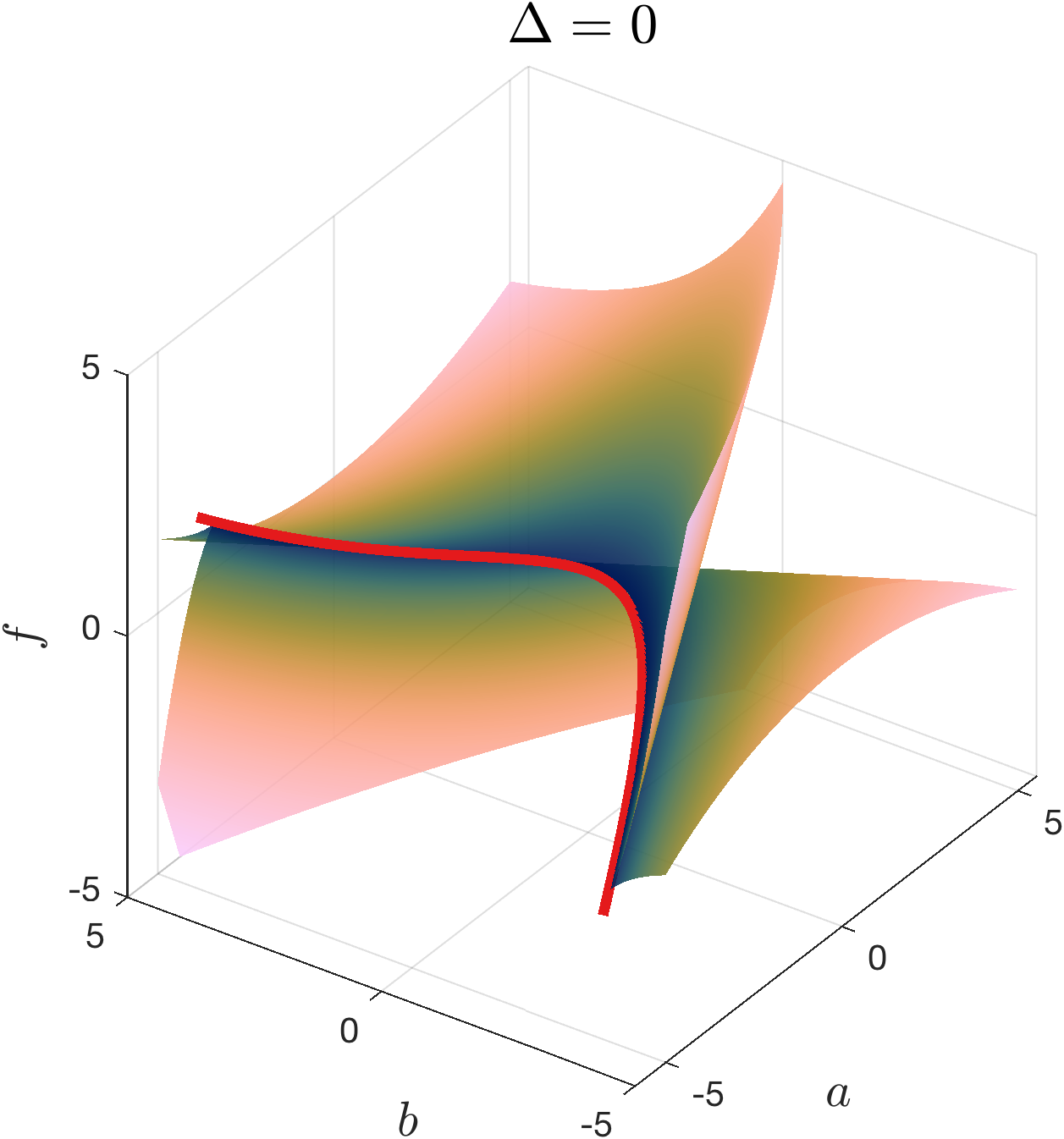}
    \caption{Regimes of the deterministic nonlinear cubic model in $(a,f)$ parameter space for $b=0$ values (left panel). The dark shaded area is bounded by the dividing curve $f_b^\pm$ in~\cref{eq:boundary} with points on the boundary having one unstable and one stable fixed points. Boundary for the discriminant $\Delta=0$ is shown in the right panel.}
    \label{fig:regcubic}
\end{figure}

To study the dynamical regimes of the nonlinear cross sweep model in~\cref{eq:cubmodel} we consider the deterministic system with no noise and study its fixed points:
\begin{equation}
\frac{d x}{dt} = f + a x + b x^2 -c x^3.
\end{equation}
The three roots of the cubic equation $f + a x + b x^2 -c x^3 =0$   determine the equilibrium points. With $c>0$ it is straightforward to see we have three possible regimes corresponding to the nature of the three roots of the cubic: two stable and one unstable fixed points, one stable and one unstable fixed points, or one stable fixed point and two non-real complex conjugate  roots. The parameters $a,b,c,f$ determine the nature of the roots of the cubic polynomial through the discriminant. For the cubic polynomial in standard form,
\begin{equation}
f(x) = x^3 + c_2 x^2 + c_1  x + c_0 ,
\end{equation}
the discriminant is given by~\cite{janson2010}
\begin{align}
\Delta &= c_2^2 c_1^2 - 4c_1^3 - 4c_2^3 c_0 + 18 c_2c_1c_0 - 27 c_0^2 \\ &= -4 p^3 - 27 q^2 , \quad \text{where }\;  p = c_1 - \tfrac{1}{3}c_2^2, \quad q = c_0 - \tfrac{1}{3}c_2c_1 + \tfrac{2}{27}c_2^3 .\label{eq:disc}
\end{align}
The boundary between  the three possible cases  is given by the condition $\Delta=0$. The form for the discriminant in~\cref{eq:disc} allows us to explicitly determine the boundaries separating the different  cases    by setting the discriminant to zero $\Delta = 0$  and solving for $c_0$. In terms of the original system parameters, this  gives the following equation for the boundary  as a function of the other parameters
\begin{equation}\label{eq:boundary}
f_b^\pm  = -\frac{ab}{3c} - \frac{2 b^3}{27 c^2} \pm 2 c\biggl(\frac{a}{3c} + \frac{b^2}{9c^2} \biggr)^{3/2} ,
\end{equation}
where   we require $a>a_c \equiv -b^2/3c$, for $c>0$. Given  fixed  $c$ and $b$, this boundary   divides the dynamics in     the parameter space  $(a,f)$   into two regimes: a region with three equilibrium points (two stable and one unstable) when  $ f_b^- < f <f_b^+$ and $a>a_c$ and the region  outside this area  with only one stable equilibrium point, see~\cref{fig:regcubic}.

\subsection{Equilibrium density}\label{appendixa:campdf}

The stationary probability measure for the general form of the nonlinear zonal flow in~\cref{eq:cubmodel} satisfies the following Fokker-Planck equation
\begin{equation}
-\frac{\partial}{\partial x}[\bigl( ax + bx^2 - c x^3 + f\bigr) p_U(x)] + \frac{1}{2}\frac{\partial^2}{\partial x^2}[\bigl( (Bx -A)^2 + \sigma_u^2 \bigr) p_U(x)] = 0
\end{equation}
The equilibrium PDF that solves this  can be shown to given by (see~\cite{yuan2011} for details)
\begin{equation}
p_u(u) = \frac{N_0}{((B x - A)^2 + \sigma_u^2)^{a_1}} \exp\biggl( d \arctan\biggl(\frac{Bx - A}{\sigma_u} \biggr)\biggl) \exp\biggl( \frac{-c_1 x^2 + b_1 x}{B^4}\biggr),
\end{equation}
where $N_0$ is a normalization constant. The coefficients $a_1,b_1,c_1,d$ are provided in.
\begin{align}
a_1 &= 1- \frac{-3 A^2c + aB^2 + 2AbB + c\sigma_u^2}{B^4},\\
b_1 &= 2b B^2 - 4cAB,\\
c_1 &= cB^2,\\
d &= \frac{d_1}{\sigma_u} + d_2 \sigma_u,\\
d_1 &= 2 \frac{A^2bB-A^3c+AaB^2+B^3f}{B^4},\\
d_2 &= 2 \frac{3 cA - bB}{B^4}.
\end{align}
 
\clearpage

\section{Mathematical Proofs}\label{sec:appendixproofs}
\subsection{\Cref{prop:trajsol}}
Integrating the equation for $\widehat T_{k,t}$ by  using~\cref{eq:shearfouriersol} we find
\begin{align}
\widehat T_{k,t} &= -\alpha \int_0^t \exp(-\gamma_{T,k}(t-s) + i \omega_{T,k}[s,t])\hat v_{k,s} \, ds \\
&= -\alpha \sigma_{v,k} \int_0^t \int_0^s \exp(-\gamma_{T,k}(t-s) -\gamma_{v,k}(s-r) +  + i \omega_{T,k}[s,t] +  i \omega_{v,k}[r,s]) \, d B_{k,r} \, ds  \\
&=  -\alpha \sigma_{v,k}  \int_0^t   \int_r^t \exp(-\gamma_{T,k}(t-s) -\gamma_{v,k}(s-r) +  + i \omega_{T,k}[s,t] +  i \omega_{v,k}[r,s])   \, ds \, d B_{k,r}.
\end{align}
Fubini's theorem is used in the last equality to exchange the order of integration.
\subsection{\Cref{prop:condvar}}
The derivation of the variance of trajectory solutions conditioned on a zonal flow trajectory is given by
\begin{align}
    \Sigma_{k,t\mid u} &=   \alpha^2 \sigma_{v,k}^2 \int_0^t \biggl\lvert \int_r^t  \exp(-\gamma_{T,k}(t-s) -\gamma_{v,k}(s-r) +  + i \omega_{T,k}[s,t] +  i \omega_{v,k}[r,s])   \, ds  \biggr\rvert^2  \, dr \\
     &=\alpha^2 \sigma_{v,k}^2 \int_0^t \biggl\lvert \int_r^t  \exp(-\gamma_{T,k}t + \gamma_{v,k} r + \gamma_{R,k} s + i \omega_{T,k}[s,t] +  i \omega_{v,k}[r,t] -   i \omega_{v,k}[s,t]) )  \, ds  \biggr\rvert^2  \, dr\\
     &=\alpha^2 \sigma_{v,k}^2 \int_0^t \biggl\lvert  \int_r^t  \exp(-\gamma_{T,k}t + \gamma_{v,k} r  +  i \omega_{v,k}[r,t]  + \gamma_{R,k} s + i \omega_{R,k}[s,t]    \, ds  \biggr\rvert^2  \, dr \\
      &=\alpha^2 \sigma_{v,k}^2 \int_0^t \biggl\lvert   \exp(-\gamma_{T,k}t + \gamma_{v,k} r  +  i \omega_{v,k}[r,t] ) \int_r^t  \exp( \gamma_{R,k} s + i \omega_{R,k}[s,t] )   \, ds  \biggr\rvert^2  \, dr  \\
      &=\alpha^2 \sigma_{v,k}^2 \int_0^t  \exp(-2\gamma_{T,k}t + 2\gamma_{v,k} r)  \biggl\lvert   \int_r^t  \exp( \gamma_{R,k} s + i \omega_{R,k}[s,t] )   \, ds  \biggr\rvert^2  \, dr ,
\end{align}
where  $\omega_{R,k} \coloneqq \omega_{T,k} - \omega_{v,k} = -(a_k+k)u_t - b_k$
    and  $\gamma_{R,k} \coloneqq \gamma_{T,k} - \gamma_{v,k}$.
Alternatively, we can express the variance as:
\begin{equation}\label{eq:integralcomplete}
{\Sigma_{k,t\mid u} = \alpha^2 \sigma_{v,k}^2 \int_0^t  \exp(-2\gamma_{v,k} (t-r) )  \biggl\lvert   \int_r^t  \exp( -\gamma_{R,k} (t-s) + i \omega_{R,k}[s,t]  )  \, ds  \biggr\rvert^2  \, dr.}
\end{equation}

\subsection{\Cref{cor:varbound}}
We can   find an upper bound for the conditional variance as follows.
Start from~\cref{eq:altintegral}  and first use $\abs{\int z} \le \int \abs{z} = \int r$, where $z = r e^{i \theta}$, for the inner integral to obtain:
\begin{align}
 \biggl\lvert   \int_r^t  \exp( \gamma_{R,k} s + i \omega_{R,k}[s,t] )   \, ds \biggr\rvert^2 \le \biggl(\int_r^t e^{ \gamma_R s} \, ds\biggr)^2 =  \frac{1}{\gamma_R^2}(e^{\gamma_R t} - e^{\gamma_R r})^2\le \frac{1}{\gamma_R^2} (e^{2 \gamma_R t} + e^{2 \gamma_R r}),
\end{align}
\begin{align}
\Sigma_{k,t\mid u} &=\alpha^2 \sigma_{v,k}^2 \int_0^t  \exp(-2\gamma_{T,k}t + 2\gamma_{v,k} r)  \biggl\lvert   \int_r^t  \exp( \gamma_{R,k} s + i \omega_{R,k}[s,t] )   \, ds  \biggr\rvert^2  \, dr\\
&\le \alpha^2 \sigma_{v,k}^2 \int_0^t  \exp(-2\gamma_{T,k}t + 2\gamma_{v,k} r)  \frac{1}{\gamma_{R,k}^2} (e^{2 \gamma_{R,k} t} + e^{2 \gamma_{R,k} r})  \, dr\\ &= \alpha^2 \sigma_{v,k}^2 \int_0^t   \frac{1}{\gamma_{R,j}^2} (e^{-2 \gamma_{v,k}( t-r)} + e^{-2 \gamma_{T,k} (t-r)})  \, dr \\ &= \frac{\alpha^2 \sigma_{v,k}^2 }{\gamma_{R,k}^2}\biggl(\frac{1 - e^{-2\gamma_{v,k}t}}{2 \gamma_{v,k}}  +  \frac{1 - e^{-2\gamma_{T,k}t}}{2 \gamma_{T,k}} \biggr)
\end{align}
We also find the bound in the long time limit
\begin{equation}
    \lim_{t \to \infty} \Sigma_{k,t\mid u} \leq \frac{\alpha^2 \sigma_{v,k}^2}{\gamma_{R,k}^2} \biggl( \frac{1}{2\gamma_{v,k}} + \frac{1}{2\gamma_{T,k}} \biggr)
\end{equation}

\subsection{\Cref{prop:tracermodepdf}}
Starting from~\cref{eq:integralcomplete} for the rescaled system in~\cref{cor:rescaled},
\begin{equation}\label{eq:condvar_appendix}
    {\Sigma_{k,t\mid u} = \epsilon^{-2}\alpha^2 \sigma_{v,k}^2 \int_0^t  \exp(-2\gamma_{v,k} (t-r) )  \biggl\lvert   \int_r^t  \exp( -\gamma_{R,k} (t-s) + i \epsilon^{-1} \omega_{R,k}[s,t]  )  \, ds  \biggr\rvert^2  \, dr,}
    \end{equation}
where $\gamma_{R,k} = \epsilon^{-1}  \gamma_{T,k} - \gamma_{v,k}$. Define the inner integral
\begin{equation}
    I(r) \coloneq   \int_r^t \exp\Bigl(-\bigl(\epsilon^{-1} \gamma_{T,k}  - \gamma_{v,k}\bigr)(t-s) + i \epsilon^{-1} \omega_{R,k}[s,t] \Bigr) \, ds,
\end{equation}
where $\omega_{R,k}[s,t] =\int_s^t \omega_{R,k}(u)\, du$. Consider the change of variables $u = t-s$,
\begin{equation}
    I(r) =   \int_0^{t-r}    \exp\Bigl(- \epsilon^{-1} \gamma_{T,k} u \Bigr)
    \exp\Bigl(  \gamma_{v,k} u  + i\epsilon^{-1} \omega_{R,k}[t-u,t] \Bigr) \, du,
\end{equation}
In the small $\epsilon$ limit, most of the contribution to this integral comes from when $u$ is small.  As a result $\omega_{R,k}[t-u,t] \approx \omega_{R,k}(t) u$:
\begin{equation}
    I(r) \approx   \int_0^{t-r} \exp\bigl( \epsilon^{-1}(- {\gamma_{T,k}}       + i  \omega_{R,k}(t) ) u \bigr) \, du,
\end{equation}
This integral is of the form
\begin{equation}
    \int_0^{t-r} \exp(-\lambda u) du = \frac{1}{\lambda}(1-\exp(-\lambda(t-r))), \quad \text{where } \lambda =  \epsilon^{-1}( {\gamma_{T,k}} - i  \omega_{R,k}(t)
\end{equation}
Taking the modulus square and keeping only leading order terms we find
\begin{equation}
    \lvert I(r) \rvert^2 \approx \frac{\epsilon^2}{ \gamma_{T,k}^2  +   \omega_{R,k}(t) ^2}.
\end{equation}
Using this result in~\cref{eq:condvar_appendix} we obtain
\begin{align}
    \Sigma_{k,t\mid u} &= \epsilon^{-2}\alpha^2 \sigma_{v,k}^2 \int_0^t  \exp(-2\gamma_{v,k} (t-r) )  \frac{\epsilon^2}{ \gamma_{T,k}^2  +   \omega_{R,k}(t) ^2}  \, dr  \\
    &=  \frac{\alpha^2\sigma_{v,k}^2 }{2\gamma_{v,k} \bigl(\gamma_{T,k}^2  +   \omega_{R,k}(t) ^2\bigr)}\bigl(1- \exp(-2\gamma_{v,k}  t ) \bigr).
\end{align}
As $t\to \infty$, the conditional variance $\Sigma_{k,t\mid u}$   converges to the stationary value
\begin{equation}
    \tilde \Sigma_{k}(u) = \frac{\alpha^2 E_{v,k}}{\gamma_{T,k}^2 + \omega_{R,k}(u)^2},   \quad \text{where } E_{v_k} =  \frac{\sigma^2_{v,k}}{2\gamma_{k,v}}.
\end{equation}

\clearpage

\section{Supplementary Numerical Simulations}\label{sec:supplementary_numerical_simulations}

\subsection{Side-by-side comparison of random and advective shear flows}\label{sub:side_by_side_comparison_of_random_and_advective_flow_no_dispersion}
\begin{figure}[htbp!]
    \centering

    \begin{minipage}{.49\textwidth}
    \centering
    \subcaption{Random Shear Flow}
    \end{minipage}%
    \begin{minipage}{.49\textwidth}
    \centering
    \subcaption{Advective Shear Flow}
    \end{minipage}%

    \vspace{0.5em}
    {\scriptsize\itshape Zonal Flow: Linear (matched resonance)}
    \vspace{0.5em}

    \begin{minipage}{.49\textwidth}
    \centering
    \includegraphics[scale=0.35]{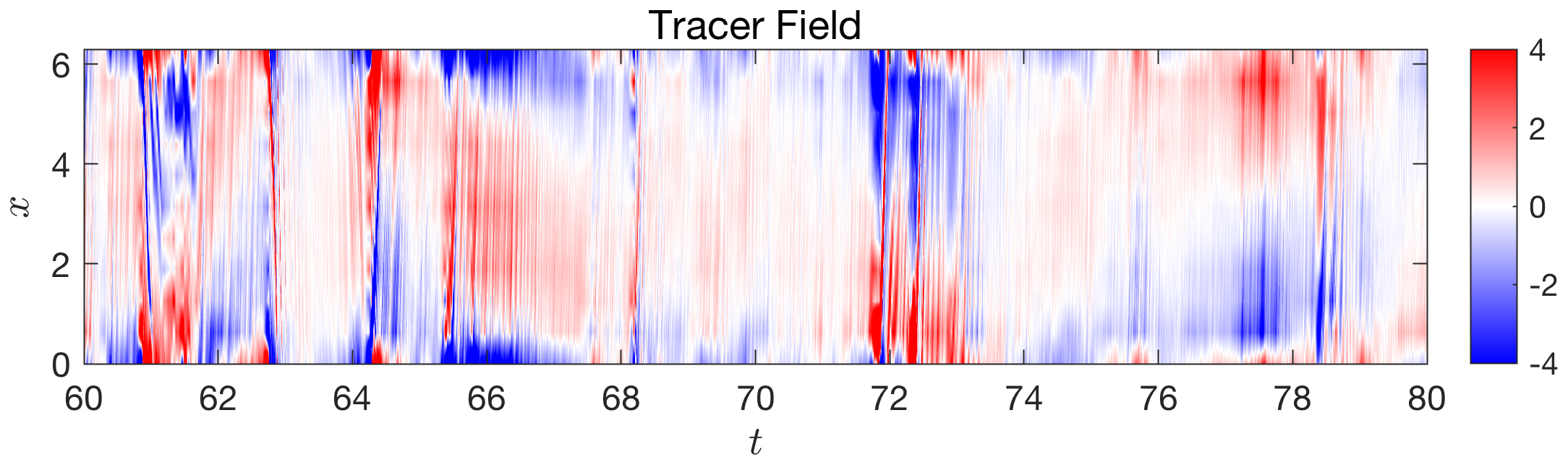}
    \end{minipage}%
    \begin{minipage}{.49\textwidth}
    \centering
    \includegraphics[scale=0.35]{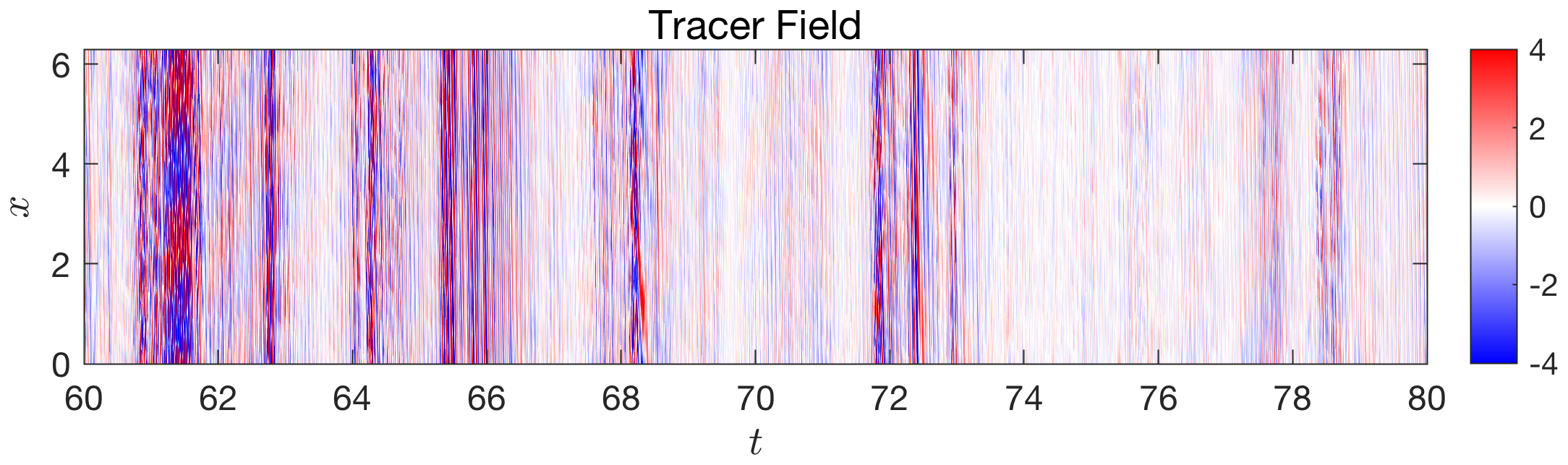}
    \end{minipage}%

    \vspace{0.5em}
    {\scriptsize\itshape Zonal Flow: Nonlinear, $f=0, B = 0$}
    \vspace{0.5em}

    \begin{minipage}{.49\textwidth}
    \centering
    \includegraphics[scale=0.35]{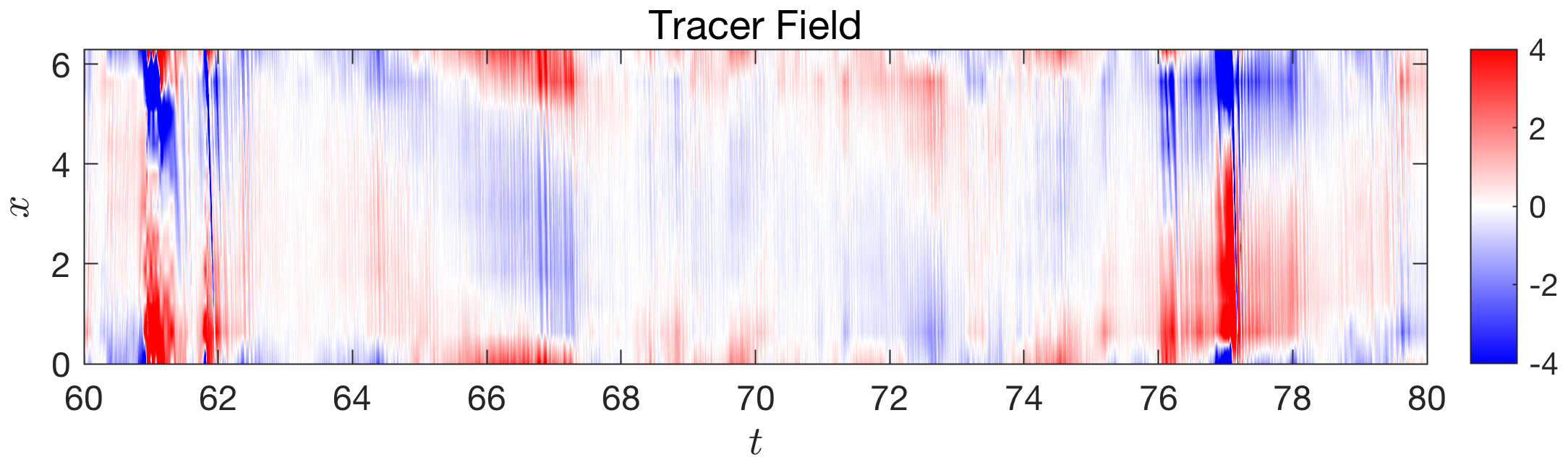}
    \end{minipage}%
    \begin{minipage}{.49\textwidth}
    \centering
    \includegraphics[scale=0.35]{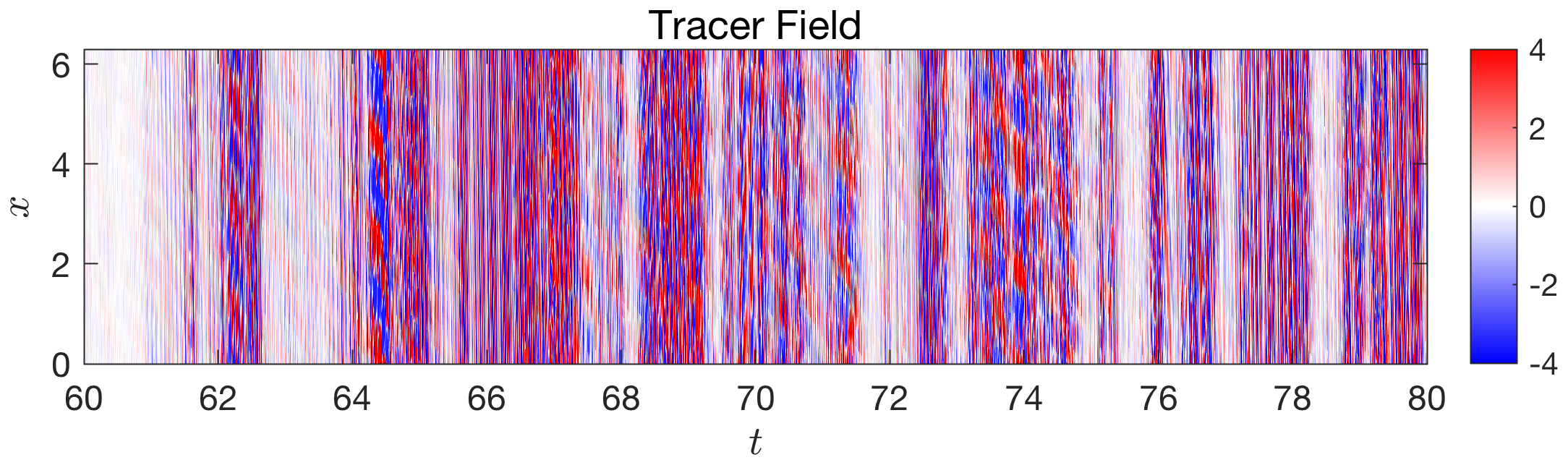}
    \end{minipage}%

    \vspace{0.5em}
    {\scriptsize\itshape Zonal Flow: Nonlinear, $f=0, B = 2.5$}
    \vspace{0.5em}

    \begin{minipage}{.49\textwidth}
    \centering
    \includegraphics[scale=0.35]{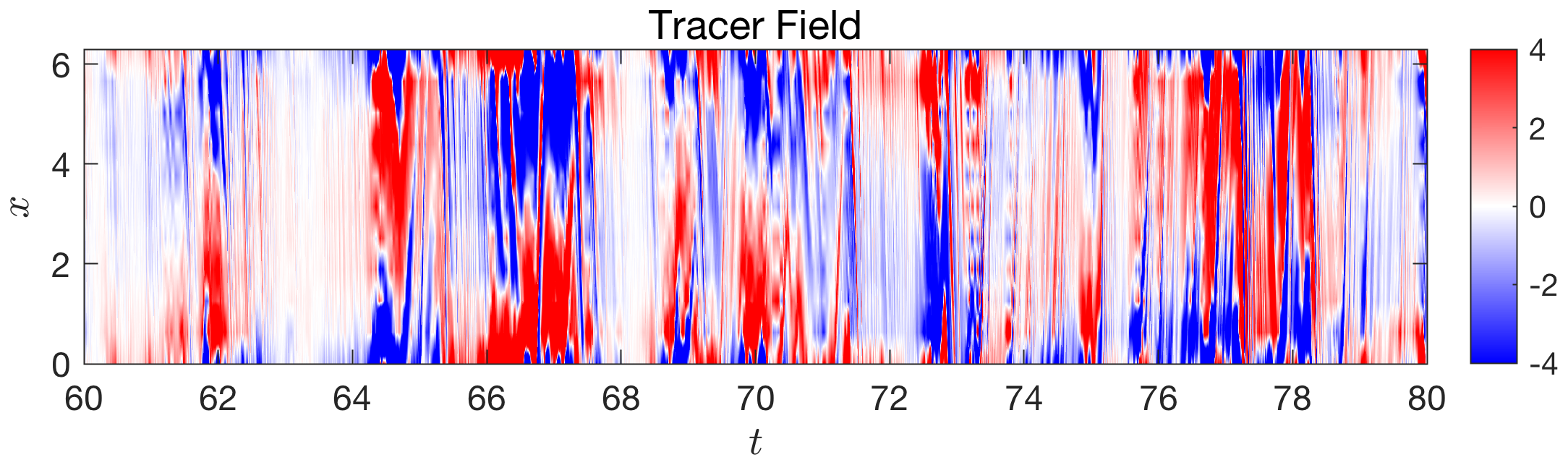}
    \end{minipage}%
    \begin{minipage}{.49\textwidth}
    \centering
    \includegraphics[scale=0.35]{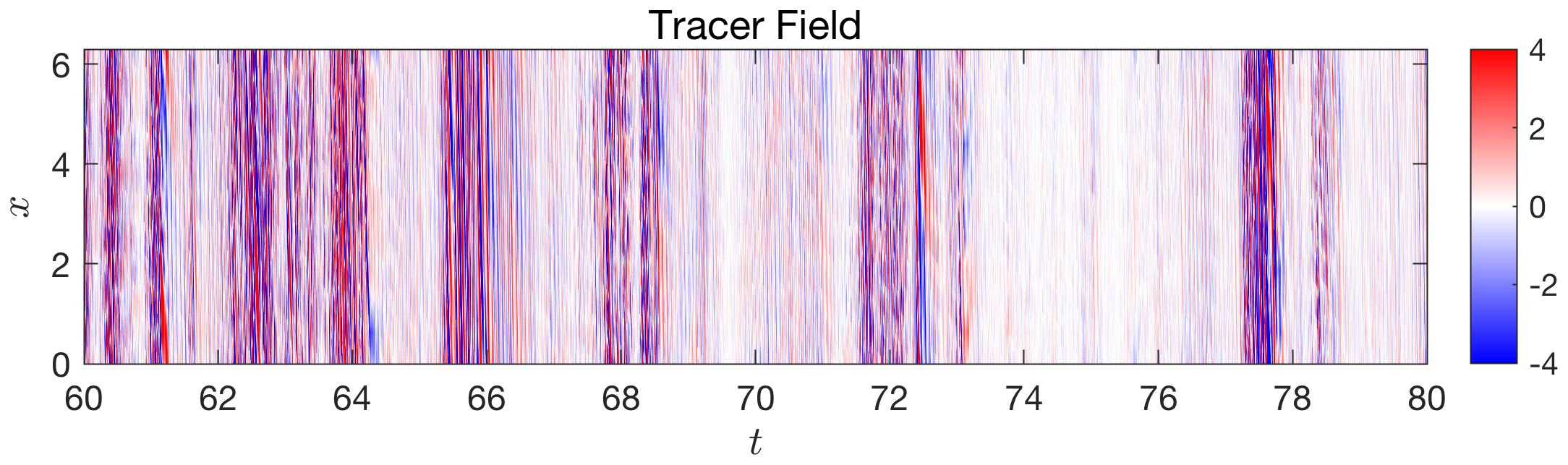}
    \end{minipage}%

    \vspace{0.5em}
    {\scriptsize\itshape Zonal Flow: Nonlinear, $f=1.0, B = 0$}
    \vspace{0.5em}

    \begin{minipage}{.49\textwidth}
    \centering
    \includegraphics[scale=0.35]{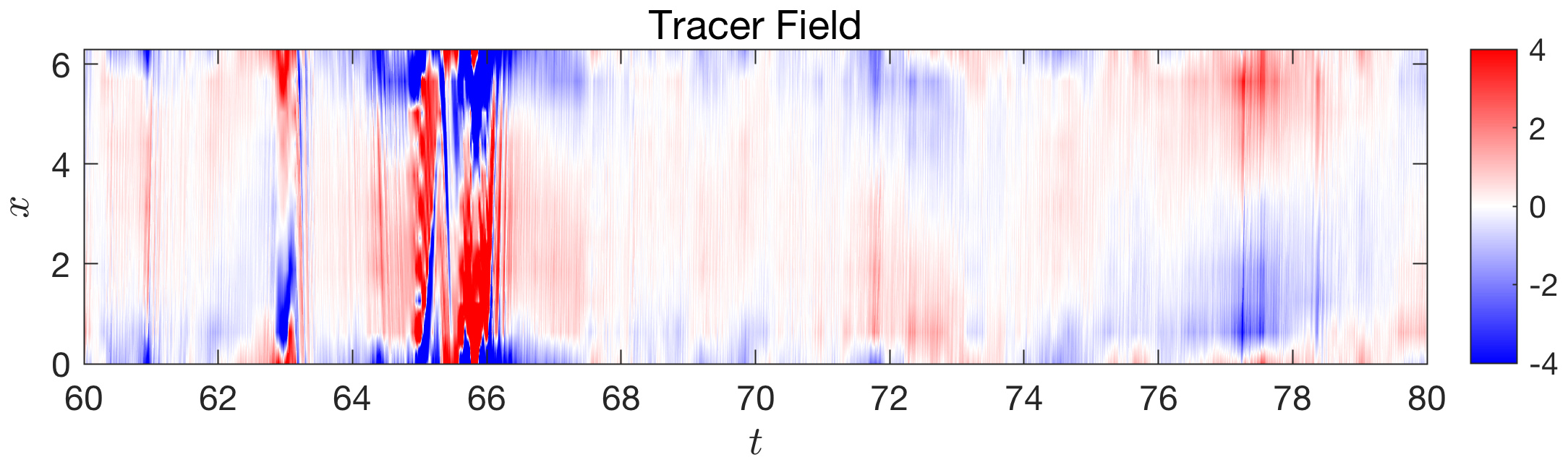}
    \end{minipage}%
    \begin{minipage}{.49\textwidth}
    \centering
    \includegraphics[scale=0.35]{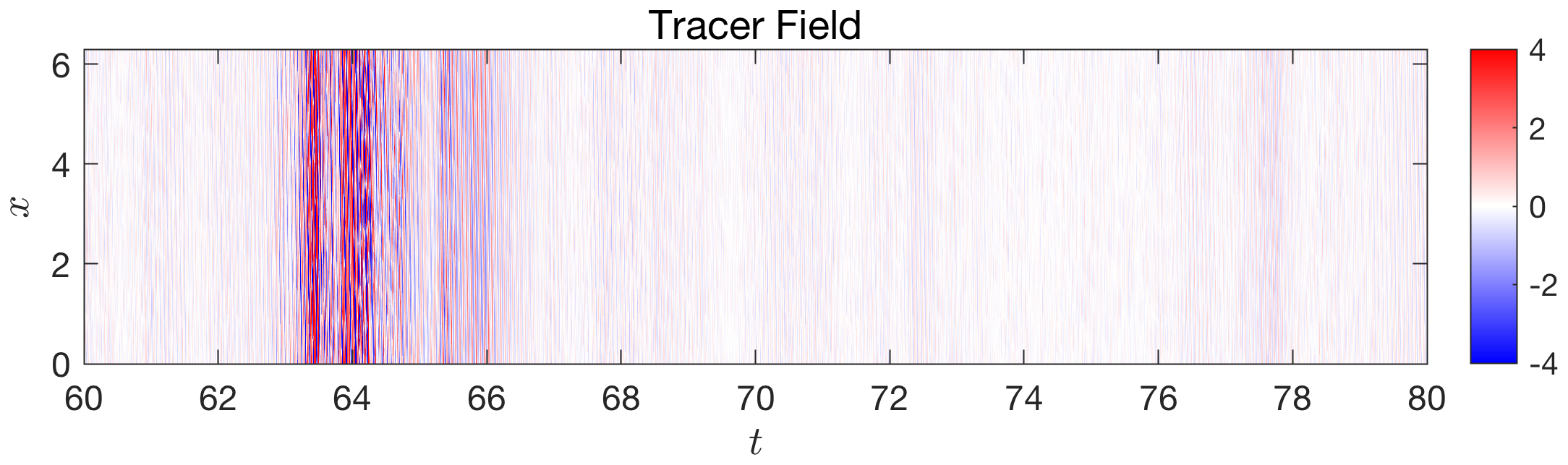}
    \end{minipage}%

    \vspace{0.5em}
    {\scriptsize\itshape Zonal Flow: Nonlinear, $f=1.0, B = 2.5$}
    \vspace{0.5em}

    \begin{minipage}{.49\textwidth}
    \centering
    \includegraphics[scale=0.35]{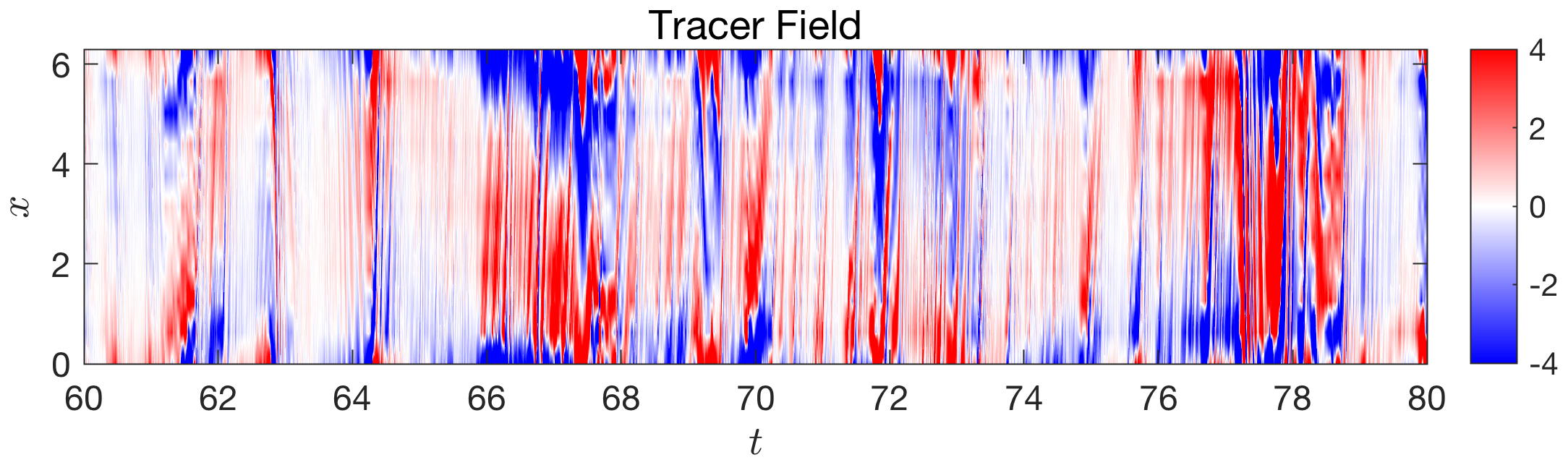}
    \end{minipage}%
    \begin{minipage}{.49\textwidth}
    \centering
    \includegraphics[scale=0.35]{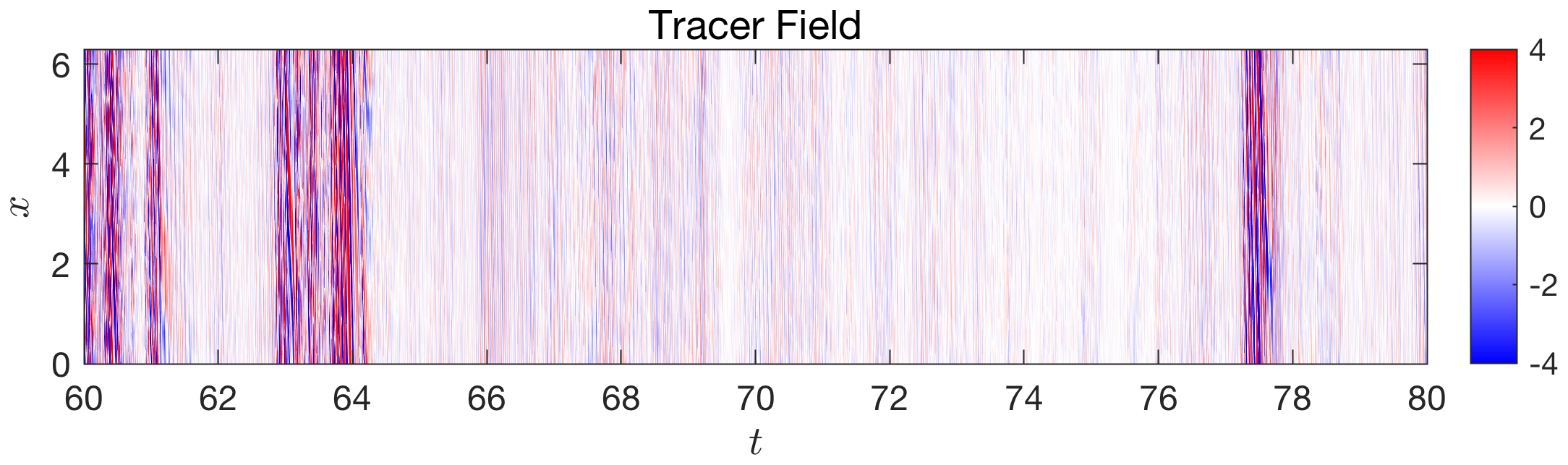}
    \end{minipage}%

    \caption{Comparison of spatio-temporal evolution of the tracer field under different shear flow models for {equipartition}.}
    \label{fig:set1_nodispappendix_field}
\end{figure}

\begin{figure}[htbp!]
    \centering
    \begin{minipage}{.49\textwidth}
    \centering
    \subcaption{Random Shear Flow}
    \end{minipage}%
    \begin{minipage}{.49\textwidth}
    \centering
    \subcaption{Advective Shear Flow}
    \end{minipage}%
    
    \vspace{0.5em}
    {\scriptsize\itshape Zonal Flow: Linear (matched resonance)}
    \vspace{0.5em}

    \begin{minipage}{.49\textwidth}
    \centering
    \includegraphics[scale=0.35]{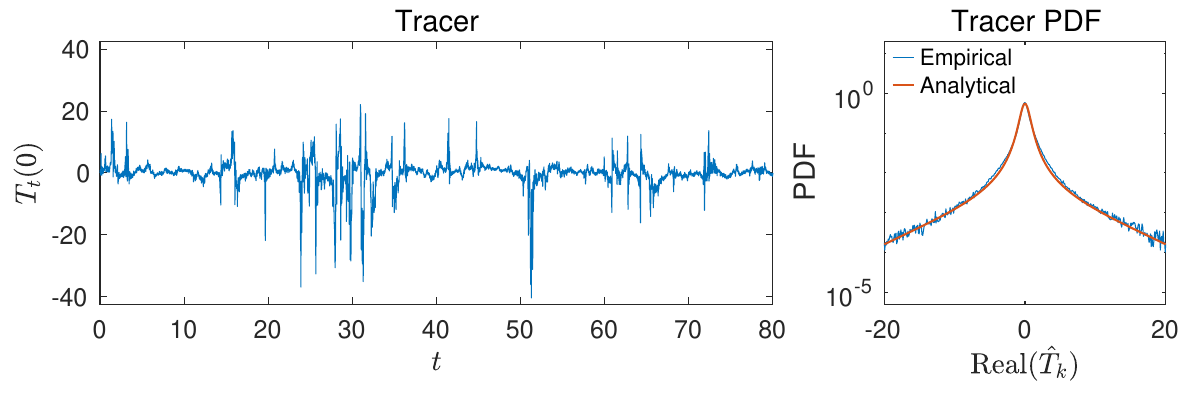}
    \end{minipage}%
    \begin{minipage}{.49\textwidth}
    \centering
    \includegraphics[scale=0.35]{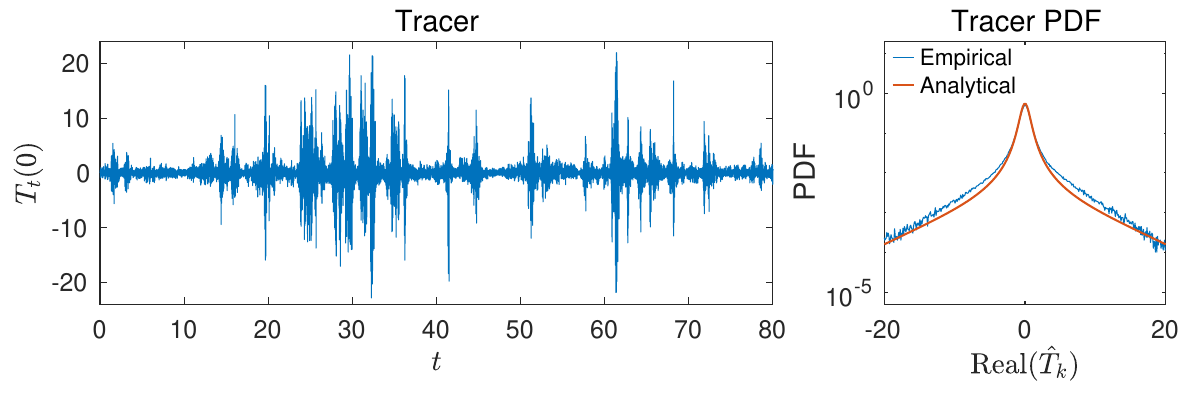}
    \end{minipage}%
   
    \vspace{0.5em}
    {\scriptsize\itshape Zonal Flow: Nonlinear, $f=0, B = 0$}
    \vspace{0.5em}
   
    \begin{minipage}{.49\textwidth}
    \centering
    \includegraphics[scale=0.35]{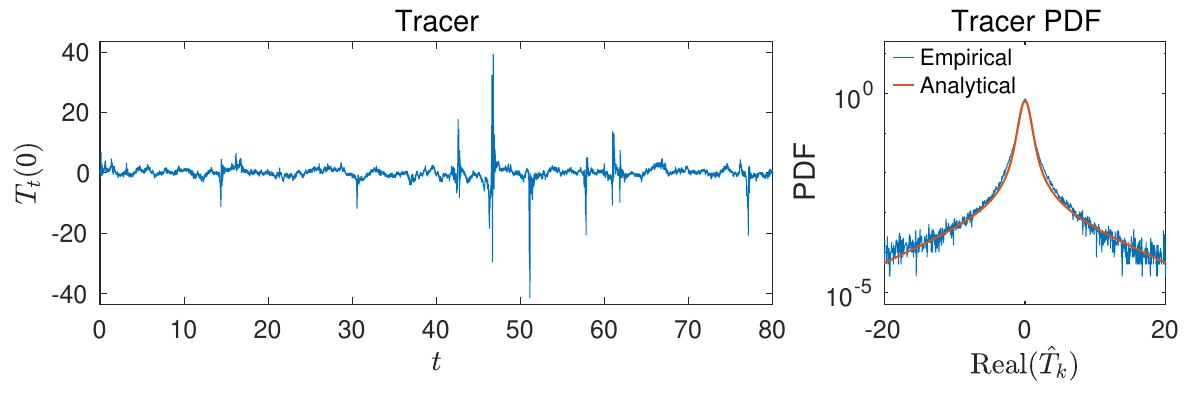}
    \end{minipage}%
    \begin{minipage}{.49\textwidth}
    \centering
    \includegraphics[scale=0.35]{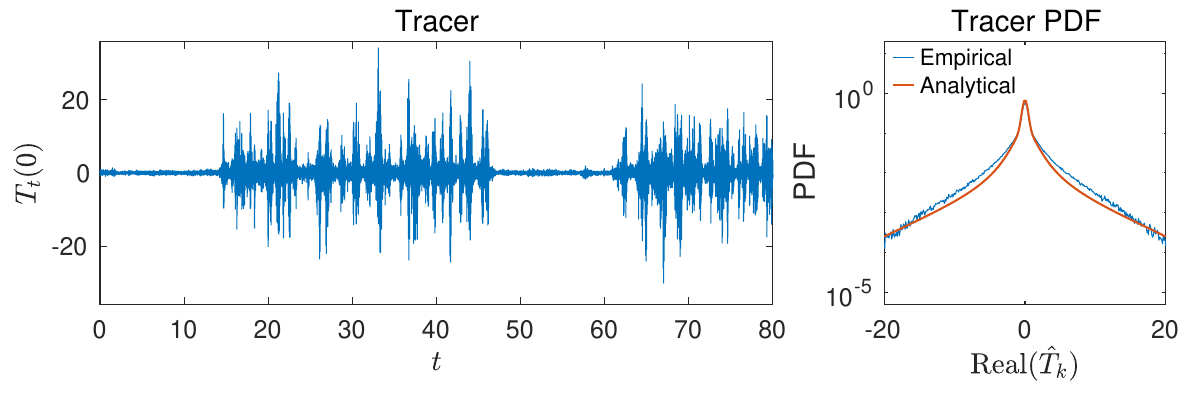}
    \end{minipage}%
   
    \vspace{0.5em}
    {\scriptsize\itshape Zonal Flow: Nonlinear, $f=0, B = 2.5$}
    \vspace{0.5em}
   
    \begin{minipage}{.49\textwidth}
    \centering
    \includegraphics[scale=0.35]{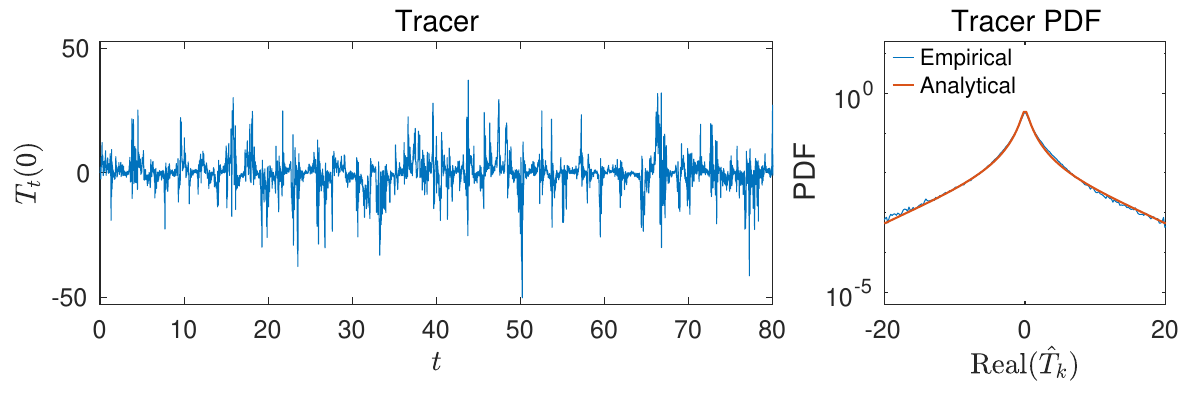}
    \end{minipage}%
    \begin{minipage}{.49\textwidth}
    \centering
    \includegraphics[scale=0.35]{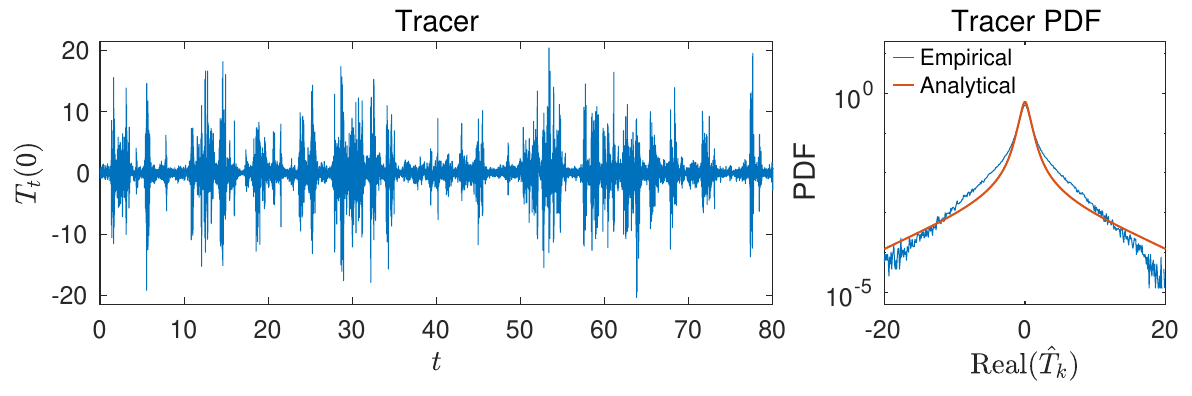}
    \end{minipage}%
    
    \vspace{0.5em}
    {\scriptsize\itshape Zonal Flow: Nonlinear, $f=1.0, B = 0$}
    \vspace{0.5em}

    \begin{minipage}{.49\textwidth}
    \centering
    \includegraphics[scale=0.35]{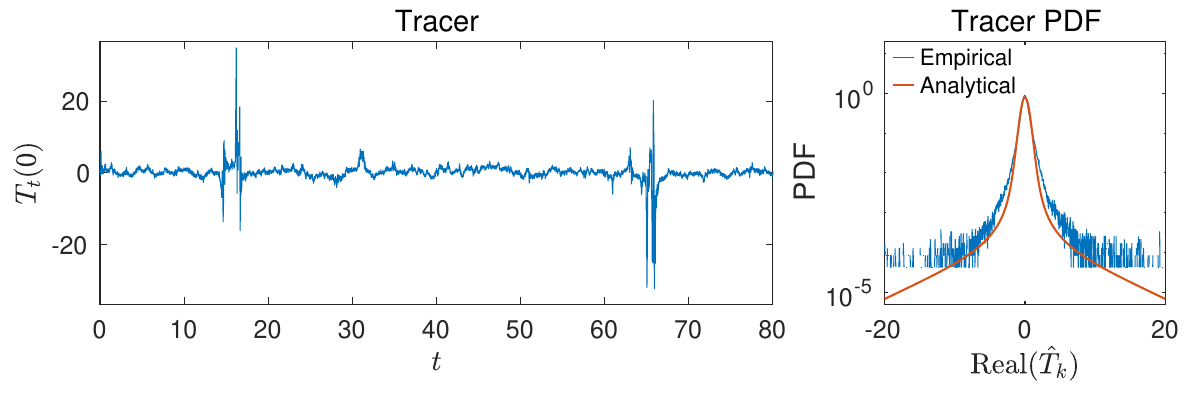}
    \end{minipage}%
    \begin{minipage}{.49\textwidth}
    \centering
    \includegraphics[scale=0.35]{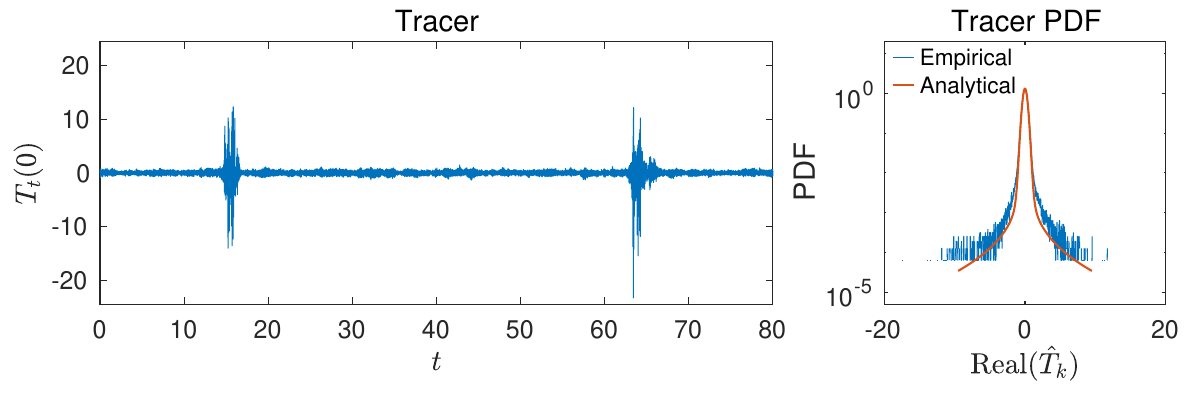}
    \end{minipage}%
    
    \vspace{0.5em}
    {\scriptsize\itshape Zonal Flow: Nonlinear, $f=1.0, B = 2.5$}
    \vspace{0.5em}
    
    \begin{minipage}{.49\textwidth}
    \centering
    \includegraphics[scale=0.35]{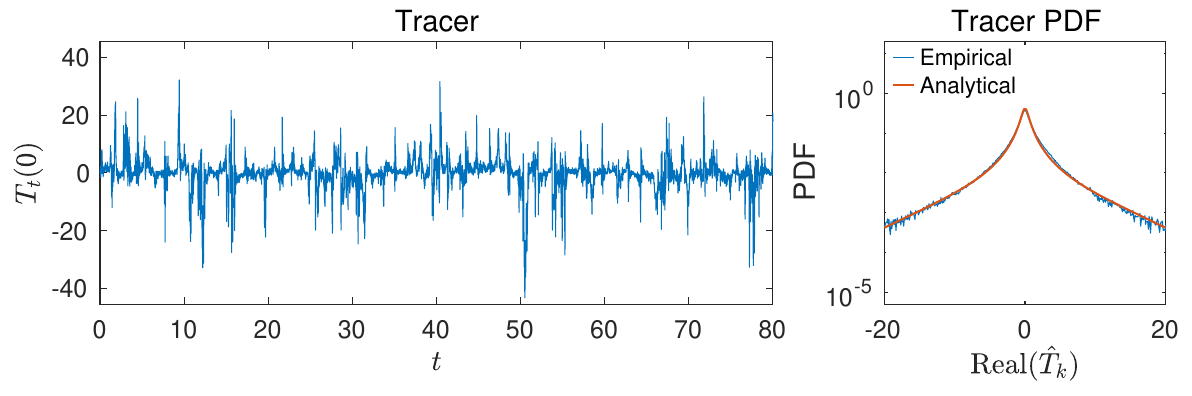}
    \end{minipage}%
    \begin{minipage}{.49\textwidth}
    \centering
    \includegraphics[scale=0.35]{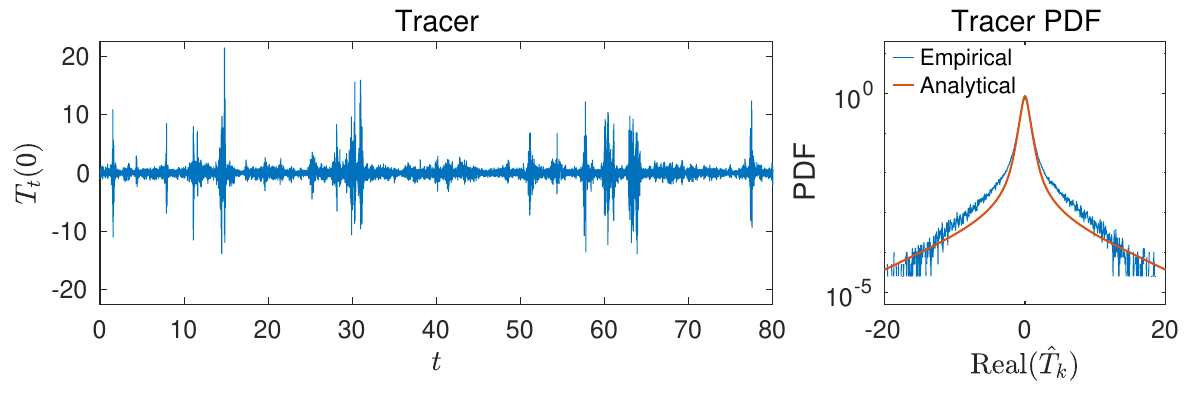}
    \end{minipage}

    \caption{Comparison of evolution of the tracer at $T_t(0)$ and the stationary PDF under different shear flow models.}
    \label{fig:set1_nodispappendix_pdf}
\end{figure}

\begin{figure}[htbp!]
    \centering
    \begin{minipage}{.49\textwidth}
    \centering
    \subcaption{Advective Shear Flow}
    \end{minipage}%
    \begin{minipage}{.49\textwidth}
    \centering
    \subcaption{Dispersive (QG) Shear Flow}
    \end{minipage}%
    
    \vspace{0.5em}
    {\scriptsize\itshape Zonal Flow: Linear (matched resonance)}
    \vspace{0.5em}
    
    \begin{minipage}{.49\textwidth}
    \centering
    \includegraphics[scale=0.31]{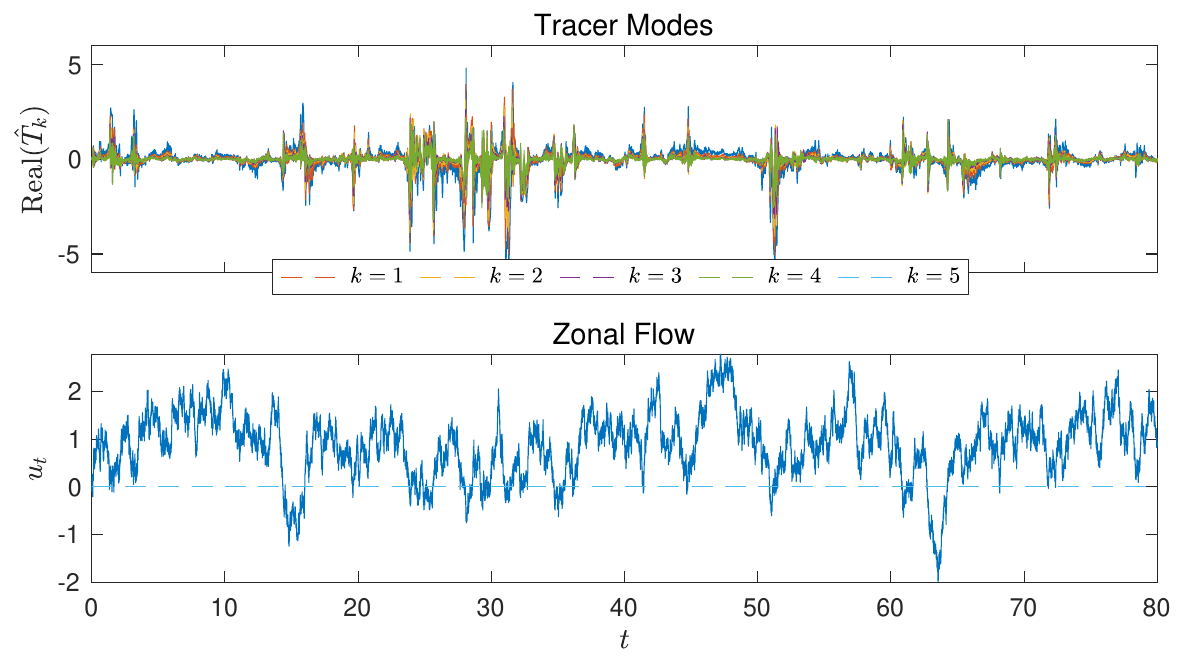}
    \end{minipage}%
    \begin{minipage}{.49\textwidth}
    \centering
    \includegraphics[scale=0.31]{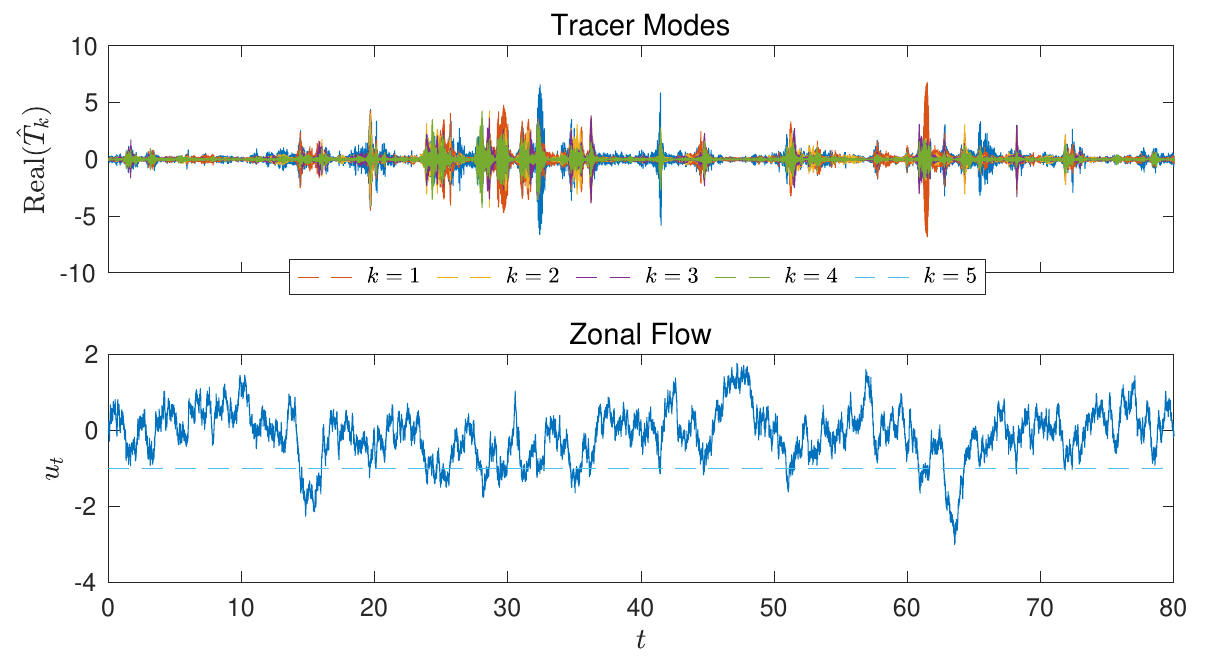}
    \end{minipage}%
    
    \vspace{0.5em}
    {\scriptsize\itshape Zonal Flow: Nonlinear, $f=0, B = 0$}
    \vspace{0.5em}
    
    \begin{minipage}{.49\textwidth}
    \centering
    \includegraphics[scale=0.31]{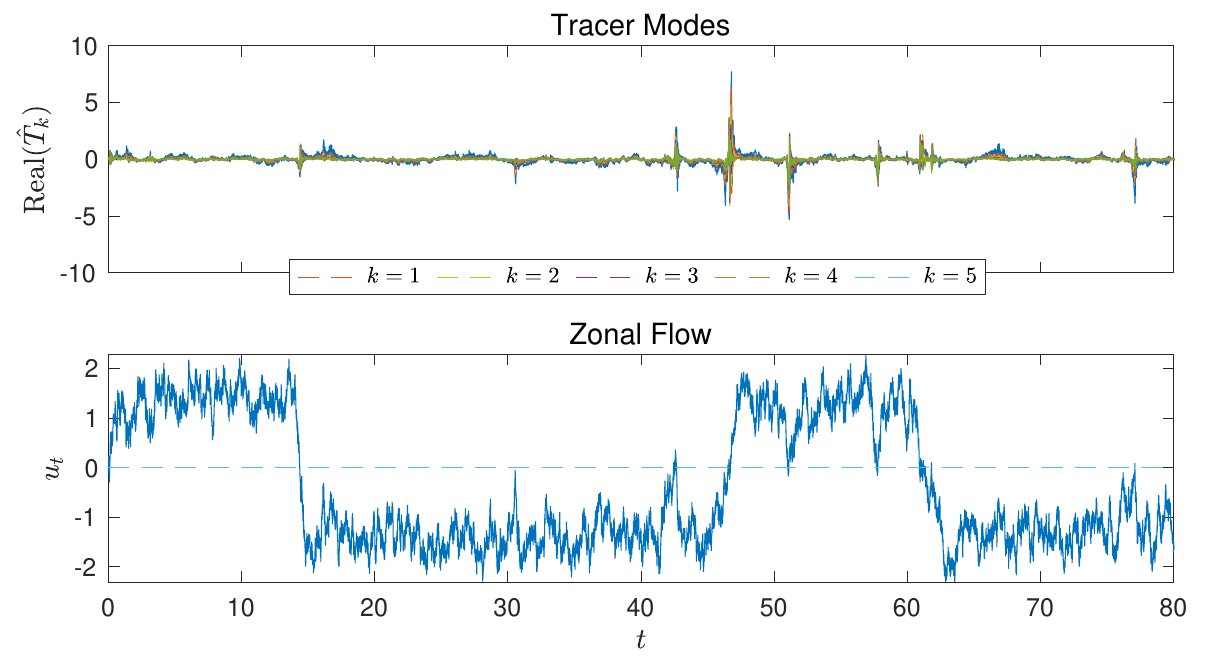}
    \end{minipage}%
    \begin{minipage}{.49\textwidth}
    \centering
    \includegraphics[scale=0.31]{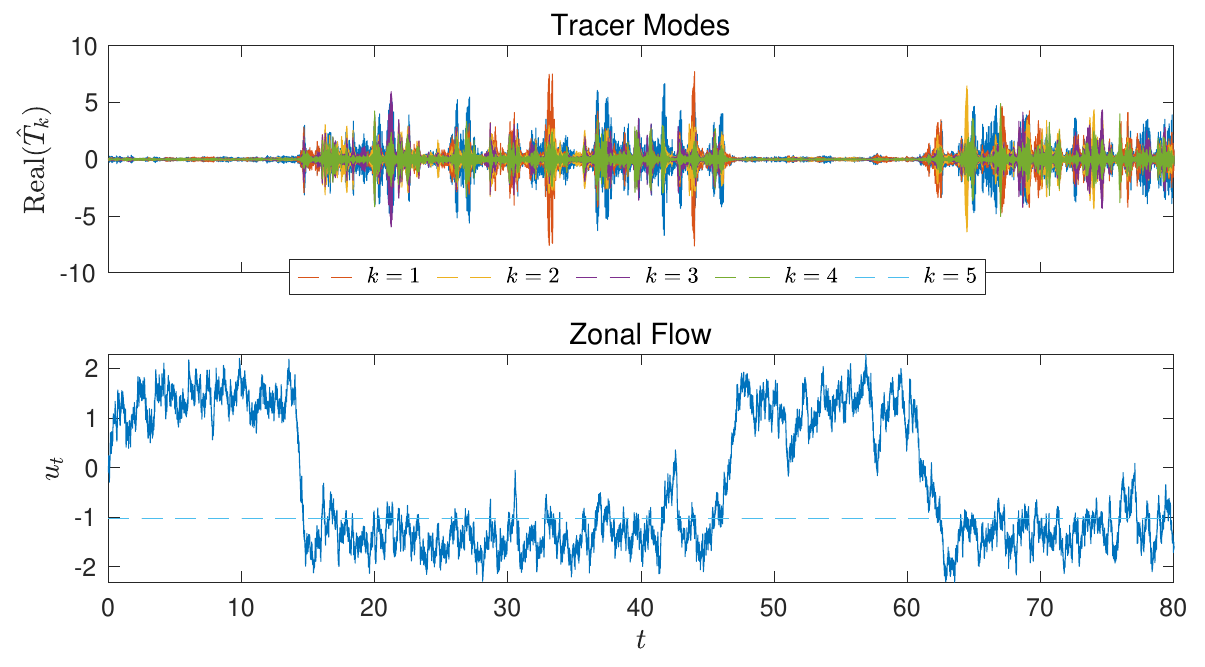}
    \end{minipage}%
    
    \vspace{0.5em}
    {\scriptsize\itshape Zonal Flow: Nonlinear, $f=0, B = 2.5$}
    \vspace{0.5em}
    
    \begin{minipage}{.49\textwidth}
    \centering
    \includegraphics[scale=0.31]{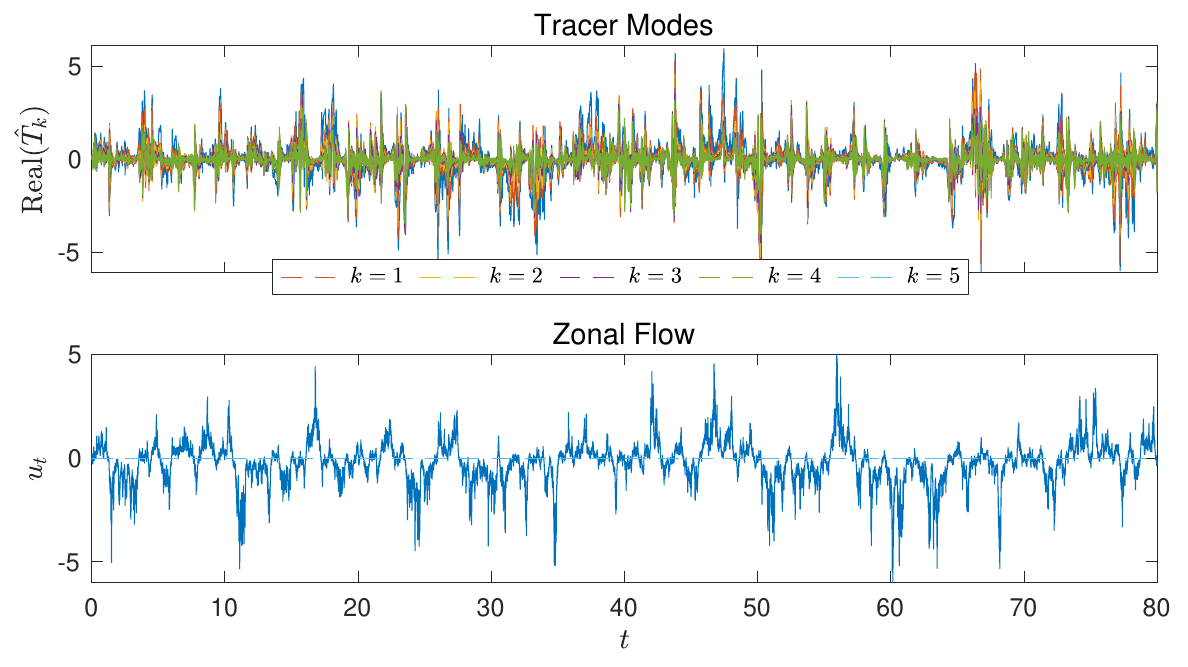}
    \end{minipage}%
    \begin{minipage}{.49\textwidth}
    \centering
    \includegraphics[scale=0.31]{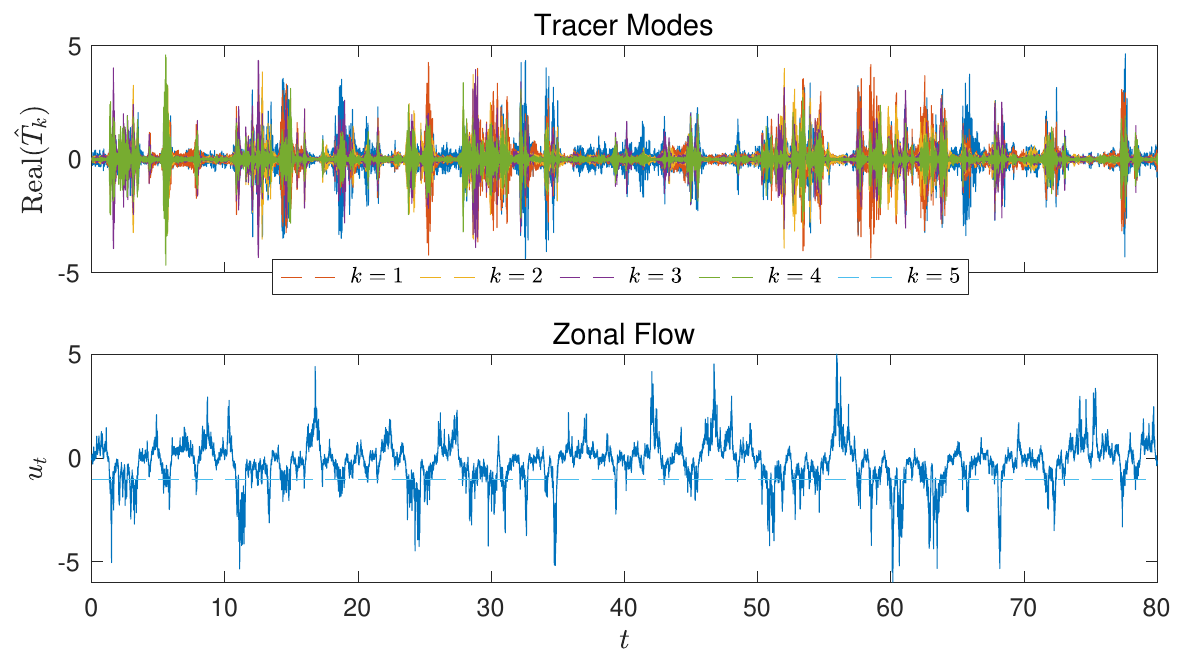}
    \end{minipage}%
    
    \vspace{0.5em}
    {\scriptsize\itshape Zonal Flow: Nonlinear, $f=1.0, B = 0$}
    \vspace{0.5em}
    
    \begin{minipage}{.49\textwidth}
    \centering
    \includegraphics[scale=0.31]{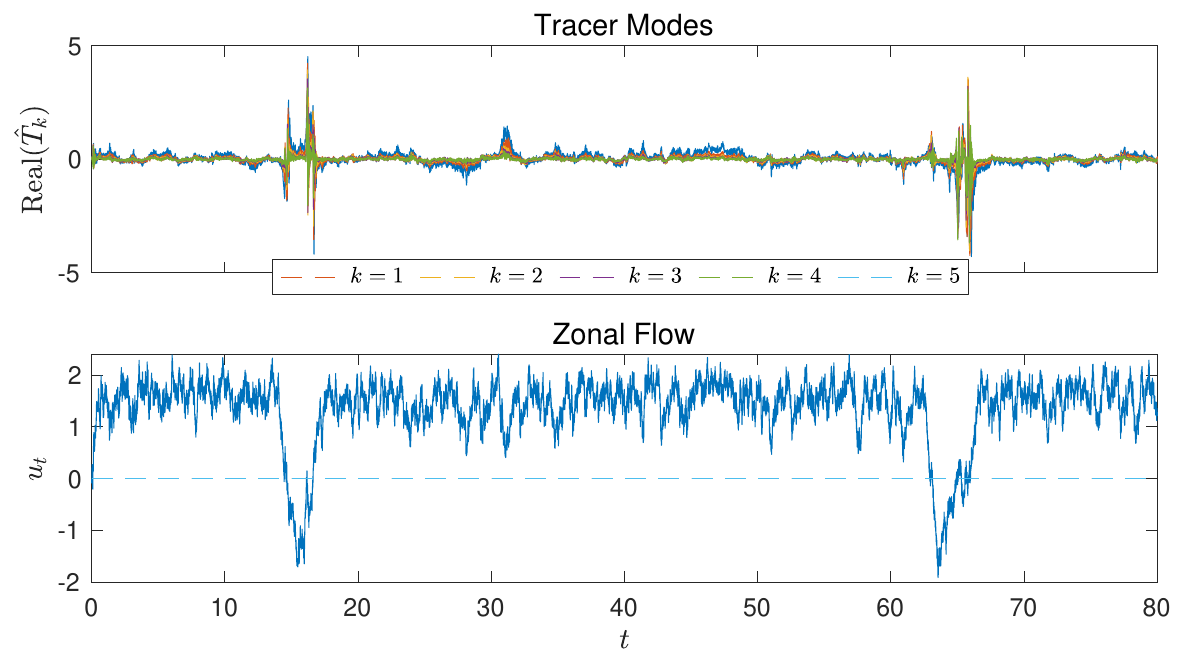}
    \end{minipage}%
    \begin{minipage}{.49\textwidth}
    \centering
    \includegraphics[scale=0.31]{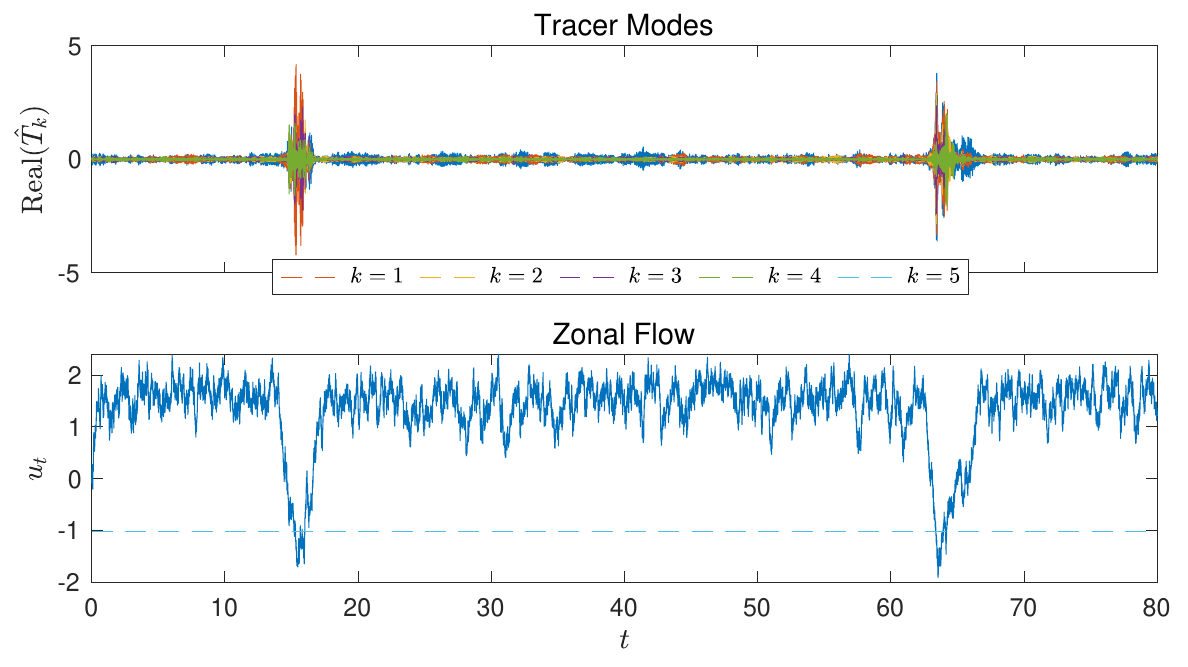}
    \end{minipage}%
    
    \vspace{0.5em}
    {\scriptsize\itshape Zonal Flow: Nonlinear, $f=1.0, B = 2.5$}
    \vspace{0.5em}

    \begin{minipage}{.49\textwidth}
    \centering
    \includegraphics[scale=0.31]{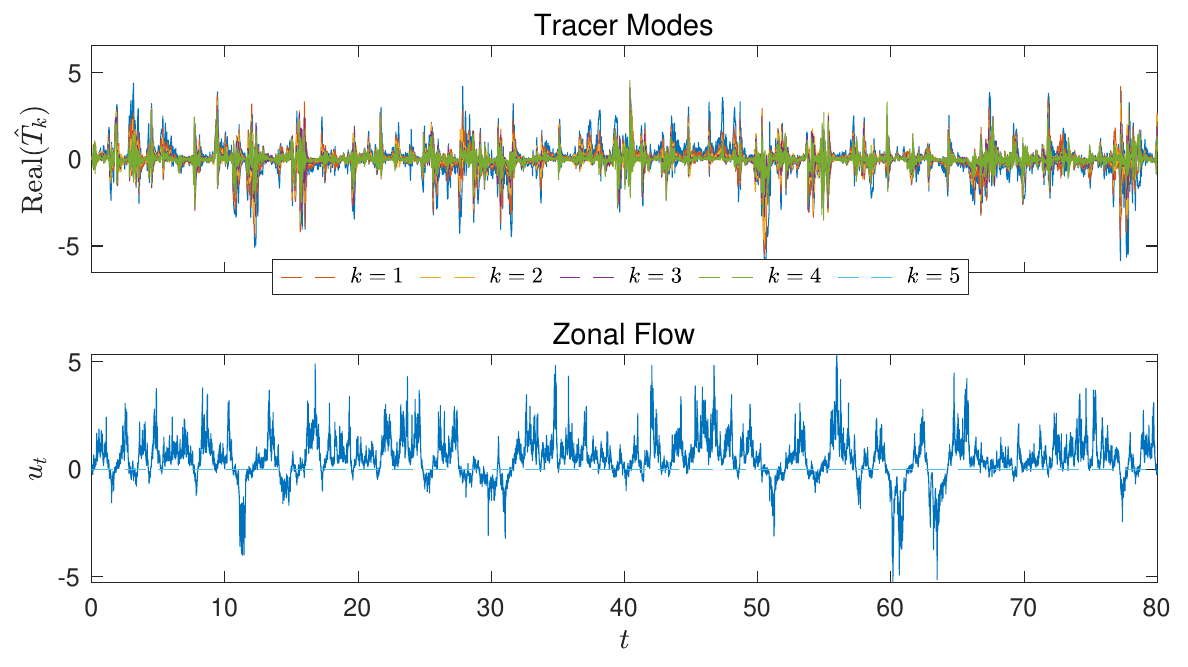}
    \end{minipage}%
    \begin{minipage}{.49\textwidth}
    \centering
    \includegraphics[scale=0.31]{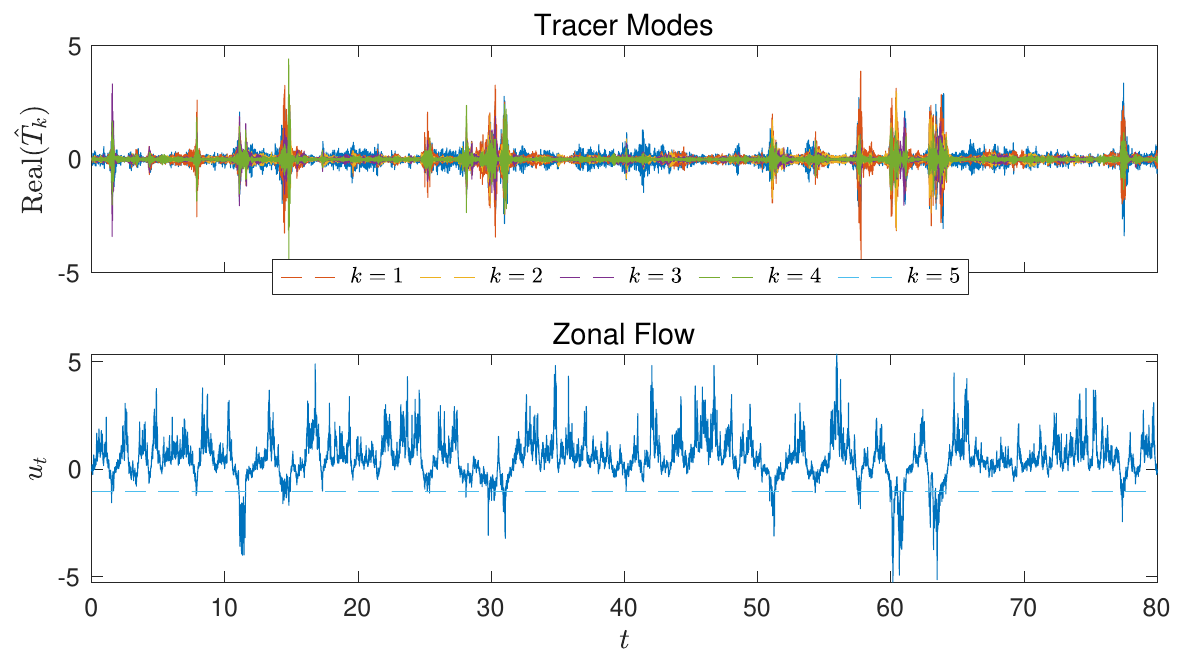}
    \end{minipage}
    \caption{Comparison of evolution of the tracer modes and zonal flow.}
    \label{fig:set1_nodispappendix_modes}
\end{figure}

\Cref{fig:set1_nodispappendix_field,fig:set1_nodispappendix_pdf,fig:set1_nodispappendix_modes}

\subsection{Side-by-side comparison of advective  and dispersive shear flows}\label{sub:side_by_side_comparison_of_all_three_shear_models}
\begin{figure}[htbp!]
    \centering

    \begin{minipage}{.49\textwidth}
    \centering
    \subcaption{Advective Shear Flow}
    \end{minipage}%
    \begin{minipage}{.49\textwidth}
    \centering
    \subcaption{Dispersive (QG) Shear Flow}
    \end{minipage}%

    \vspace{0.5em}
    {\scriptsize\itshape Zonal Flow: Linear, $f=0$}
    \vspace{0.5em}

    \begin{minipage}{.49\textwidth}
    \centering
    \includegraphics[scale=0.35]{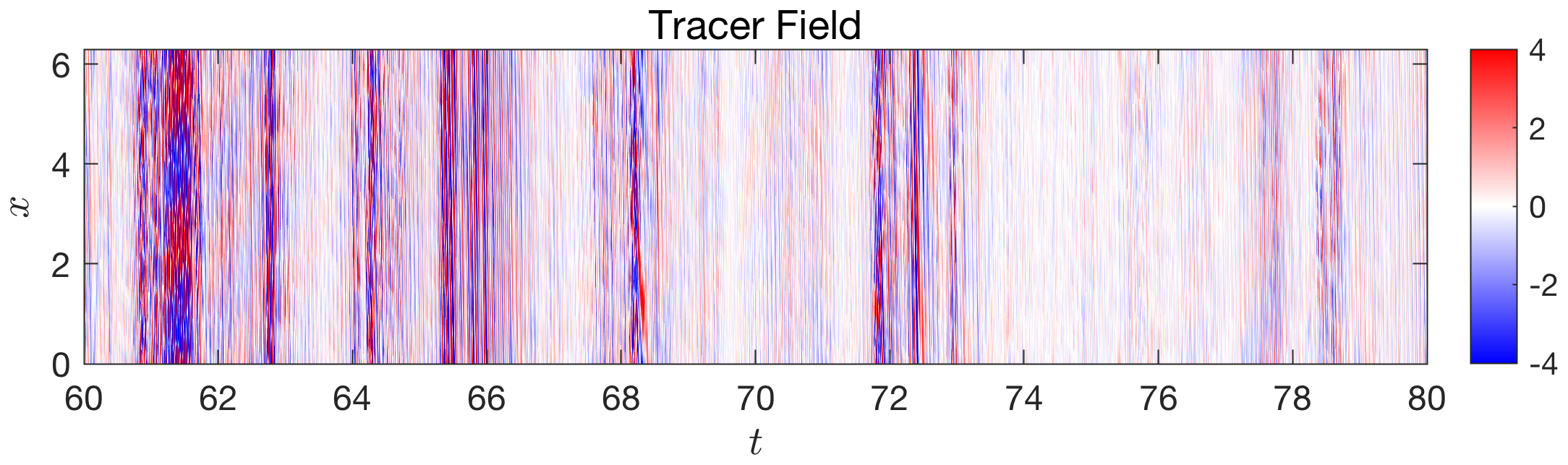}
    \end{minipage}%
    \begin{minipage}{.49\textwidth}
    \centering
    \includegraphics[scale=0.35]{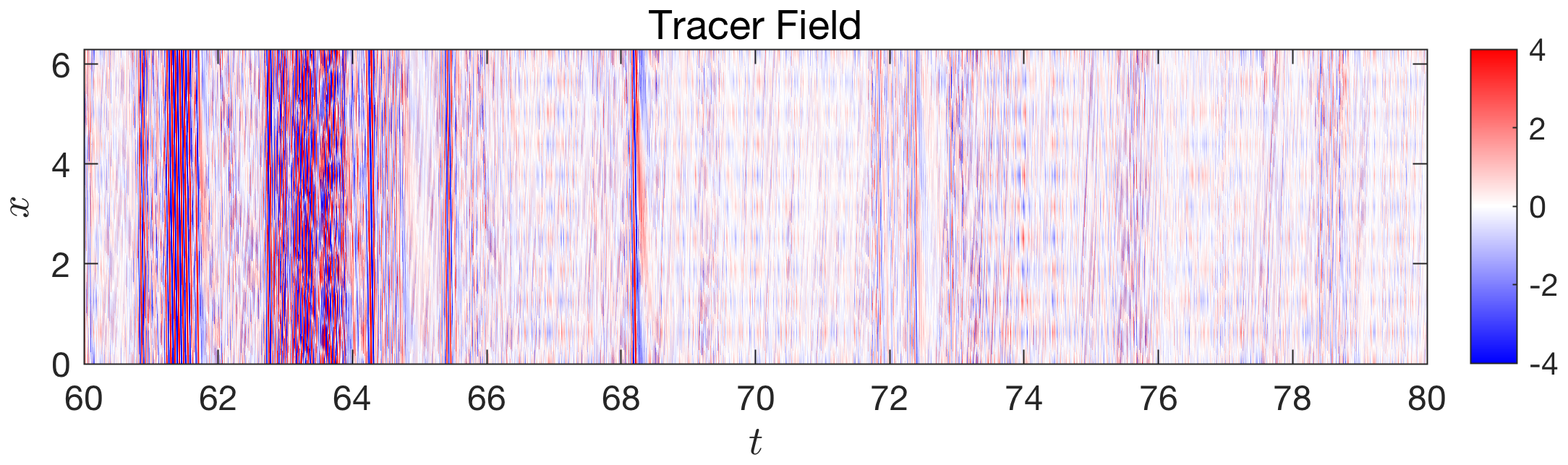}
    \end{minipage}%

    \vspace{0.5em}
    {\scriptsize\itshape Zonal Flow: Nonlinear, $f=0, B = 0$}
    \vspace{0.5em}

    \begin{minipage}{.49\textwidth}
    \centering
    \includegraphics[scale=0.35]{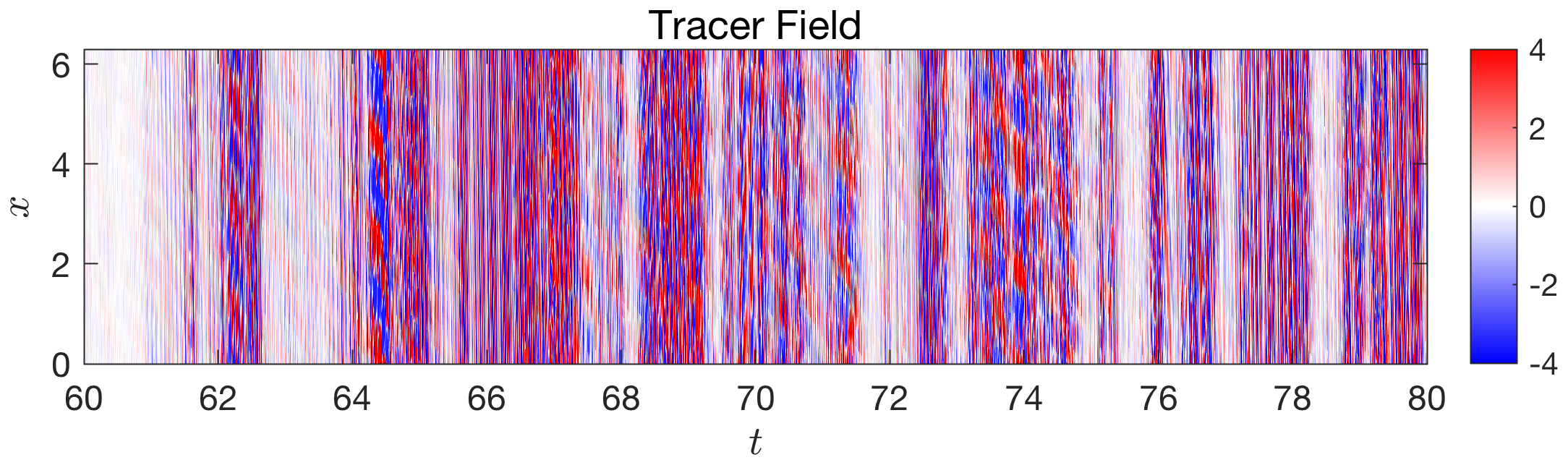}
    \end{minipage}%
    \begin{minipage}{.49\textwidth}
    \centering
    \includegraphics[scale=0.35]{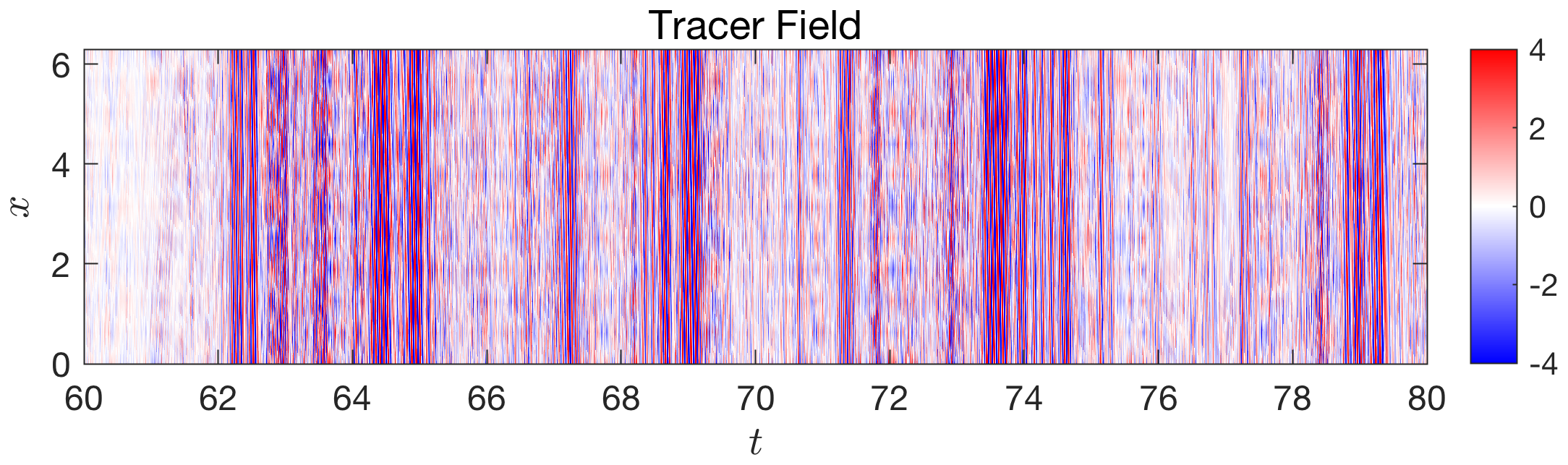}
    \end{minipage}%

    \vspace{0.5em}
    {\scriptsize\itshape Zonal Flow: Nonlinear, $f=0, B = 2.5$}
    \vspace{0.5em}

    \begin{minipage}{.49\textwidth}
    \centering
    \includegraphics[scale=0.35]{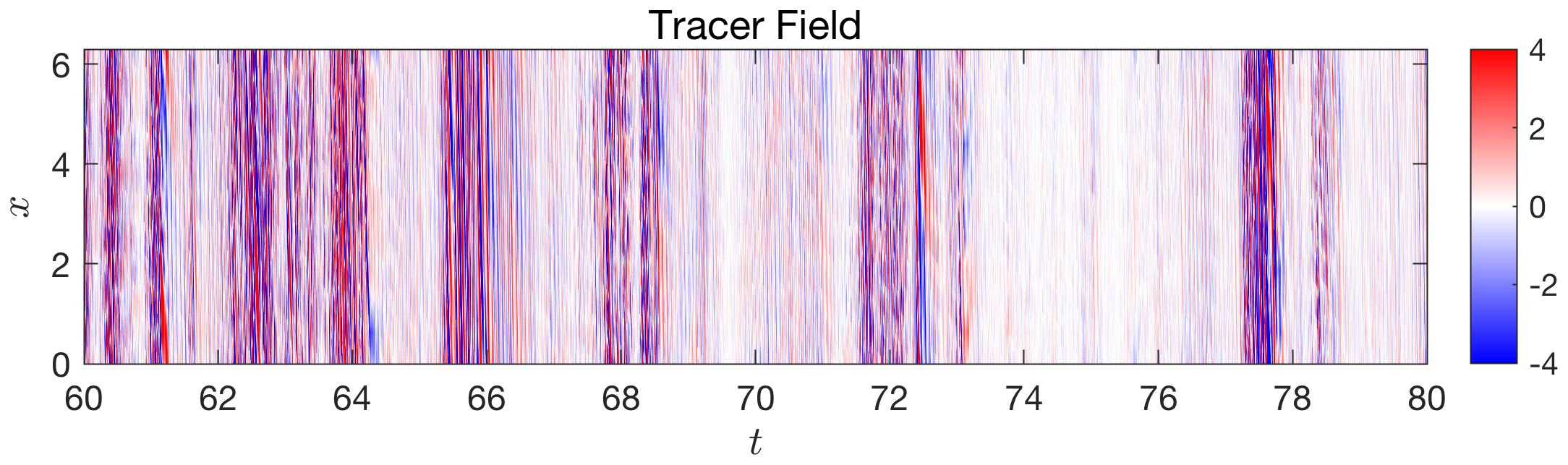}
    \end{minipage}%
    \begin{minipage}{.49\textwidth}
    \centering
    \includegraphics[scale=0.35]{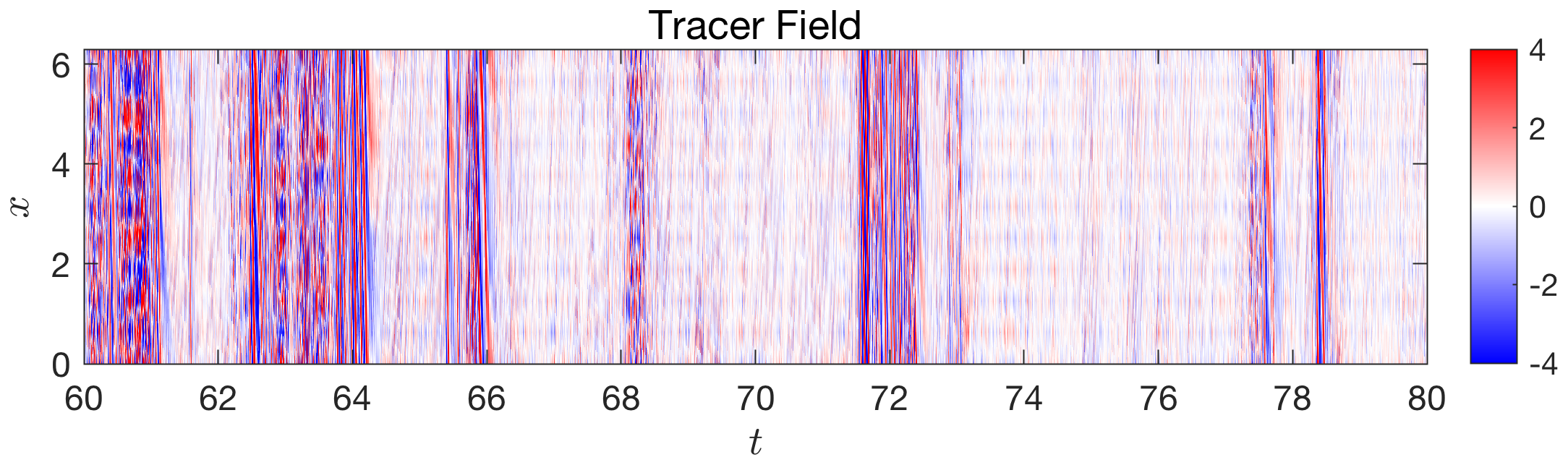}
    \end{minipage}%

    \vspace{0.5em}
    {\scriptsize\itshape Zonal Flow: Nonlinear, $f=1.0, B = 0$}
    \vspace{0.5em}

    \begin{minipage}{.49\textwidth}
    \centering
    \includegraphics[scale=0.35]{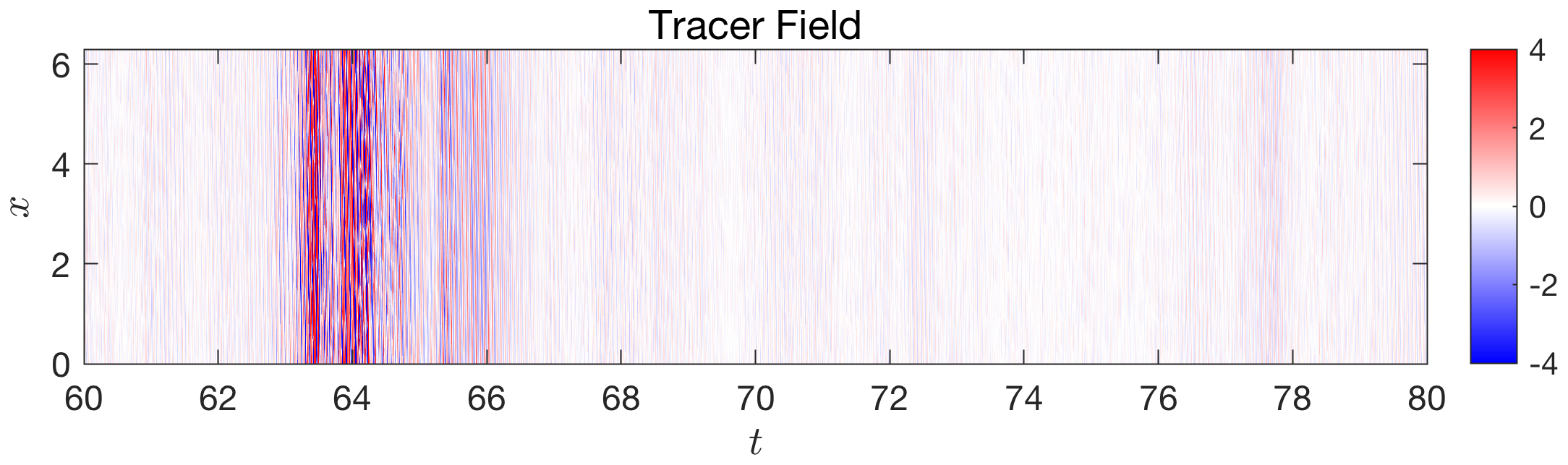}
    \end{minipage}%
    \begin{minipage}{.49\textwidth}
    \centering
    \includegraphics[scale=0.35]{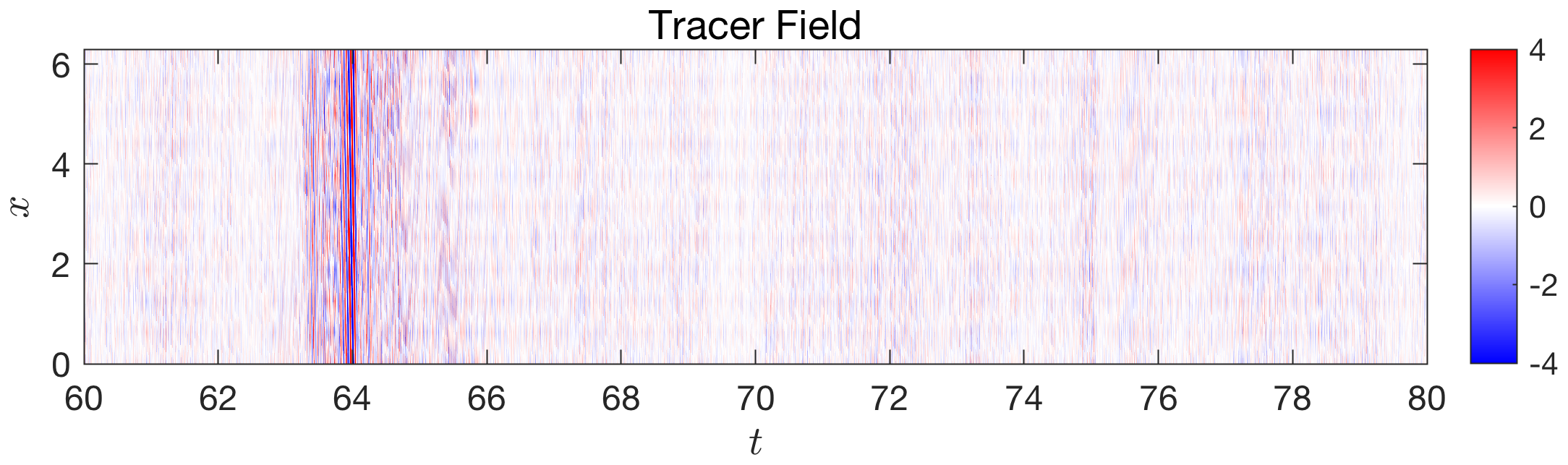}
    \end{minipage}%

    \vspace{0.5em}
    {\scriptsize\itshape Zonal Flow: Nonlinear, $f=1.0, B = 2.5$}
    \vspace{0.5em}

    \begin{minipage}{.49\textwidth}
    \centering
    \includegraphics[scale=0.35]{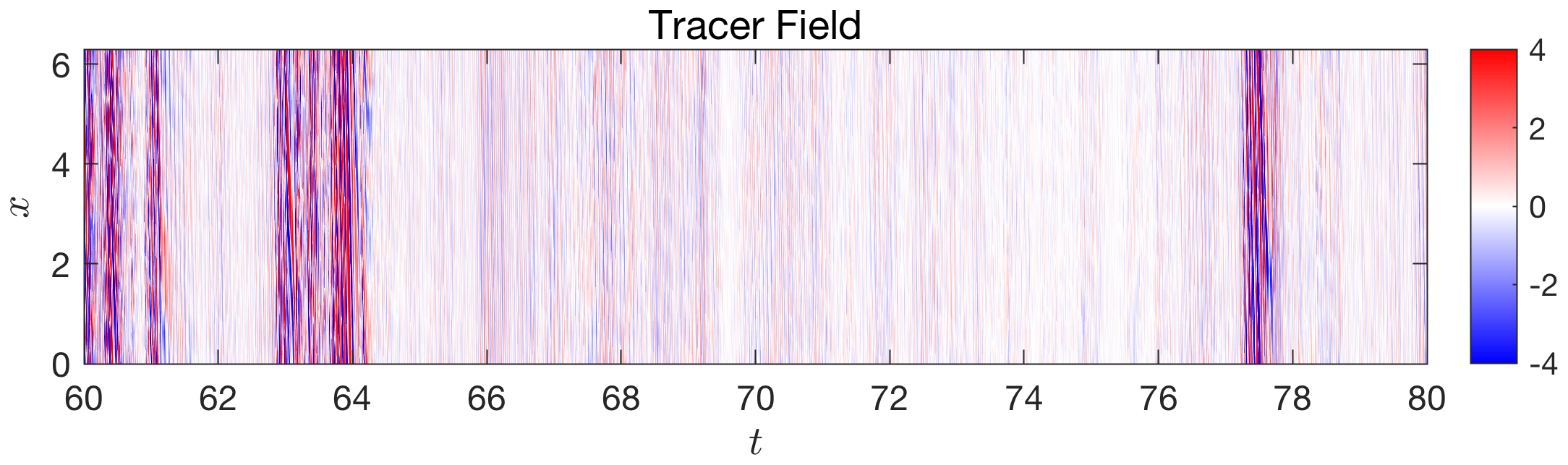}
    \end{minipage}%
    \begin{minipage}{.49\textwidth}
    \centering
    \includegraphics[scale=0.35]{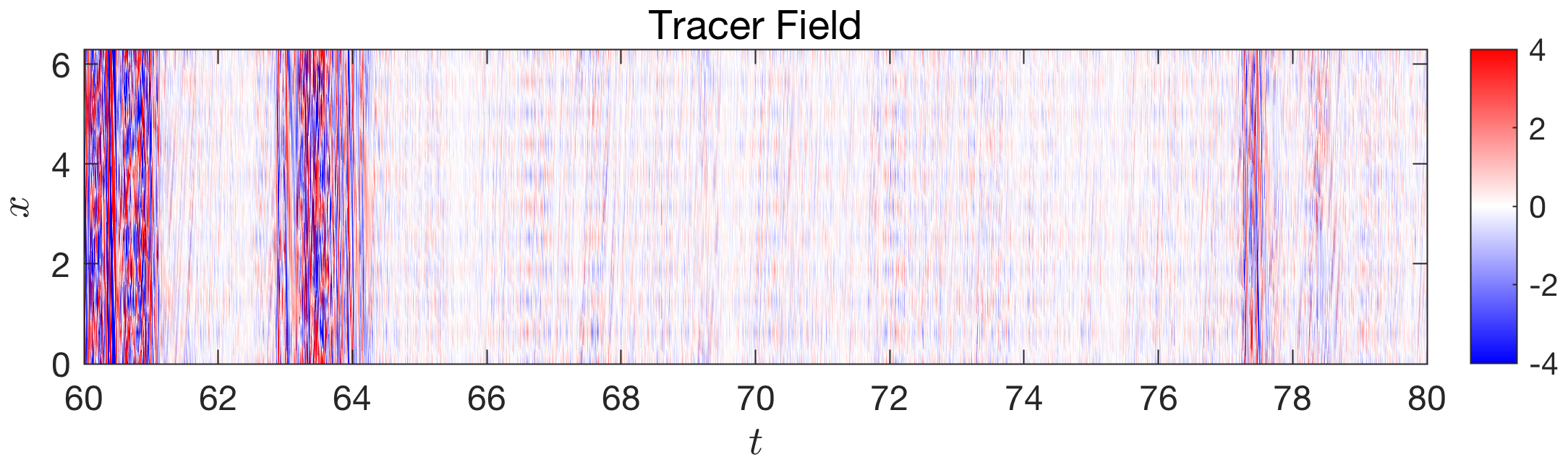}
    \end{minipage}%

    \caption{Comparison of spatio-temporal evolution of the tracer field under different shear flow models for {equipartition}.}
    \label{fig:set1_dispappendix_field}
\end{figure}

\begin{figure}[htbp!]
    \centering
    \begin{minipage}{.49\textwidth}
    \centering
    \subcaption{Advective Shear Flow}
    \end{minipage}%
    \begin{minipage}{.49\textwidth}
    \centering
    \subcaption{Dispersive (QG) Shear Flow}
    \end{minipage}%
    
    \vspace{0.5em}
    {\scriptsize\itshape Zonal Flow: Linear, $f=0$}
    \vspace{0.5em}

    \begin{minipage}{.49\textwidth}
    \centering
    \includegraphics[scale=0.35]{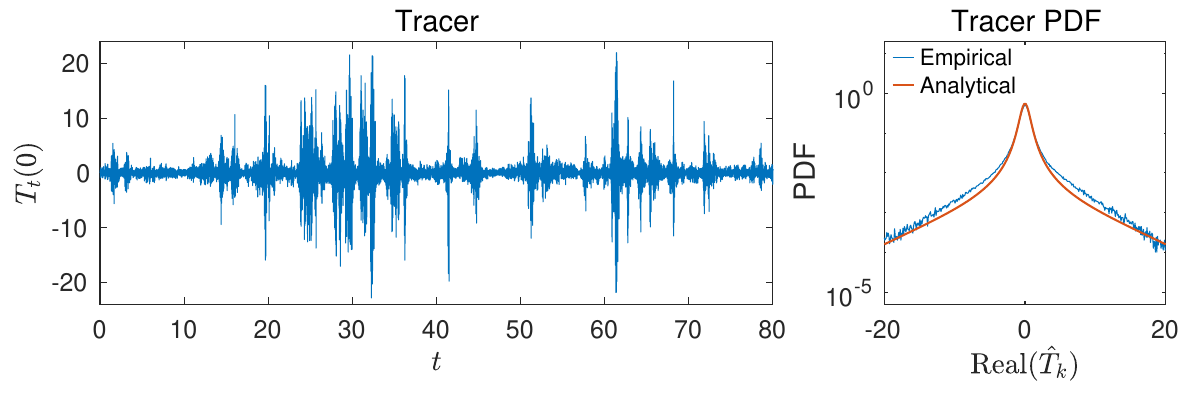}
    \end{minipage}%
    \begin{minipage}{.49\textwidth}
    \centering
    \includegraphics[scale=0.35]{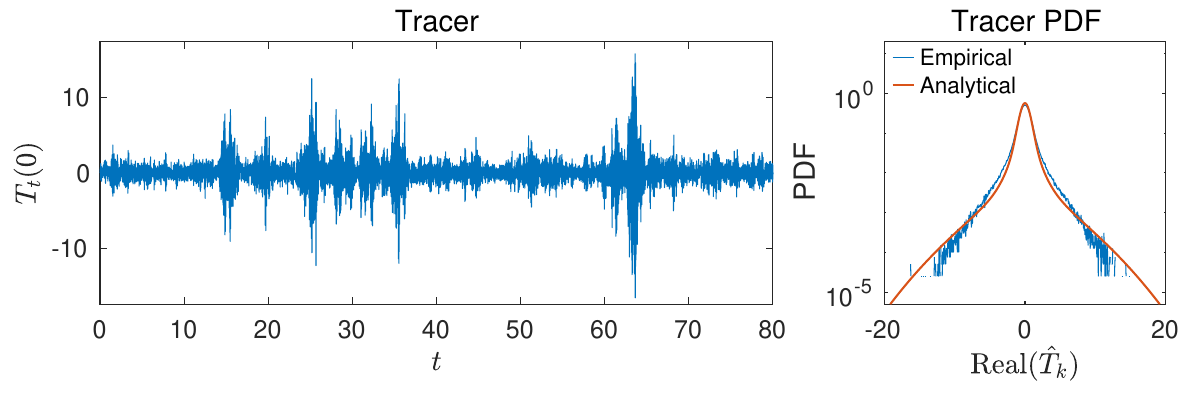}
    \end{minipage}%
   
    \vspace{0.5em}
    {\scriptsize\itshape Zonal Flow: Nonlinear, $f=0, B = 0$}
    \vspace{0.5em}
   
    \begin{minipage}{.49\textwidth}
    \centering
    \includegraphics[scale=0.35]{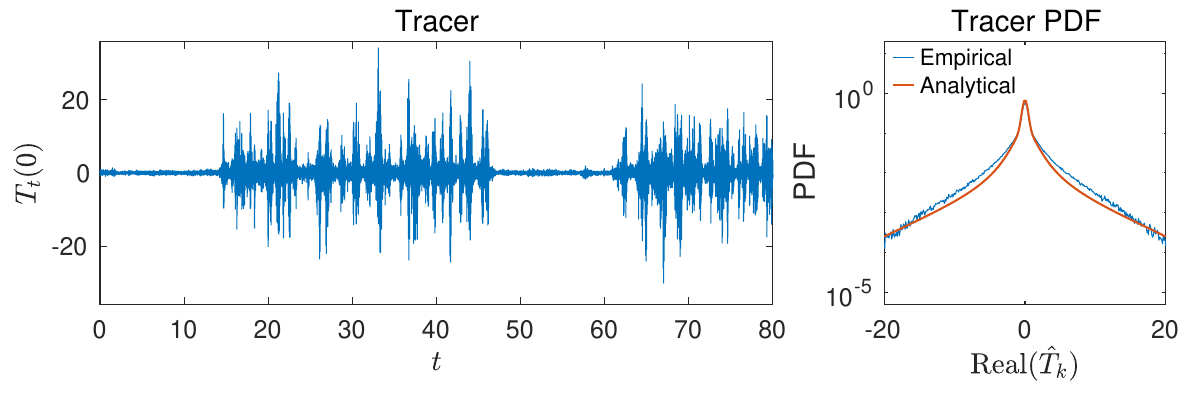}
    \end{minipage}%
    \begin{minipage}{.49\textwidth}
    \centering
    \includegraphics[scale=0.35]{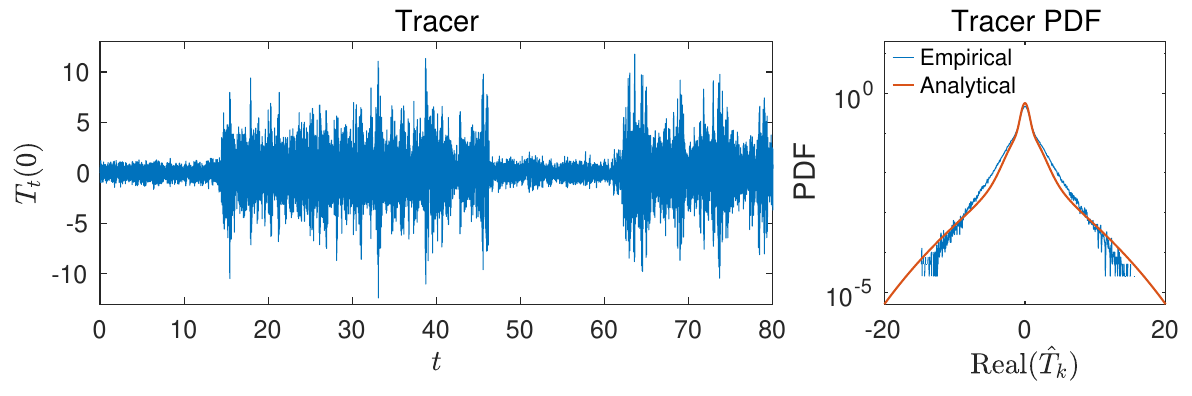}
    \end{minipage}%
   
    \vspace{0.5em}
    {\scriptsize\itshape Zonal Flow: Nonlinear, $f=0, B = 2.5$}
    \vspace{0.5em}
   
    \begin{minipage}{.49\textwidth}
    \centering
    \includegraphics[scale=0.35]{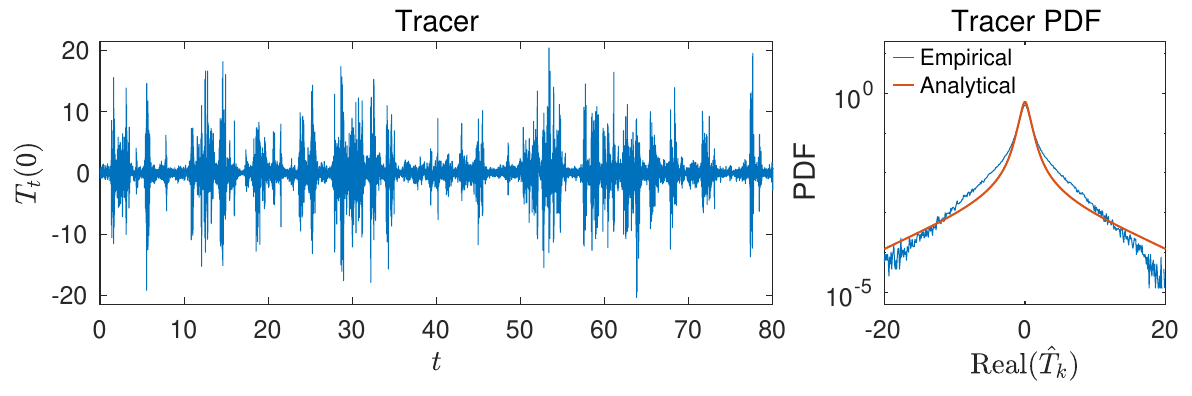}
    \end{minipage}%
    \begin{minipage}{.49\textwidth}
    \centering
    \includegraphics[scale=0.35]{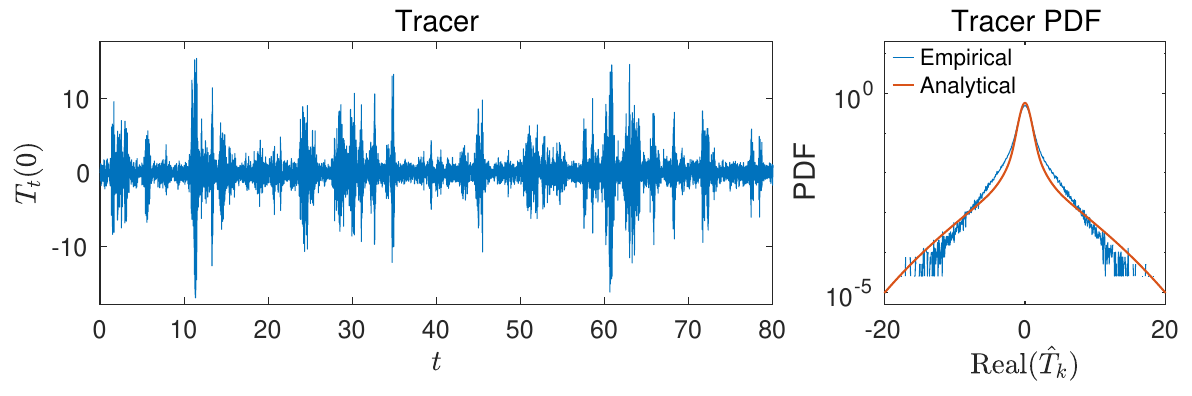}
    \end{minipage}%
    
    \vspace{0.5em}
    {\scriptsize\itshape Zonal Flow: Nonlinear, $f=1.0, B = 0$}
    \vspace{0.5em}

    \begin{minipage}{.49\textwidth}
    \centering
    \includegraphics[scale=0.35]{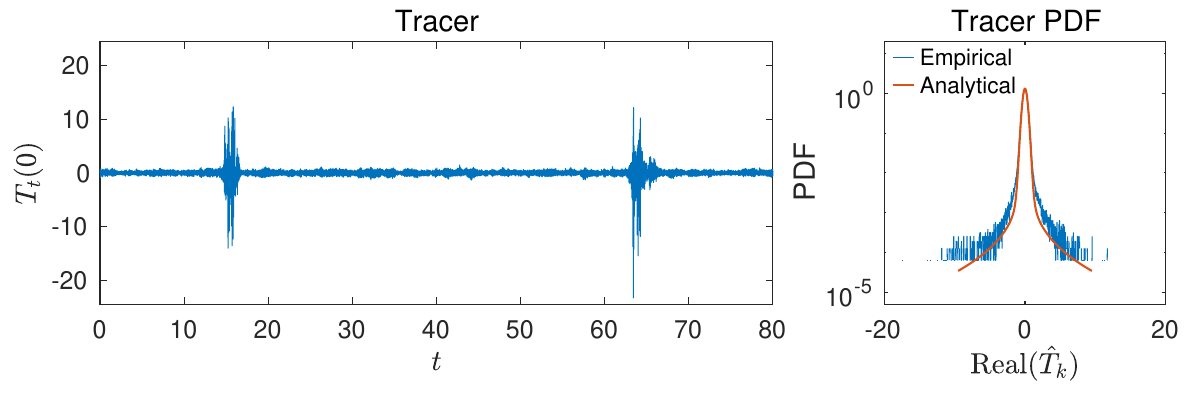}
    \end{minipage}%
    \begin{minipage}{.49\textwidth}
    \centering
    \includegraphics[scale=0.35]{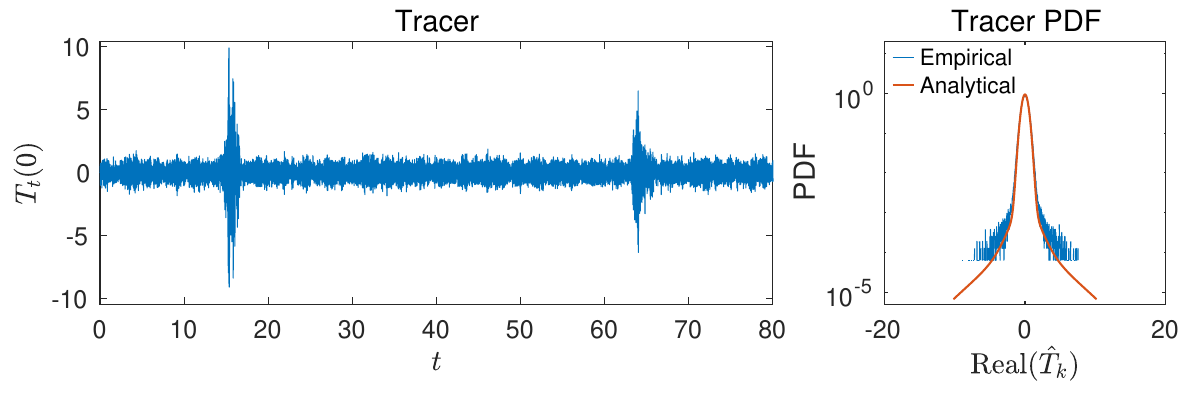}
    \end{minipage}%
    
    \vspace{0.5em}
    {\scriptsize\itshape Zonal Flow: Nonlinear, $f=1.0, B = 2.5$}
    \vspace{0.5em}
    
    \begin{minipage}{.49\textwidth}
    \centering
    \includegraphics[scale=0.35]{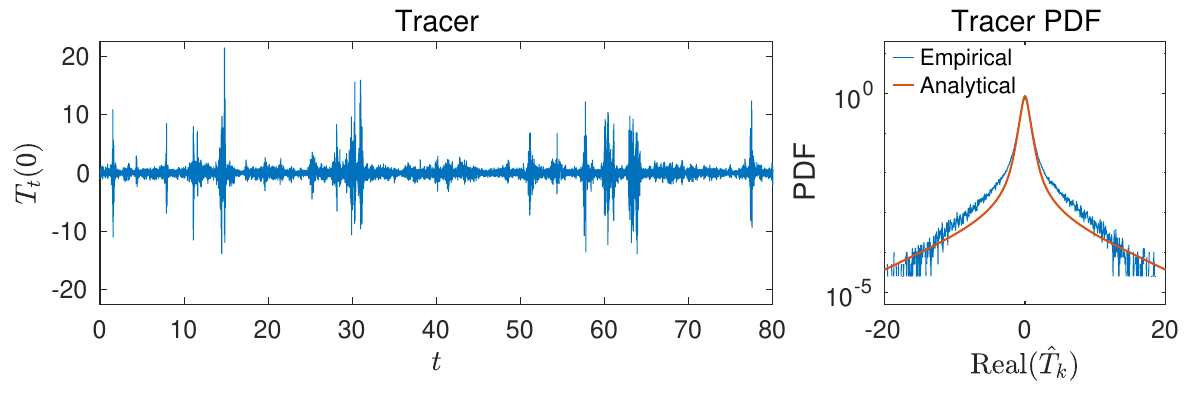}
    \end{minipage}%
    \begin{minipage}{.49\textwidth}
    \centering
    \includegraphics[scale=0.35]{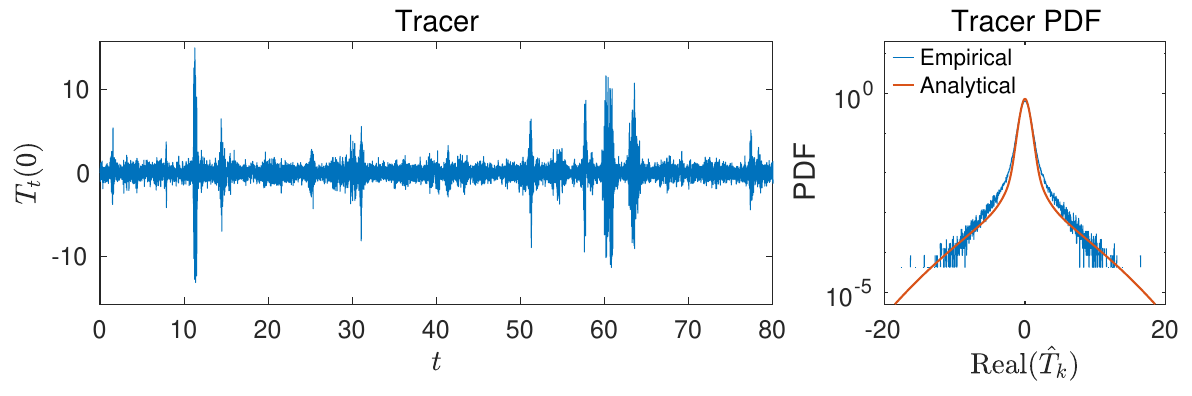}
    \end{minipage}

    \caption{Comparison of evolution of the tracer  at $T_t(0)$ and the stationary PDF under different shear flow models.}
    \label{fig:set1_dispappendix_field_pdf}
\end{figure}

\begin{figure}[htbp!]
    \centering
    \begin{minipage}{.49\textwidth}
    \centering
    \subcaption{Advective Shear Flow}
    \end{minipage}%
    \begin{minipage}{.49\textwidth}
    \centering
    \subcaption{Dispersive (QG) Shear Flow}
    \end{minipage}%
    
    \vspace{0.5em}
    {\scriptsize\itshape Zonal Flow: Linear, $f=0$}
    \vspace{0.5em}
    
    \begin{minipage}{.49\textwidth}
    \centering
    \includegraphics[scale=0.31]{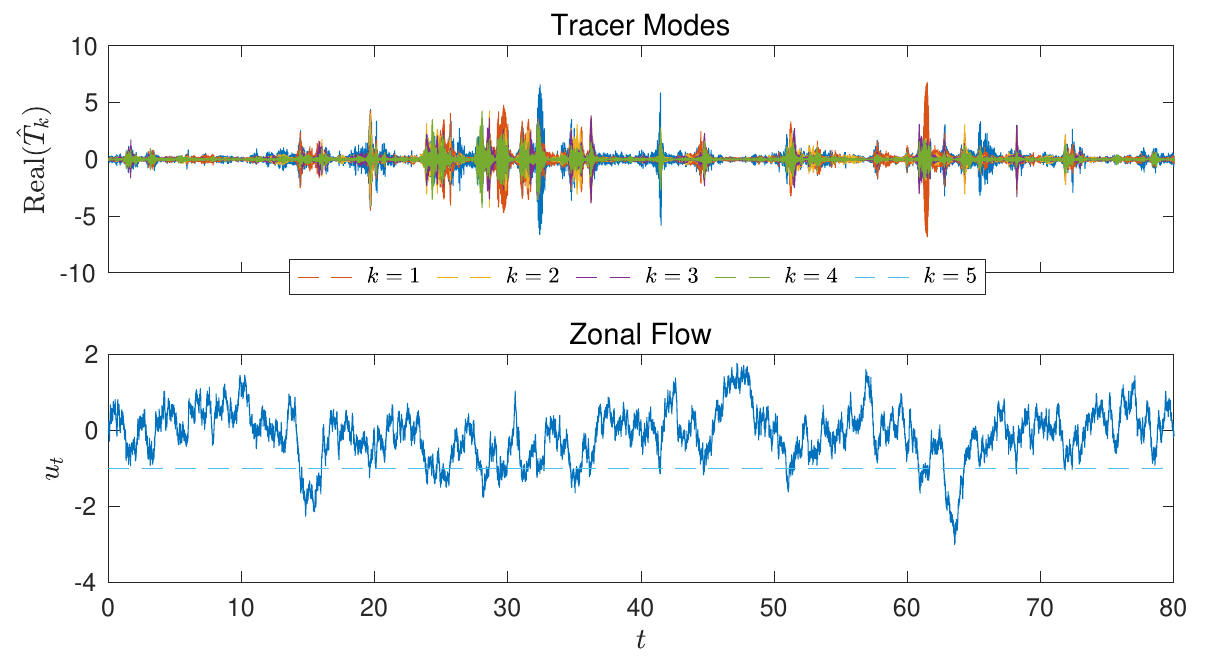}
    \end{minipage}%
    \begin{minipage}{.49\textwidth}
    \centering
    \includegraphics[scale=0.31]{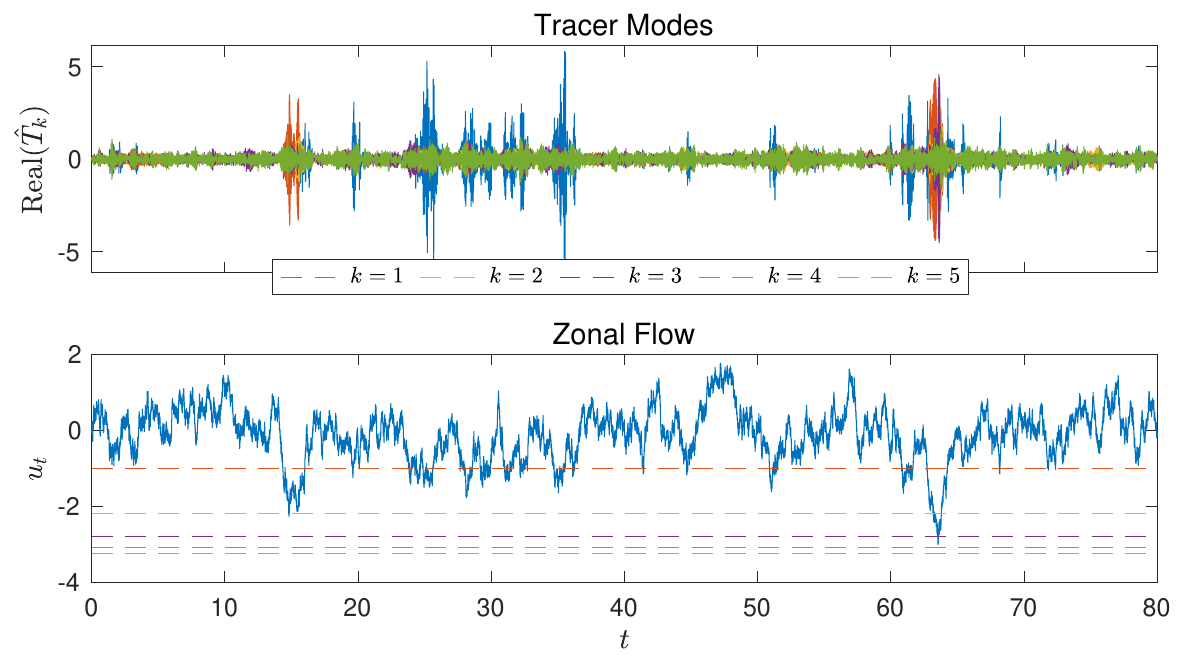}
    \end{minipage}%
    
    \vspace{0.5em}
    {\scriptsize\itshape Zonal Flow: Nonlinear, $f=0, B = 0$}
    \vspace{0.5em}
    
    \begin{minipage}{.49\textwidth}
    \centering
    \includegraphics[scale=0.31]{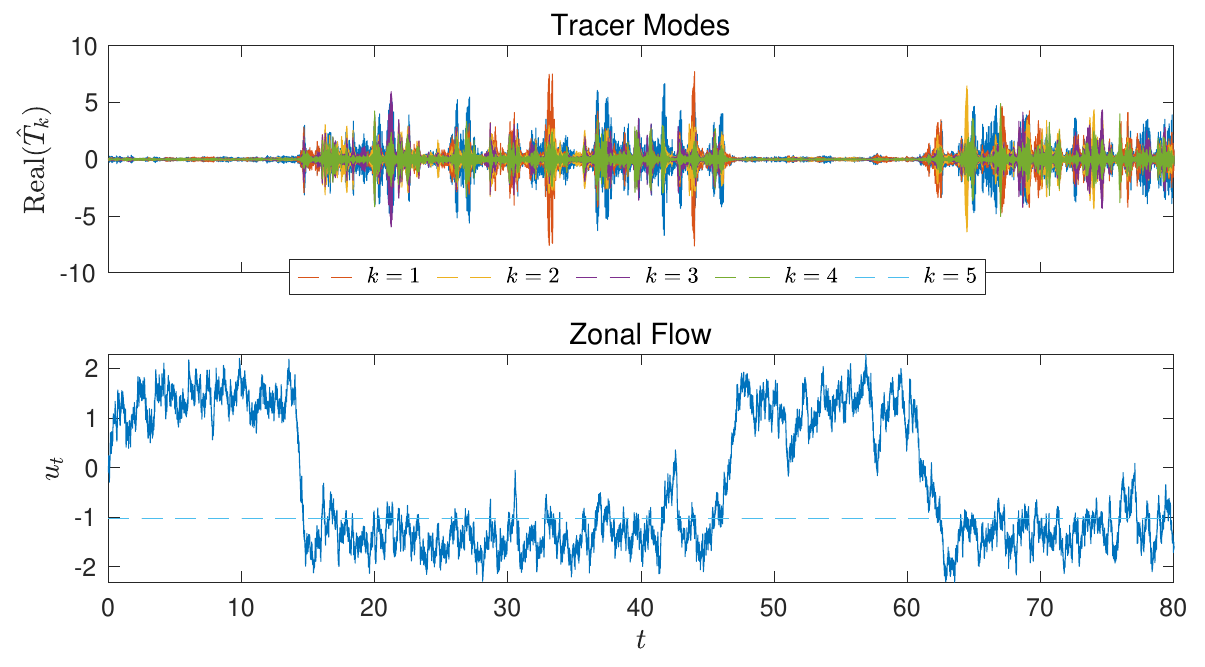}
    \end{minipage}%
    \begin{minipage}{.49\textwidth}
    \centering
    \includegraphics[scale=0.31]{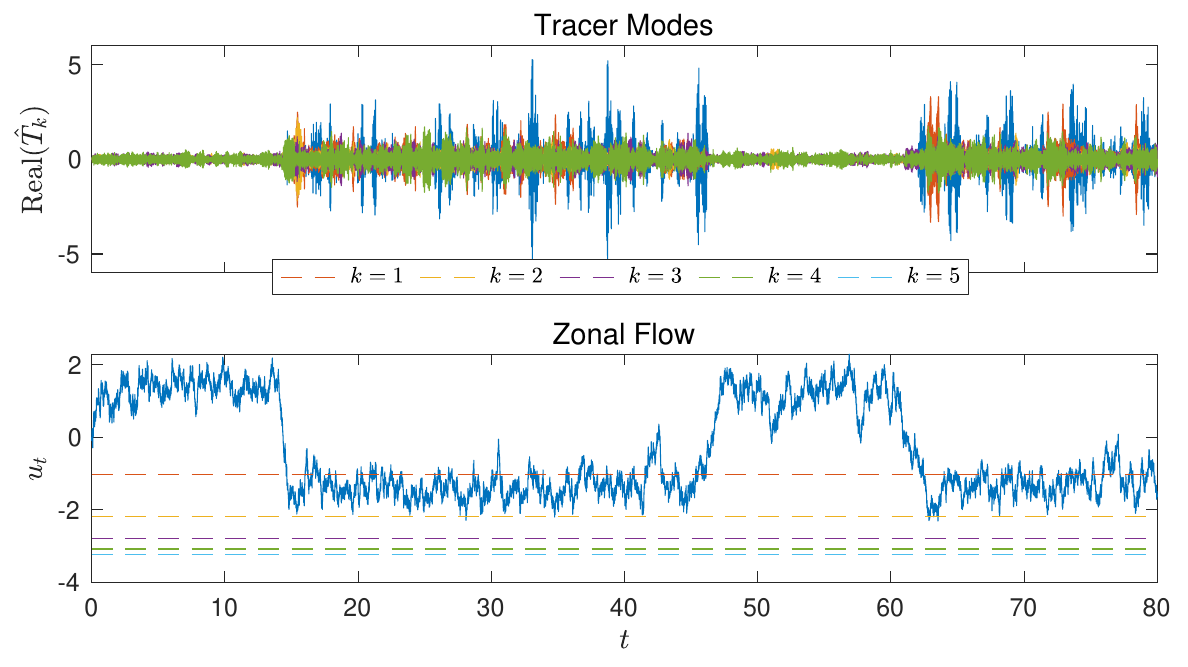}
    \end{minipage}%
    
    \vspace{0.5em}
    {\scriptsize\itshape Zonal Flow: Nonlinear, $f=0, B = 2.5$}
    \vspace{0.5em}
    
    \begin{minipage}{.49\textwidth}
    \centering
    \includegraphics[scale=0.31]{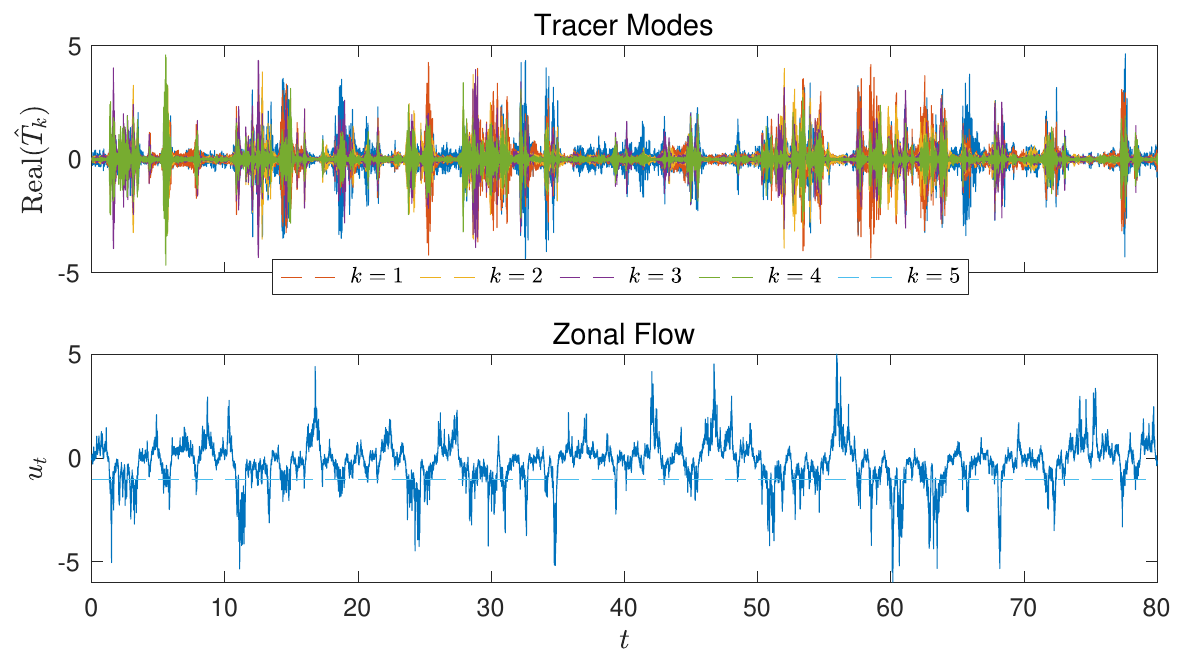}
    \end{minipage}%
    \begin{minipage}{.49\textwidth}
    \centering
    \includegraphics[scale=0.31]{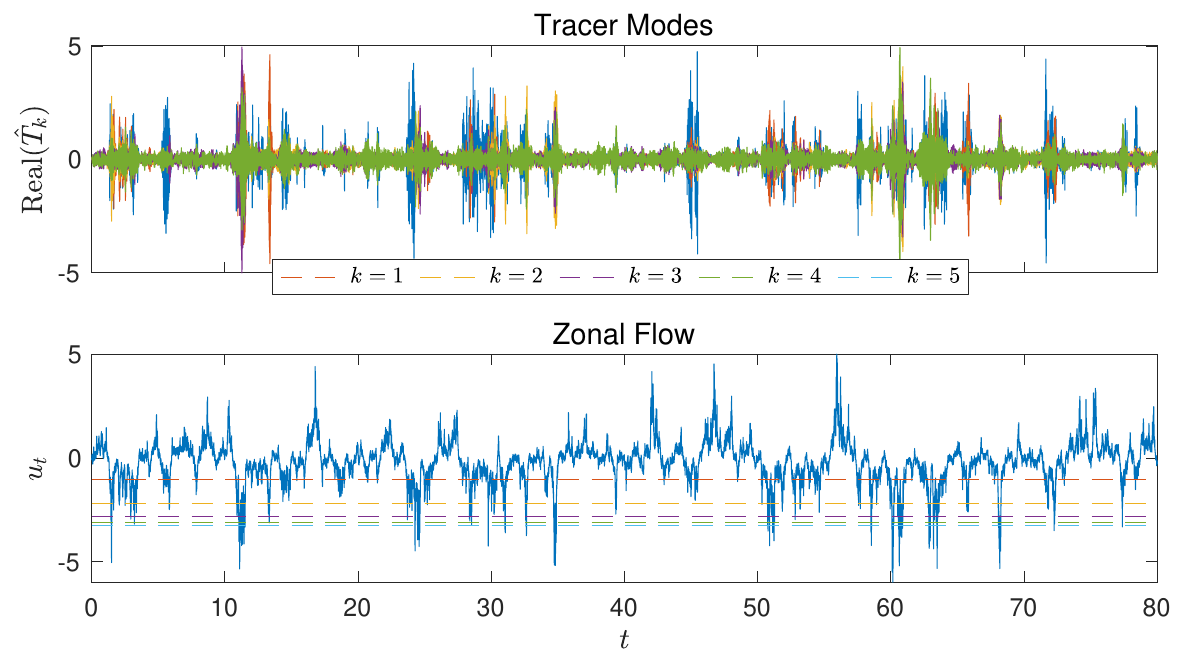}
    \end{minipage}%
    
    \vspace{0.5em}
    {\scriptsize\itshape Zonal Flow: Nonlinear, $f=1.0, B = 0$}
    \vspace{0.5em}
    
    \begin{minipage}{.49\textwidth}
    \centering
    \includegraphics[scale=0.31]{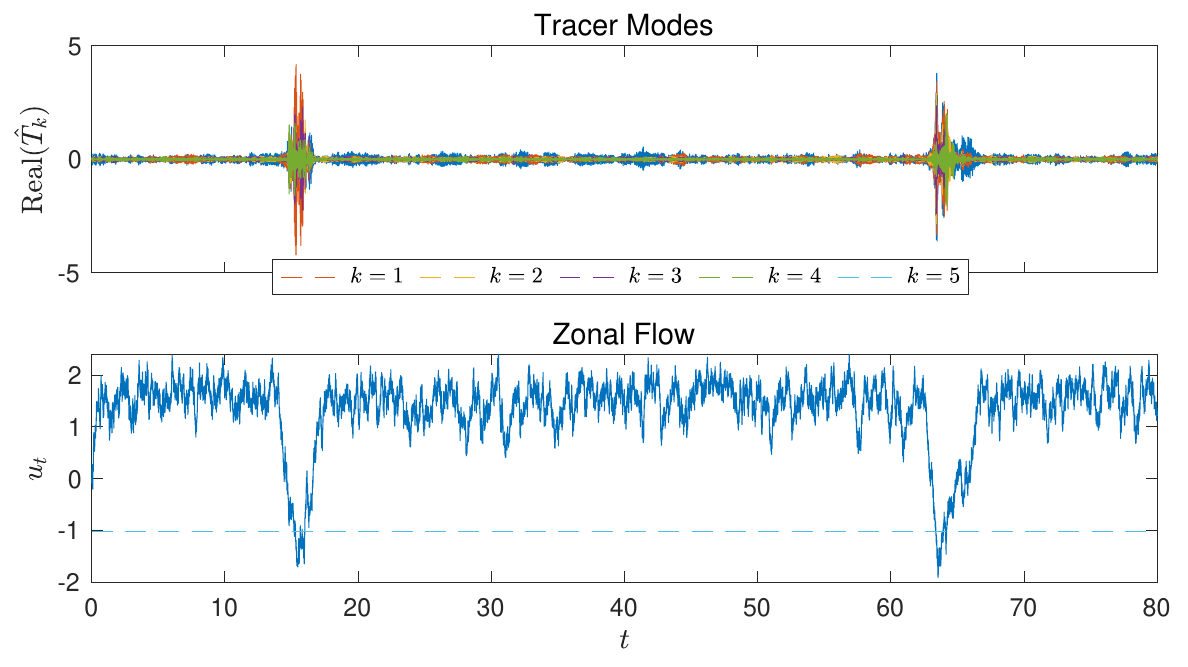}
    \end{minipage}%
    \begin{minipage}{.49\textwidth}
    \centering
    \includegraphics[scale=0.31]{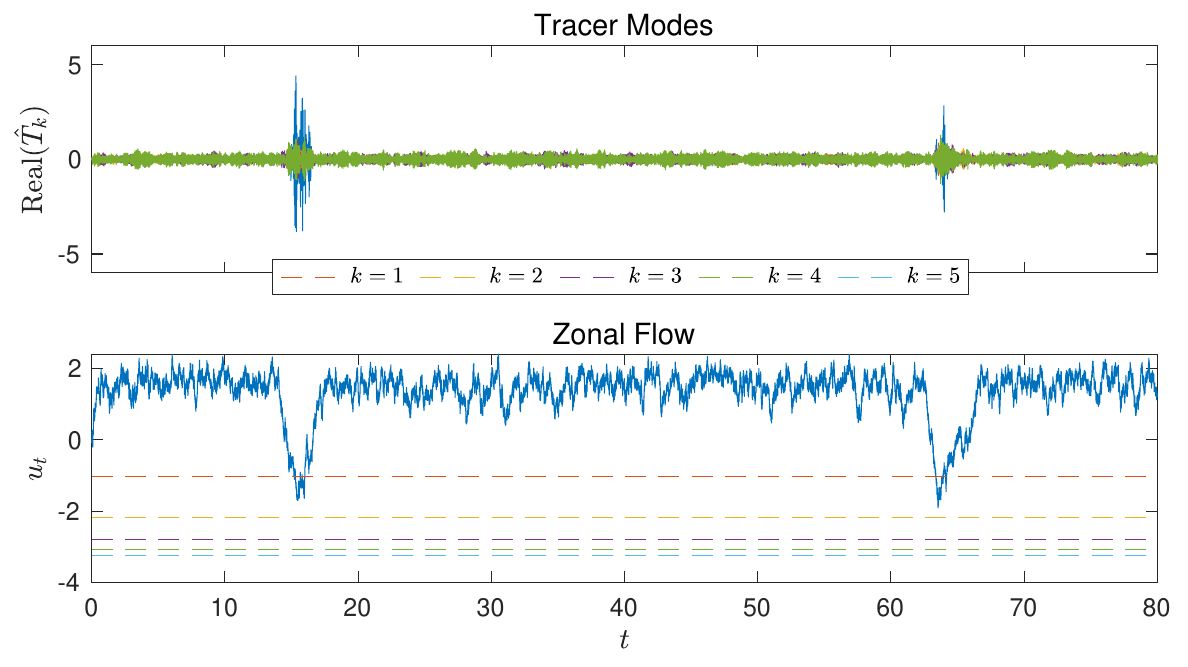}
    \end{minipage}%
    
    \vspace{0.5em}
    {\scriptsize\itshape Zonal Flow: Nonlinear, $f=1.0, B = 2.5$}
    \vspace{0.5em}

    \begin{minipage}{.49\textwidth}
    \centering
    \includegraphics[scale=0.31]{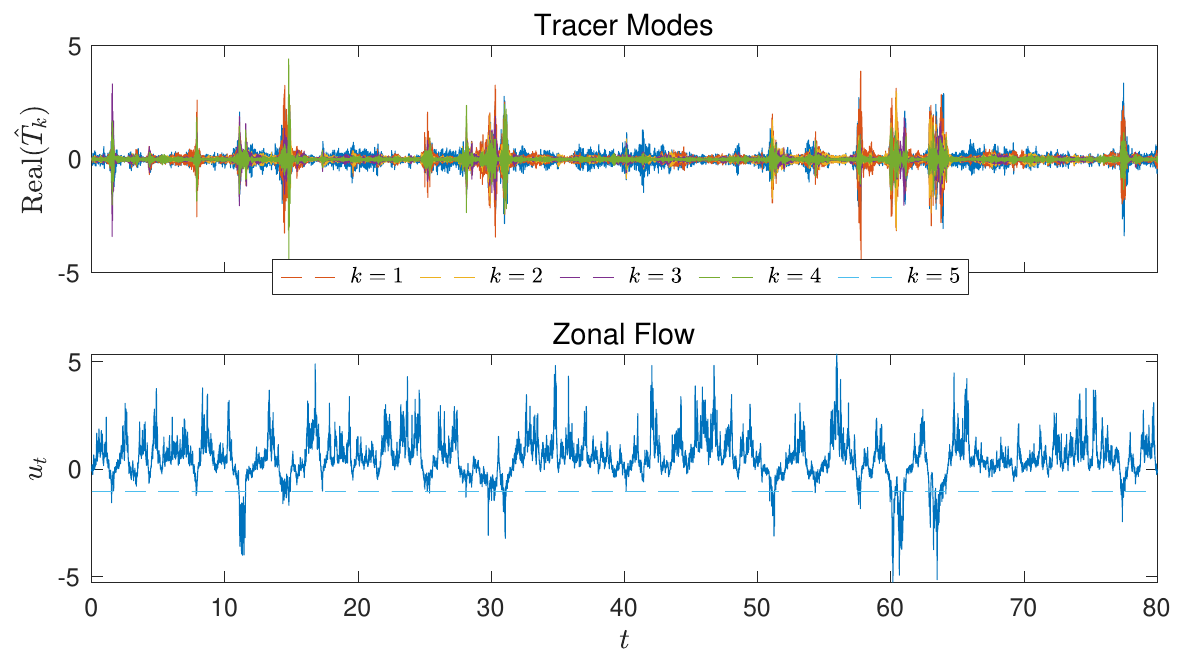}
    \end{minipage}%
    \begin{minipage}{.49\textwidth}
    \centering
    \includegraphics[scale=0.31]{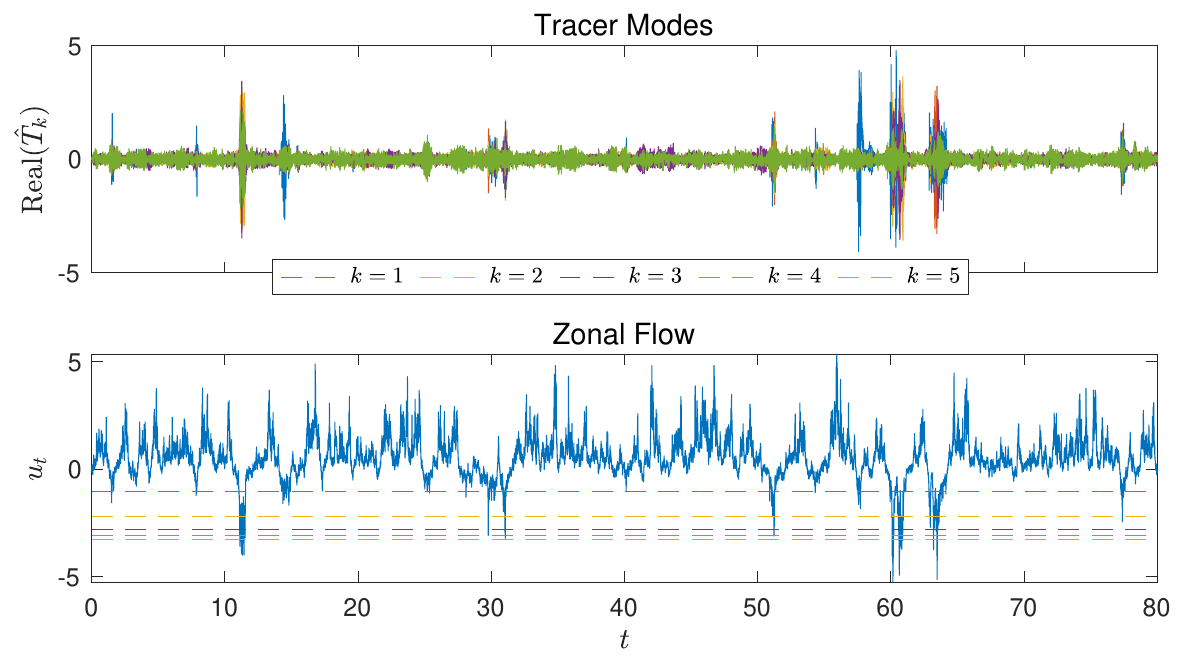}
    \end{minipage}
    \caption{Comparison of evolution of the tracer modes and zonal flow.}
    \label{fig:set1_dispappendix_modes}
\end{figure}

\Cref{fig:set1_dispappendix_field,fig:set1_dispappendix_field_pdf,fig:set1_dispappendix_modes}
 
\clearpage



\end{document}